\documentclass[fleqn,usenatbib]{mnras}

\usepackage{newtxtext,newtxmath}

\usepackage[T1]{fontenc}

\DeclareRobustCommand{\VAN}[3]{#2}
\let\VANthebibliography\thebibliography
\def\thebibliography{\DeclareRobustCommand{\VAN}[3]{##3}\VANthebibliography}

\usepackage[normalem]{ulem}
\usepackage[dvipsnames]{xcolor}
\usepackage{graphicx}	
\usepackage{subfig}
\usepackage{amsmath}	
\usepackage{amsmath,siunitx}
\usepackage{gensymb}
\usepackage{booktabs}   
\usepackage{multirow} 
\DeclareUnicodeCharacter{2212}{-}
\usepackage{enumitem}

\title[sGRB host galaxies]{A deep survey of short GRB host galaxies over $z\sim0-2$: implications for offsets, redshifts, and environments}

\author[O'Connor et al. (2021)]{
B. O'Connor$^{1,2,3,4}$\thanks{E-mail: oconnorb@gwu.edu}, E. Troja$^{4,5}$, S. Dichiara$^{6}$, P. Beniamini$^{7}$, S.~B. Cenko$^{4,8}$, C. Kouveliotou$^{1,2}$,  \newauthor 
J. Becerra Gonz\'{a}lez$^{9,10}$, J. Durbak$^{3,4}$, P. Gatkine$^{11}$, A. Kutyrev$^{3,4}$, T. Sakamoto$^{12}$,
\newauthor 
R. S\'{a}nchez-Ram\'{i}rez$^{13,14}$, S. Veilleux$^{3,8}$
\\ 
$^{1}$Department of Physics, The George Washington University, 725 21st Street NW, Washington, DC 20052, USA\\
$^{2}$Astronomy, Physics and Statistics Institute of Sciences (APSIS), The George Washington University, Washington, DC 20052, USA\\
$^{3}$Department of Astronomy, University of Maryland, College Park, MD 20742-4111, USA \\
$^{4}$Astrophysics Science 
Division, NASA Goddard Space Flight Center, 8800 Greenbelt Rd, Greenbelt, MD 20771, USA\\
$^{5}$University of Rome Tor Vergata, Department of Physics, via della Ricerca Scientifica 1, 00100, Rome, IT\\
$^{6}$Department of Astronomy and Astrophysics, The Pennsylvania State University, 525 Davey Lab, University Park, PA 16802, USA\\
$^{7}$Astrophysics Research Center of the Open University (ARCO), The Open University of Israel, P.O Box 808, Ra’anana 43537, Israel \\
$^{8}$Joint Space-Science Institute, University of Maryland, College Park, MD 20742 USA \\
$^{9}$Instituto de Astrof\'{i}sica de Canarias, E-38200 La Laguna, Tenerife, Spain\\
$^{10}$Universidad de La Laguna, Dpto. Astrof\'{i}sica, E-38206 La Laguna, Tenerife, Spain\\
$^{11}$Department of Astronomy, California Institute of Technology, Pasadena, CA, USA \\
$^{12}$Department of Physics and Mathematics, Aoyama Gakuin University, 5-10-1 Fuchinobe, Chuo-ku, Sagamihara-shi Kanagawa 252-5258, Japan \\
$^{13}$INAF-Istituto di Astrofisica e Planetologia Spaziali, via Fosso del Cavaliere, 100, I-00133 Rome RM, Italy\\
$^{14}$Instituto de Astrof\'isica de Andaluc\'ia (IAA-CSIC), Glorieta de la Astronom\'ia s/n, E-18008, Granada \\
}

\date{Accepted XXX. Received YYY; in original form ZZZ}

\pubyear{2021}

\begin{document}
\label{firstpage}
\pagerange{\pageref{firstpage}--\pageref{lastpage}}
\maketitle

\begin{abstract}

A significant fraction (30\%) of well-localized short gamma-ray bursts (sGRBs) lack a coincident host galaxy. This leads to two main scenarios: \textit{i}) that the progenitor system merged outside of the visible light of its host, or \textit{ii}) that the sGRB resided within a faint and distant galaxy that was not detected by follow-up observations.  Discriminating between these scenarios has important implications for constraining the formation channels of neutron star mergers, the rate and environments of gravitational wave sources, and the production of heavy elements in the Universe. In this work, we present the results of our observing campaign targeted at 31 sGRBs that lack a putative host galaxy. Our study effectively doubles the sample of well-studied sGRB host galaxies, now totaling 72 events of which $28\%$ lack a coincident host to deep limits ($r$\,$\gtrsim$\,$26$ or $F110W$\,$\gtrsim$\,$27$ AB mag), and represents the largest homogeneously selected catalog of sGRB offsets to date. We find that 70\% of sub-arcsecond localized sGRBs occur within 10 kpc of their host's nucleus, with a median projected physical offset of $5.6$ kpc. Using this larger population, we discover an apparent redshift evolution in their locations: bursts at low-$z$ occur at $2\times$ larger offsets compared to those at $z$\,$>$\,$0.5$. This evolution could be due to a physical evolution of the host galaxies themselves or a bias against faint high-$z$ galaxies. 
Furthermore, we discover a sample of hostless sGRBs at $z$\,$\gtrsim$\,$1$ that are indicative of a larger high-$z$ population, constraining the redshift distribution and disfavoring log-normal delay time models.

\end{abstract}

\begin{keywords}
gamma-ray bursts -- transients: neutron star mergers -- stars: jets 
\end{keywords}



\section{Introduction}
\label{sec: introduction}


Short duration gamma-ray bursts (sGRBs) are bright, brief flashes of gamma rays \citep[$<$\,$2$ s;][]{Chryssa93} produced by the coalescence of two compact objects \citep[][]{Eichler1989,Narayan1992}, either a binary neutron star system (BNS; \citealt{Ruffert1999, Rosswog2003}) 
or a neutron star and a black hole (NS-BH; \citealt{Faber2006,Shibata2011}). Beginning in the era of the \textit{Neil Gehrels Swift Observatory} (subsequently \textit{Swift}; \citealt{Gehrels04}), sGRBs were, for the first time, localized to arcsecond accuracy based on the detection of their X-ray afterglows \citep{Gehrels2005,Barthelmy2005}, and shortly thereafter, their optical afterglows \citep{Fox2005,Hjorth2005sgrb,Villasenor2005}. These accurate localization's allowed for the identification of their host galaxies, and, in turn, their redshifts. Nevertheless, $\sim$\,$20$\,$-$\,$30\%$ of sub-arcsecond localized sGRBs are classified as hostless, hereafter \textit{observationally} hostless, due to their lack of a coincident galaxy to deep limits ($\gtrsim$\,26 mag; \citealt{Stratta2007,Perley2009,Rowlinson2010,Berger2010a,FongBerger2013,Tunnicliffe2014}) or multiple galaxies with a similar probability of chance coincidence \citep{Bloom2002,Berger2010a}.

Although these events lack a coincident galaxy, a number of low-$z$ candidate hosts have been identified at large physical offsets (out to $\sim$\,$75$ kpc; \citealt{Bloom2007,Stratta2007,Troja2008,Rowlinson2010,Berger2010a}) localizing the sGRBs to well outside of the galaxy's light and potentially in tenuous (low density) environments. 
Furthermore, some events with secure host associations have been discovered within the outskirts of their galaxies at $>$\,$15$ kpc from their host's nucleus \citep{DAvanzo2009,Rowlinson2010,Troja2019,Lamb2019}, while others are found at $<$\,$1$ kpc \citep{Antonelli2009,DAvanzo2009,Levesque2010,Troja2016,OConnor2021}. The diverse environments of sGRBs could be an indicator of multiple progenitor formation channels within the observed population: \textit{i}) a primordial (isolated) formation  channel \citep{PortegiesZwart1998,Voss2003,OShaughnessy2005,Belczynski2006,Belczynski2008,Tauris2017,Abbott2017kick,Kruckow2018,VignaGomez2018,Zevin2019}, \textit{ii)} dynamical formation in a globular cluster \citep{Phinney1991,Davies1995,Grindlay2006,Hopman2006,Salvaterra2008,Guetta2009,Salvaterra2010,Lee2010,Church2011,Bae2014,Andrews2019a,Adhikari2020,Ye2020,Stegmann2021}, or \textit{iii}) even formation in a galaxy cluster environment \citep{Niino2008,Salvaterra2010}. Thus, identifying events formed through these multiple channels impacts our understanding of stellar formation and evolution and provides useful insight for population synthesis studies.

In the primordial formation channel, these large offsets are expected due to a change in velocity (a natal kick) imparted to the system, following mass ejection from the second supernova explosion \citep{Lyne1994,Hansen1997,Bloom1999,Fryer1999,Wex2000,Hobbs2005,Belczynski2006}. Combined with the long merger delay times ($10^7$\,$-$\,$10^{11}$ yr) predicted for BNS systems \citep{Zheng2007,Zemp2009}, a large natal kick can allow the binary to reach substantial distances and even escape its birth galaxy. However, a binary escaping its galaxy, denoted as \textit{physically} hostless, is theorized to occur in an extremely low density ($n$\,$<$\,$10^{-4}$ cm$^{-3}$) intergalactic medium (IGM) environment, making detection of an afterglow unlikely \citep[][]{Panaitescu2001,Salvaterra2010,Duque2019}. Moreover, by studying their early X-ray afterglow lightcurves, \citet{OConnor2020} found that $\lesssim$\,$16\%$ of sGRBs are consistent with such low densities, including only a single observationally hostless event (GRB 080503; \citealt{Perley2009}). Nevertheless, this does not exclude sGRBs with large offsets from having occurred within the halo's of their host galaxies or within a dense globular cluster environment \citep{Salvaterra2010}.


An alternative explanation for observationally hostless bursts is that these sGRBs occurred in faint, undetected host galaxies at higher redshifts (i.e., $z$\,$\gtrsim$\,$1$\,$-$\,$2$; \citealt{Berger2010a,Tunnicliffe2014}). Such high-$z$ events 
suggest progenitors that formed through a primordial channel with short merger delay times \citep[e.g.,][]{Andrews2019b,BeniaminiPiran2019}, indicating that BNS systems may have formed early enough to pollute the early Universe with heavy metals  \citep{Ji2016a,Ji2016b,Roederer2016,Hansen2017,Safarzadeh2017,Beniamini2018,Safarzadeh2019,Zevin2019}. 
Furthermore, our understanding of the environments and formation channels of sGRBs has fundamental implications for inferring the rate of detectable gravitational wave (GW) sources and for the follow-up of their electromagnetic (EM) counterparts, as the quick localization of the EM counterpart depends on inferences (such as, e.g., stellar mass, star formation rate, offset) from the known population of sGRB host galaxies \citep[][]{Nissanke2013,Gehrels2016,Arcavi2017,Artale2020gw,Ducoin2020} and on targeted searches using catalogs of nearby galaxies \citep[][]{White2011,Dalya2016,Cook2019}.

Disentangling between the different scenarios is observationally challenging. Due to the faintness of sGRB afterglows, redshift measurements from afterglow spectroscopy are rarely successful \citep[e.g.,][]{deUgarte2014,AguiFernandez2021}.
Therefore, deep imaging and spectroscopic observations from the most sensitive telescopes are required to identify the GRB host galaxy and estimate its distance scale. In this work, we targeted a sample of 31 sGRBs that lack a putative host galaxy with large-aperture telescopes to search for faint, coincident galaxies. Our facilities include: the Lowell Discovery Telescope (LDT), the Keck Observatory, the Gemini Observatory, the Gran Telescopio Canarias (GTC), the Very Large Telescope (VLT), and the \textit{Hubble Space Telescope} (\textit{HST}).

The paper is outlined as follows. In \S \ref{sec: observations}, we define our sample selection criteria, and the optical and near-infrared (nIR) imaging analysis techniques used in this work. In \S \ref{sec: methods}, we describe the methods employed to detect, localize, and compute photometry of the host galaxies, as well as the probabilistic criteria used for host assignment. In \S \ref{sec: Results}, we present the results and discuss the demographics of sGRB offsets, host galaxies, and environments. We present a discussion of these results in \S \ref{sec: discussion} and conclude in \S \ref{sec: conclusions}.  We present a detailed summary of the individual events analyzed in this work in Appendix \ref{sec: appendixsampleanalysis}.

We adopt the standard $\Lambda$-CDM cosmology with parameters $H_0=67.4$, $\Omega_M=0.315$, and $\Omega_\Lambda=0.685$ \citep{Planck2018}. 
 All confidence intervals are at the $1\sigma$ level and upper limits at the $3\sigma$ level, unless otherwise stated. 
 All reported magnitudes are in the AB system, and are corrected for Galactic extinction \citep{Schlafly2011}.
 Throughout the paper we adopt the convention $F_\nu\propto t^{-\alpha}\nu^{-\beta}$.

\section{Observations and Analysis}
\label{sec: observations}

\subsection{Sample selection}
\label{sec: sampleselection}


The association of a GRB with a host galaxy relies on the accurate localization of its afterglow. 
Therefore, we consider the sample of short GRBs detected with \textit{Swift} and localized by the X-ray Telescope \citep[XRT;][]{Burrows2005} to arcsecond accuracy. 
We include both GRBs with a short duration\footnote{\url{https://swift.gsfc.nasa.gov/results/batgrbcat/}}, defined as $T_{90}$\,$<$\,$2$ s \citep{Chryssa93}, and GRBs with a temporally extended emission (hereafter sGRBEE), as defined by \citet{Norris2006}.

\textit{GRB classification --} As of May 2021, the \textit{Swift} Burst Alert Telescope \citep[BAT;][]{Barthelmy2005} has detected 127 short duration GRBs of which 91 (72\%) have an X-ray afterglow localization. These X-ray localized events form the basis of our sample.  
Short duration bursts with soft spectra (i.e., a hardness ratio  $S_{50-100\,\textrm{keV}}/S_{25-50\,\textrm{keV}}$\,$<$\,$1$, where $S$ represents the gamma-ray fluence in a given energy range;  \citealt{Lien2016}) or non-negligible spectral lag \citep{Norris2006}  were flagged as ``possibly short'' (see, e.g., \citealt{Lien2016}) as some of these events may be produced by collapsar progenitors (see, e.g., GRB 040924, \citealt{Huang2005,Soderberg2006sn,Wiersema2008}; and 200826A, \citealt{Ahumada21,Rossi2021,Zhang2021}). 
In addition, we include sGRBEEs and candidate sGRBEEs identified by \citet{Lien2016}, \citet{Dichiara2021}, and GCN Circulars. 
We note that a classification as sGRBEE can be highly subjective due to the fact that they share properties of both short hard bursts and long GRBs (see, e.g., GRBs 060614, \citealt{DellaValle2006,Gehrels2006,Gal-Yam2006}; and 211211A, \citealt{Troja2022,Rastinejad2022, Gompertz2022, Yang2022}). One example is GRB~170728B which displays a short pulse ($<$\,2 s) followed by visibly extended emission ($T_{90}$\,$=$\,$48\pm27$ s). However, the spectrum of the initial short pulse is quite soft with $E_\textrm{peak}$\,$\sim$\,$80-175$ keV. 
Not having any additional information on, e.g., the spectral lag, host galaxy, or supernova, we label GRB~170728B a candidate sGRBEE. 

Other events which display the characteristic features of sGRBEE, such as a spectrally hard initial pulse with negligible spectral lag \citep{Norris2006}, can be more confidently assigned to this class. 
In total, we identify 32 sGRBEE (including 18 candidate sGRBEE\footnote{GRBs 051210, 120804A and 181123B satisfy $T_{90}$\,$<$\,2 s but also display evidence for extended emission, see \citet{Dichiara2021} for details. We therefore include these in the sample of candidate sGRBEEs.}) of which 29 (90\%) have an X-ray localization. Therefore, our initial sample totals 159 events which are either classical sGRBs ($T_{90}$\,$<$\,2 s) or sGRBEEs.


\begin{figure}
\includegraphics[width =\columnwidth, trim= 0 0 0 35, clip]{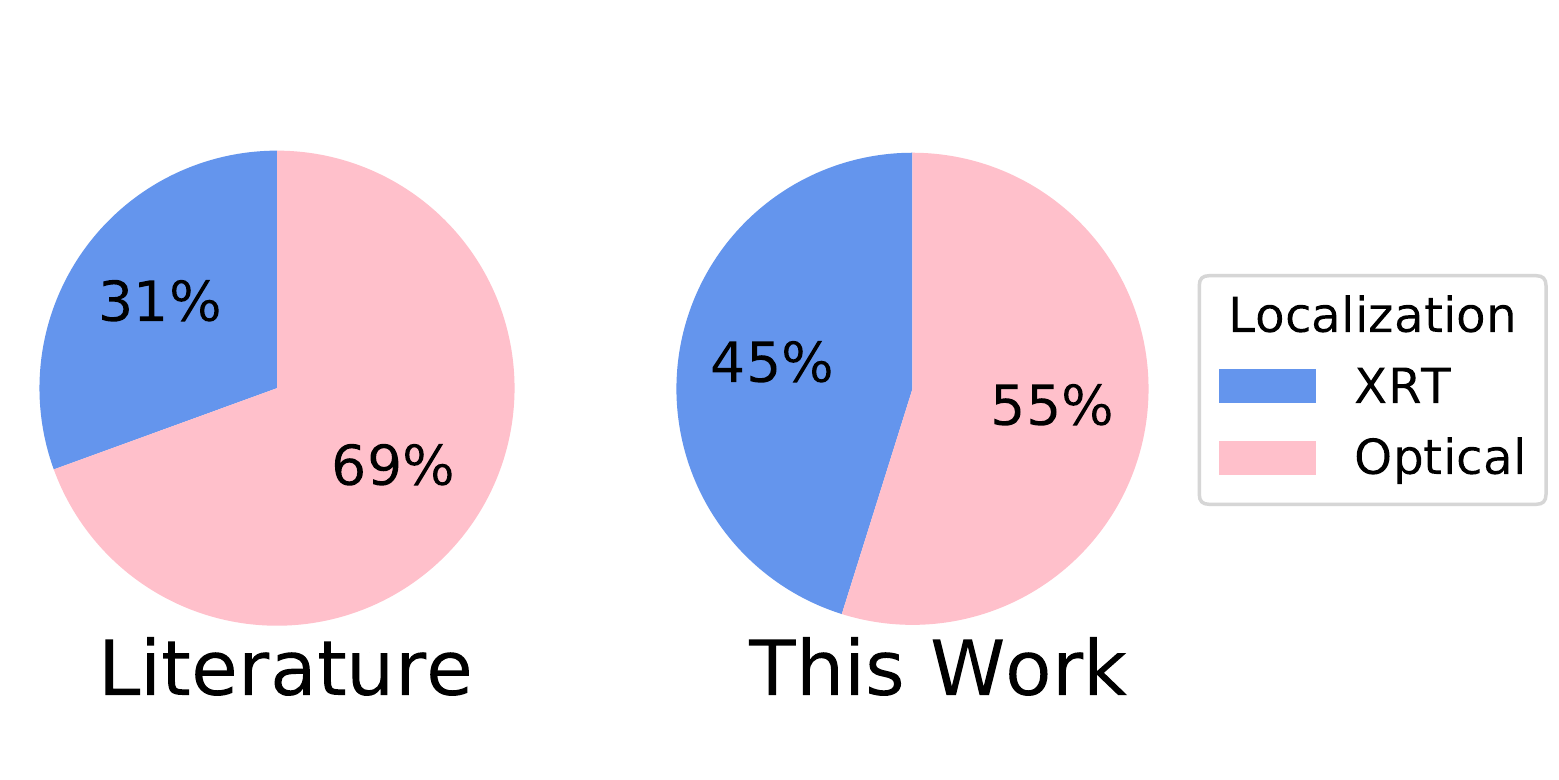}
\vspace{-8mm}
\caption{The distribution of short GRB localization methods between X-ray and optical for the sample of 31 events analyzed in this work and the sample of 36 events in \citet{Fong2013}. 
}
\label{fig: localization}
\end{figure}

\begin{figure}
\includegraphics[width =\columnwidth, trim= 0 0 0 25, clip]{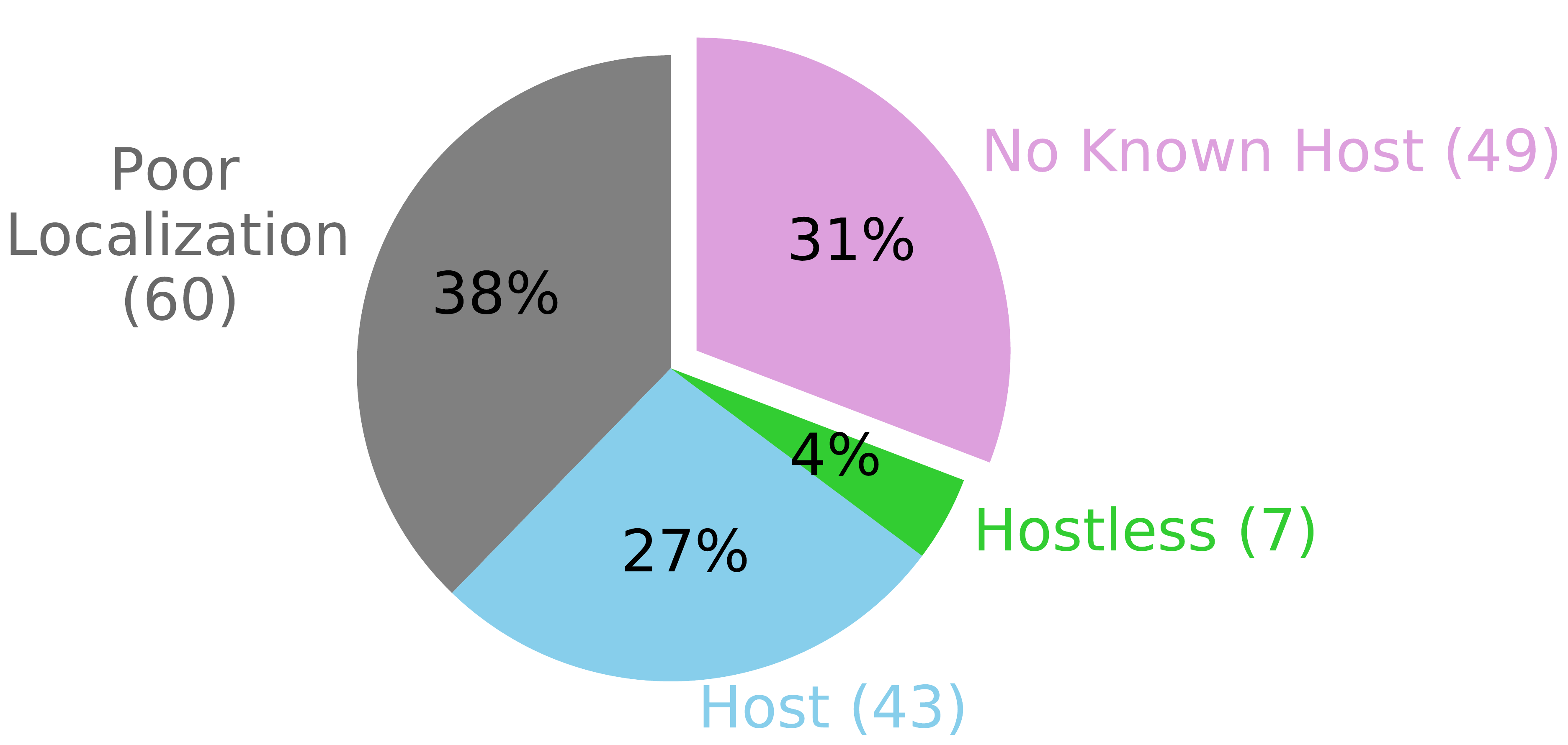}
\caption{Breakdown of host classification for the \textit{Swift}/BAT sample of 159 short GRBs used in this work: GRBs with a published host galaxy in the literature are shown in blue, those classified as hostless are shown in green, and those with no published host galaxy are displayed in purple. Poorly localized short GRBs, such as those with only BAT detections or a large positional uncertainty based on their afterglow $\sigma_\textrm{AG}$\,$>$\,$4\arcsec$, are shown in gray, and are excluded from the sample compiled in this work.
} \label{fig: toplevel}
\end{figure}

\textit{GRB localization --} 
Past searches for the host galaxies of short GRBs \citep[e.g.,][]{Prochaska2006,DAvanzo2009,Berger2010a,FongBerger2013,Tunnicliffe2014} mainly focused on optically localized events with sub-arcsecond positions (Figure \ref{fig: localization}). 
However, an optically selected sample is potentially subject to multiple observing biases, which can affect the observed redshift and offset distributions. An optical position disfavors small offsets from the host's nucleus \citep[e.g.,][]{OConnor2021} as the afterglow light can be masked by the glare of the host galaxy, especially in the case of faint short GRB afterglows or dusty environments. In addition it may disfavor events occurring in the low-density environments expected for large-offset GRBs
\citep{Panaitescu2001,Salvaterra2010,Duque2019,OConnor2020}.

In order to mitigate potential biases due to an optical selection of the sample, we included all XRT localized events within our follow-up campaign.
Although XRT positions typically have larger uncertainties than optical, radio, or \textit{Chandra} localizations, XRT localized bursts contribute valuable information to the demographics of sGRB host galaxies in terms of redshift, stellar mass, star formation rate, and galaxy type \citep[e.g.,][]{Gehrels2005,Bloom2006}. Hereafter, we consider only the 120 events with at least an X-ray localization, of which 49 ($\sim$40\%) also have an optical localization. 

\textit{Selection criteria --}  
We adopt two additional criteria to build a homogeneous sample of bursts. 
The first is that the uncertainty on the GRB's localization is $<$\,$4\arcsec$ (90\% confidence level, hereafter CL) as bursts with a poorer localization can only be securely associated to bright ($r$\,$\lesssim$\,$21$ mag) galaxies
and would not benefit from a campaign of deep optical imaging. 
This requirement excludes 13 XRT localized events from our sample\footnote{These are: GRBs 050509B, 060502B, 061210 (EE), 090621B, 100206A, 100628A, 130313A, 140320A, 140611A, 150301A, 150728A, 161104A, and 170524A.}.
We further impose a limit of $A_V$\,$<$\,$1.5$ mag \citep{Schlafly2011} on the Galactic extinction along the GRB sightline in order to eliminate regions where host galaxy searches would be less sensitive\footnote{This condition excluded GRBs 050724A (EE), 080426A, 080702A, 081024A, 150101A, 180402A, 200907B, and 201006A.}. This cut allows us to remove crowded regions along the Galactic plane ($|b|$\,$<$\,$15^\circ$) where our search would not be meaningful due to chance alignment with foreground stars.

Among the remaining 99 short GRBs matching our criteria (see Figure \ref{fig: toplevel}), 43 are associated to a host galaxy, 7 are classified as hostless based on deep ground-based and \textit{HST} imaging \citep[see, e.g.,][]{Berger2010a,FongBerger2013}, and 49 more events lack evidence of an underlying host galaxy based on the initial ground-based follow-up reported through GCN circulars. The latter group of bursts is the focus of our study. Deep late-time imaging is crucial to determine whether the lack of a candidate host galaxy is due to the shallow depth of the initial ground-based follow-up, a high redshift, or a large angular separation due, for example, to a high natal kick velocity imparted to the progenitor.

\begin{figure}
\includegraphics[width = \columnwidth]{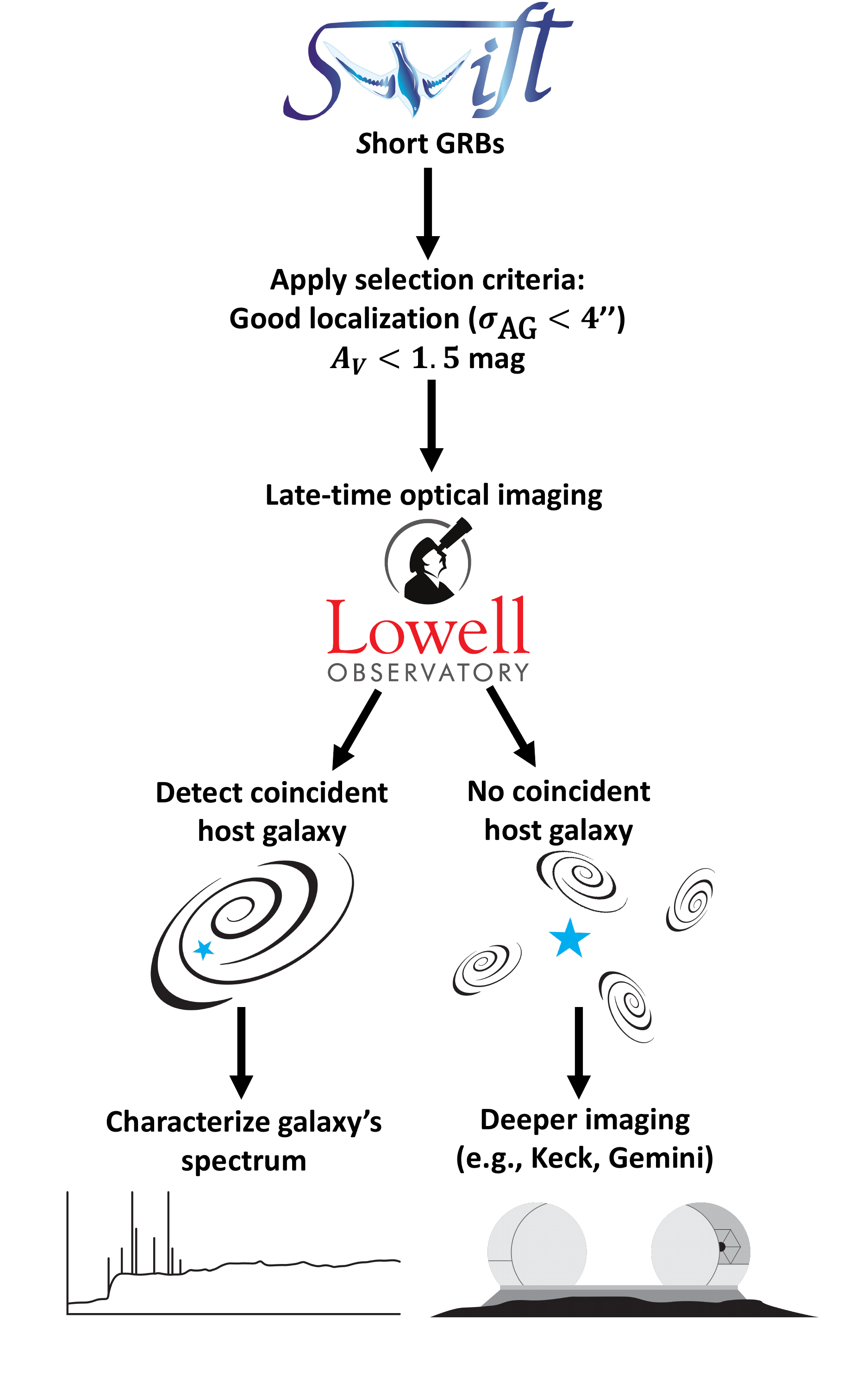}
\caption{An outline of the candidate selection process, and follow-up methodology employed in this work in order to locate and identify the host galaxies of short GRBs. Optical spectroscopy was carried out if the candidate host galaxy was brighter than $21$\,$-$\,$22$ mag, otherwise multi-color imaging was obtained in order to derive a photometric redshift.
} \label{fig: outline}
\end{figure}


\textit{Observing strategy --} As a first step (see Figure \ref{fig: outline}), we targeted these bursts with the 4.3-m LDT (PIs: Troja, Cenko,  Gatkine, Dichiara) and performed deep optical imaging, typically in $r$-band, to search for an underlying host galaxy to depth $r$\,$\gtrsim$\,$25$ mag. 
In the case of a detection, we scheduled the target for multi-color imaging in order to characterize the galaxy's spectral energy distribution (SED) and, if the galaxy's candidate was brighter than $\approx$\,$21$\,$-$\,$22$ AB mag, for optical spectroscopy in order to measure its redshift. In total, 30 out of 46 short GRBs (65\% of the sample) were followed-up with the LDT from 2014 to 2021 through our programs. Those events which were not observed by LDT were either only visible from the Southern Hemisphere or already had limits comparable to LDT's typical depth ($r$\,$\sim$\,$24.5$\,$-$\,$25$ mag). 
In all other cases, we flagged the burst for further deep imaging with large-aperture telescopes. We targeted these sGRBs as part of our programs on the twin 8.1-m Gemini telescopes (PI: Troja) and the 10-m Keck-I telescope (PI: Cenko) to search for host galaxies to deeper limits ($r$\,$\gtrsim$\,$26$\,$-$\,$28$ AB mag).  
These observations were further complemented with public archival data from the 10.4-m GTC, the Keck Observatory, the Gemini Observatory, and \textit{HST}. 

The final sample of events observed through these programs comprises 31 sGRBs (see Table \ref{tab: observations}) discovered between 2009 to 2020 (14 of which have only an XRT localization). Of these 31 events, about 20\% display extended emission. 
When compared to previous studies of sGRB host galaxies, which included 36 sGRBs discovered between 2005 to 2013 \citep[e.g.,][]{Fong2013}, our program doubles the sample of well-studied sGRB environments. A table of the X-ray and gamma-ray properties of sGRBs in our sample is shown in Table \ref{tab: XrayAGprop}.


\subsection{Optical/nIR Imaging}
\label{sec:imaging description}

Due to the isotropic distribution of GRBs on the sky and the multi-year nature of this project, the optical and near-infrared imaging obtained for our sample is heterogeneous and spans a range of observatories, filters, and exposure times. 
These observations were typically taken months to years after the explosion when contamination from the GRB afterglow is negligible.
The majority of our optical observations were carried out by the Large Monolithic Imager (LMI) on the LDT, the Gemini Multi-Object Spectographs \citep[GMOS;][]{Hook2004} on both Gemini North (GMOS-N) and Gemini South (GMOS-S), the Low Resolution Imaging Spectrometer \citep[LRIS;][]{Oke1995} at the Keck Observatory, and the Optical System for Imaging and low-Intermediate-Resolution Integrated Spectroscopy \citep[OSIRIS;][]{Cepa2000} at the GTC. We also include publicly available near-infrared observations obtained with the \textit{HST} Wide Field Camera 3 (WFC3).
A log of observations presented in this work is reported in Table \ref{tab: observations}.

\subsubsection{Lowell Discovery Telescope (LDT)}
\label{sec:LDT}

Observations with the Large Monolithic Imager (LMI) mounted on the 4.3-meter LDT at the Lowell Observatory in Happy Jack, AZ were carried out starting in 2014 as part of a long-term project (PIs: Troja, Gatkine, Dichiara) to study the afterglow and host galaxies of sGRBs. In order to have good visibility, only bursts with declination $\gtrsim$\,$-30^\circ$ were selected. 
Over 60 sGRBs were observed as part of this program, and results on
single events were presented in, e.g., \citet{Troja2016,Troja2018,Troja2019,OConnor2021,Ahumada21}. 
In this work, we present unpublished observations for 22 sGRBs in our sample.


LDT/LMI observations were carried out largely in the $r$-band with a typical exposure of $1200$\,$-$\,$1500$ s, chosen to obtain a depth of $r$\,$\gtrsim$\,$24.5$\,$-$\,$25$ mag in good observing conditions. 
However, the true image depth varies depending on the observing conditions at the time of our observations, which span multiple observing cycles across $\sim$\,$7$ years.
All images were visually inspected and those flagged as poor were re-acquired at a later date. When a candidate host galaxy was detected, we performed additional observations in the $g$, $i$, and $z$ bands in order to better characterize the galaxy's SED.

Data were reduced and analyzed using a custom pipeline \citep{Toy2016} that makes use of standard CCD reduction techniques in the IRAF\footnote{IRAF is distributed by the National Optical Astronomy Observatory, which is operated by the Association of Universities for Research in Astronomy (AURA) under cooperative agreement with the National Science Foundation (NSF).} package including bias subtraction, flat-fielding, sky subtraction, fringe correction, and cosmic ray rejection using Laplacian edge detection based on the \texttt{L.A.Cosmic} algorithm \citep{vanDokkum2001}. Following this image reduction process, the pipeline uses \texttt{SExtractor} \citep{Bertin1996} to identify sources in each frame, and then the \texttt{Software for Calibrating
AstroMetry and Photometry} \citep[\texttt{SCAMP};][]{Bertin2006} to compute the astrometric solution. The aligned frames are then stacked using the \texttt{SWarp} software \citep{Bertin2002,Bertin2010}. The absolute astrometry of the stacked image was calibrated against the astrometric system of either the Sloan Digital Sky Survey \citep[SDSS;][]{Ahumada2020} Data Release 16 or the Panoramic Survey Telescope and Rapid Response System Survey \citep[Pan-STARRS1, hereafter PS1;][]{Chambers2016} Data Release 2, likewise using the combination of \texttt{SExtractor} and \texttt{SCAMP}. The SDSS and PS1 catalogs were further used to calibrate the photometric zeropoint (using \texttt{SExtractor} aperture photometry for the magnitude determination). We selected the SDSS catalog when available, and otherwise used PS1. We ensured that the sources used for both the astrometric and photometric calibrations were isolated point sources by sorting out those which did not pass our selection criteria based on their signal-to-noise ratio (SNR), 
full width at half-maximum intensity (FWHM), ellipticity, and \texttt{SExtractor} \texttt{CLASS\_STAR} parameter.

\subsubsection{Gemini Observatory}

We carried out observations (PI: Troja) of short GRB host galaxies using the Gemini Multi-Object Spectographs (GMOS) mounted on the twin 8.1-m Gemini North and Gemini South telescopes located on Mauna Kea and Cerro Pach\'{o}n, respectively. These observations targeted 9 sGRBs (GRBs 110402A, 140930B, 151229A,  160601A, 160927A, 170127B, 171007A, 191031D, and 200411A) with deep constraints ($r\gtrsim 25$ mag) on an underlying host galaxy. The observations occurred between November 3, 2019 and February 1, 2021. We mainly selected the $r$-band and $i$-band with exposure times ranging from 900\,-\,2250 s and 355\,-\,1440 s, respectively. We supplemented our observations with archival data for GRBs 120305A, 120630A, 130822A, 140516A, 140930B, 150831A, 160408A, 180727A.

We made use of tasks within the \texttt{Gemini IRAF} package (v. 1.14) to perform bias and overscan subtraction, flat-fielding, de-fringing, and cosmic ray rejection. The individual frames were then aligned and stacked using the IRAF task \texttt{imcoadd}. We additionally performed sky subtraction using the \texttt{photutils}\footnote{\url{https://photutils.readthedocs.io/en/stable/}} package to estimate the median sky background after masking sources in the image. The world-coordinate systems were then calibrated against the astrometric systems of SDSS or PS1 using either \texttt{astrometry.net} \citep{Lang2010} or the combination of \texttt{SExtractor} and \texttt{SCAMP} outlined in \S \ref{sec:LDT}. 
For southern targets we used the Dark Energy Survey \citep[DES;][]{Abbott2021DES} Data Release 2.
Isolated field stars selected from these catalogs were used for photometric calibration.

We additionally performed observations of GRB 151229A with Flamingos-2 (hereafter, F2) at Gemini South in Cerro Pach\'{o}n, Chile on July 22, 2021. These observations were carried out in the $J$ and $K_s$ filters (see Table \ref{tab: observations}). We reduced and analyzed these data using the \texttt{DRAGONS}\footnote{\url{https://dragons.readthedocs.io/}} software \citep{Labrie2019}. The photometry was calibrated using nearby point-sources in the Two Micron All Sky Survey \citep[2MASS;][]{Skrutskie2006} catalog. We then applied a standard conversion between the Vega and AB magnitude systems.

\begin{figure*}
\includegraphics[width = 2\columnwidth]{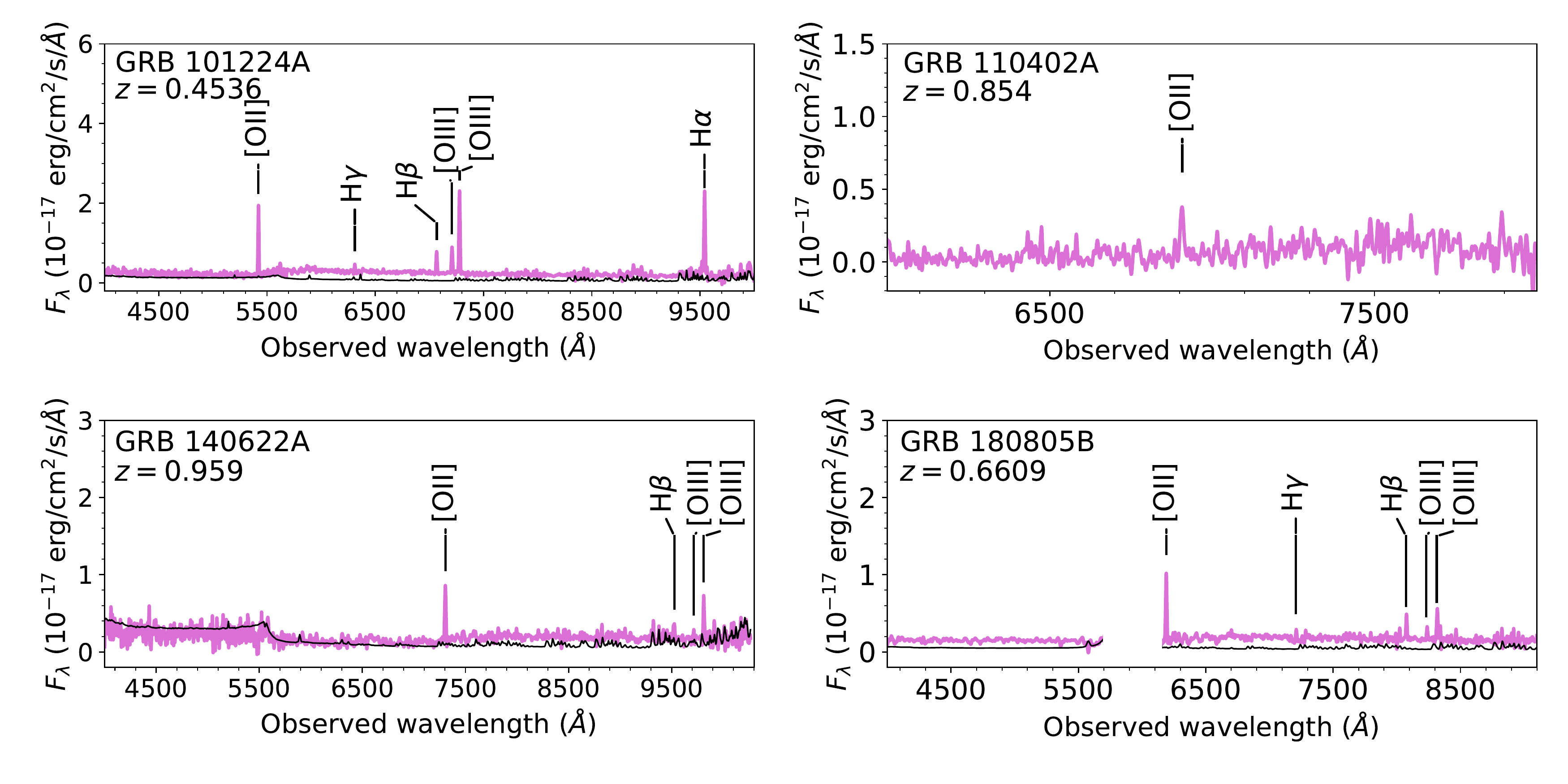}
\caption{Optical spectra of sGRB host galaxies (solid purple line) obtained with Keck/LRIS in flux units of $10^{-17}$ erg cm$^{-2}$ s$^{-1}$ \AA$^{-1}$ versus wavelength in \AA. The observed emission lines are marked by black lines, and the error spectrum is displayed as a solid black line. 
The spectra are smoothed with a Savitzky-Golay filter for display purposes. The spectra are not corrected for Galactic extinction. 
} \label{fig: spectra}
\end{figure*}

\begin{figure*}
\includegraphics[width = 1.65\columnwidth]{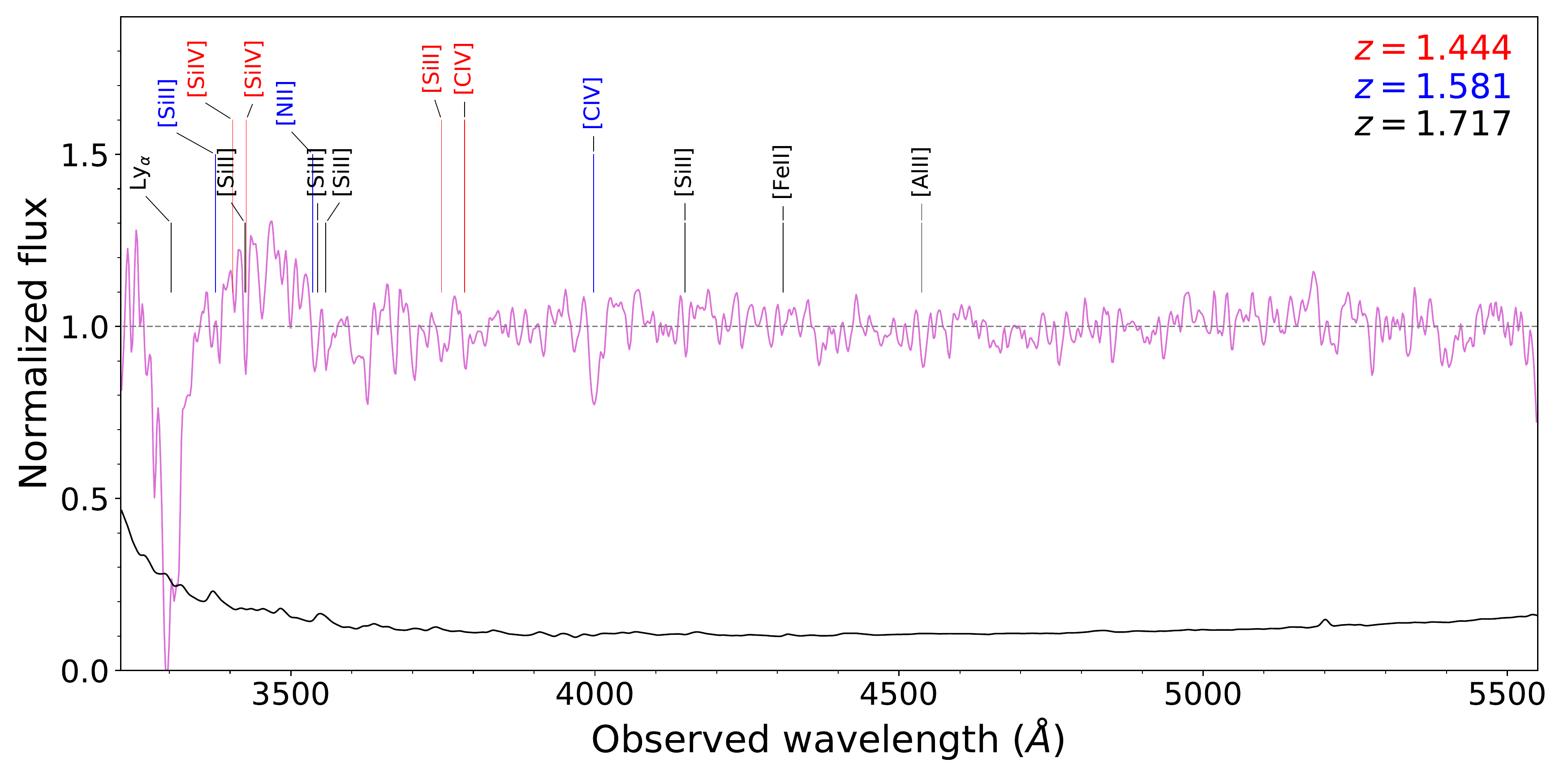}  
\caption{ 
Keck/LRIS optical spectrum of the afterglow of sGRB 160410A at $z=1.717 \pm 0.001$. The spectrum is normalized to the continuum. Absorption lines at this redshift are marked by black lines, and lines corresponding to intervening absorbers at $z=1.444$ and $1.581$ are marked by red and blue lines, respectively. The error spectrum is represented by a solid black line.} \label{fig: spectra160410A}
\end{figure*}

\subsubsection{Keck Observatory}

Through our program (PI: Cenko) on the 10-m Keck-I Telescope on Mauna Kea we obtained deep late-time imaging of GRBs 120305A, 120630A, 130822A, and 130912A. The Keck/LRIS observations took place during one half-night on October 25, 2014, and were carried out in both the $G$ and $R$ filters with exposure times of 3000 s and 2750 s, respectively. 
Observations of a fifth target (GRB~110112A) were incorrectly pointed by 0.15 deg and do not cover the GRB position (Chris Gelino, Priv. Comm.). Therefore, these data were not included.
We complemented our observations with public archival LRIS data for GRBs 110402A, 140516A, 160927A, 170127B, 170728A, and 180805B.

The data were retrieved from the Keck Observatory Archive, and analyzed using the \texttt{LPipe} 
pipeline \citep{Perley2019}. The pipeline processes raw files through standard CCD reduction techniques (e.g., bias-subtraction, flat-fielding, sky-subtraction, cosmic-ray rejection) to produce fully calibrated and stacked images. The final stacked image's absolute astrometry was calculated based on either the SDSS or PS1 catalogs. We used \texttt{astrometry.net} or the combination of \texttt{SExtractor} and \texttt{SCAMP} outlined in \S \ref{sec:LDT}. We found that \texttt{astrometry.net} provided an accurate astrometric solution for sparse fields by making use of the standard stars within the Keck field-of-view. The photometric zeropoints were likewise calibrated using unsaturated SDSS (when available) or PS1 sources.

We additionally include archival infrared imaging obtained with Keck MOSFIRE \citep{McLean2012} for GRBs 131004A, 151229A, 160601A, 170127B, and 180805B. These data were reduced using the MOSFIRE data reduction pipeline\footnote{\url{https://keck-datareductionpipelines.github.io/MosfireDRP/}}, and calibrated using point sources in the 2MASS catalog. Standard offsets were applied to convert magnitudes into the AB system.

\subsubsection{Gran Telescopio Canarias (GTC)}
\label{sec: GTC_imaging}

We obtained publicly available images of GRBs 160601A and 160927A (Table \ref{tab: observations}) taken with the 10.4-m GTC, which is located at the Roque de los Muchachos Obervatory in La Palma, Spain. The observations used the  OSIRIS instrument, and were carried out in $r$-band. 
The data were retrieved from the GTC Public Archive\footnote{\url{https://gtc.sdc.cab.inta-csic.es/gtc/}}. They were reduced and aligned using standard techniques within the \texttt{astropy} \citep{Astropy2018} software library to perform bias subtraction and flat-fielding. The individual frames were then combined to produce the final reduced image.
The absolute astrometric correction was performed using \texttt{astrometry.net}, and the photometric zeropoints were calibrated to SDSS.

\subsubsection{Very Large Telescope (VLT)}
\label{sec: vlt_imaging}

We analyzed archival images of GRBs 091109B, 150423A, and 150831A (Table \ref{tab: observations}) obtained with the 8.2-m VLT, operated by the European Southern Observatory (ESO) in Cerro Paranal, Chile. The observations were taken with the FOcal Reducer/low dispersion Spectrograph 2 (FORS2) in $R$-band for GRBs 091109B, 150423A, and 150831A and an additional $I$-band observation for GRB 150831A. The raw images were retrieved from the ESO Science Archive\footnote{\url{http://archive.eso.org/eso/eso_archive_main.html}}. The data were processed using standard tasks within \texttt{astropy} (similarly to \S \ref{sec: GTC_imaging}).

\subsubsection{Hubble Space Telescope (HST)}
\label{sec: hst imaging}

We obtained the publicly available  \textit{Hubble Space Telescope} (\textit{HST}) Wide Field Camera 3 (WFC3) data from the Mikulski Archive for Space Telescopes (MAST)\footnote{\url{https://archive.stsci.edu/index.html}} for GRBs 091109B, 110112A, 131004A, and  150423A. The observations (ObsID: 14685; PI: Fong) were taken between October 11, 2016 and February 3, 2017 in the \textit{F110W} filter with a typical exposure of 5200 s ($\sim$\,$2$ \textit{HST} orbits). 


The data were processed using standard procedures within the \texttt{DrizzlePac} package \citep{Gonzaga2012} in order to align, drizzle, and combine exposures. The observations within a single epoch were aligned to a common world-coordinate system with the \texttt{TweakReg} package. The \texttt{AstroDrizzle} software was then used to reject cosmic rays and bad pixels, and to create the final drizzled image combining all exposures within a single epoch. The final pixel scale was \ang{;;0.06}/pix using \texttt{pixfrac} $=0.8$. The \textit{HST} photometric zeropoints were determined with the photometry keywords obtained from the \textit{HST} image headers, and were corrected with the STScI tabulated encircled energy fractions.

\subsection{Optical Spectroscopy}
\label{sec:spectroscopy description}

Bright host galaxies identified through our imaging campaign were targeted for optical spectroscopy in order to constrain their distance scale. These targets include the fields of sGRBs~101224A and 140622A, observed with Keck/LRIS, and sGRBs 180618A and 191031D, observed with Gemini/GMOS-N. 
We complemented these observations with archival Keck spectroscopic data for sGRBs 110402A, 151229A, 160410A and 180805B as these bursts also match our selection criteria (\S \ref{sec: sampleselection}). 
Our spectroscopic campaign also included the candidate short GRB~060121 for which no visible trace was detected in a deep $3\times900$~s Keck/LRIS exposure. 
This was likewise the case for the archival Keck spectroscopy of sGRB 151229A. 
For sGRBs 180618A and 191031D, a weak trace was detected by the Gemini spectroscopic observations, but no obvious emission or absorption features were identified.
The log of spectroscopic observations analyzed in this work is provided in Table \ref{tab: SpecObs}.

The Gemini data were reduced and analyzed using the Gemini IRAF package (v. 1.14), whereas Keck/LRIS data were reduced using the \texttt{LPipe} software. The processed spectra are displayed in Figure \ref{fig: spectra}, and the result for each sGRB is reported in Table \ref{tab: SpecObs} and described in more detail in Section \ref{sec: Results}. 
We note that the optical spectrum obtained for sGRB 160410A is a rare case of afterglow spectroscopy (Figure \ref{fig: spectra160410A}) as discussed in \citet{AguiFernandez2021}. 


\begin{figure*}
\centering
\includegraphics[width = 1.5\columnwidth]{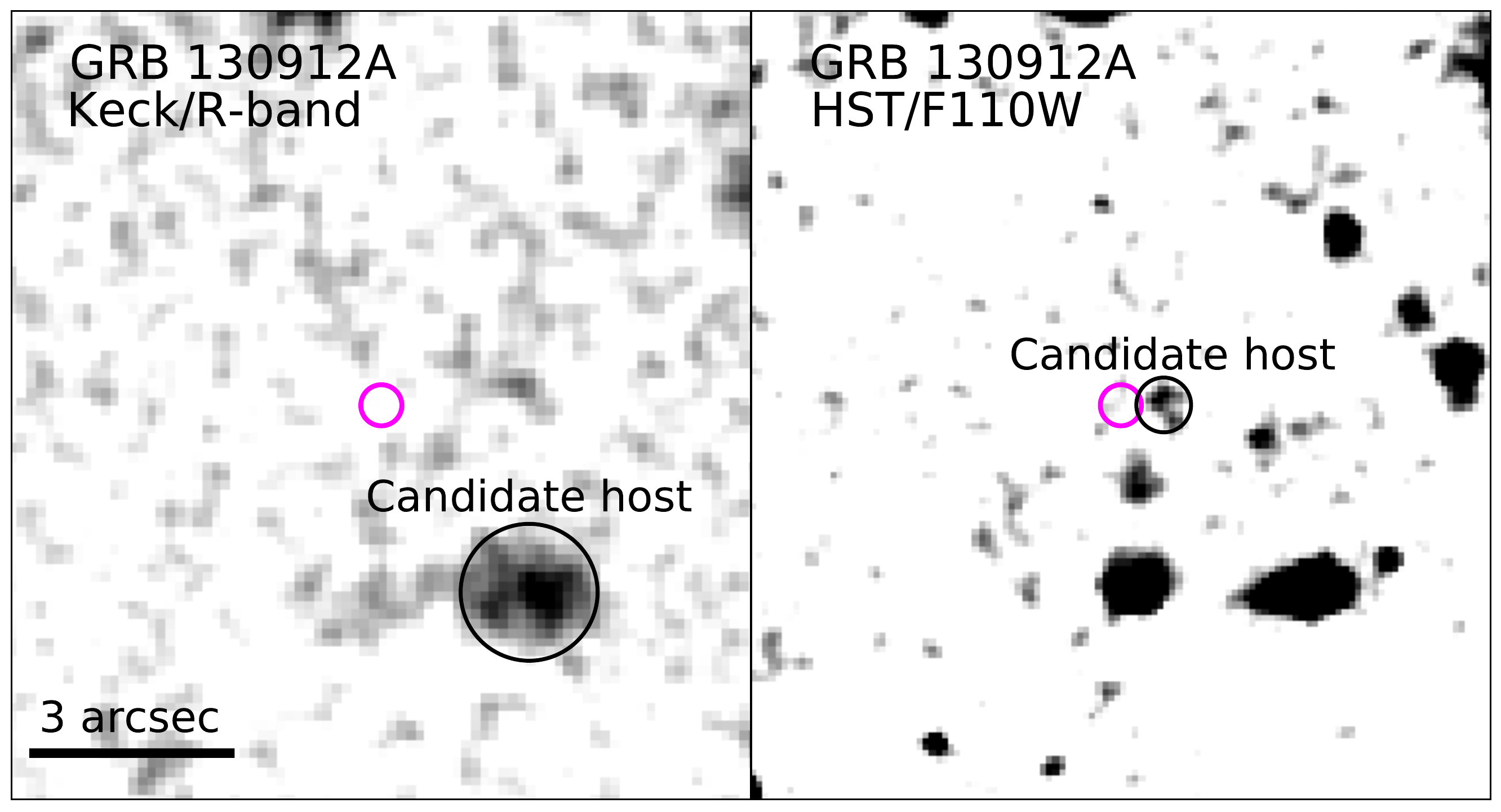} 
\caption{A comparison between ground-based Keck/LRIS imaging in $R$-band (left) and \textit{HST}/WFC3 imaging in the \textit{F110W} filter (right) for sGRB 130912A. The Keck imaging sets an upper limit of $R\gtrsim 26.2$ mag on a coincident host galaxy, whereas \textit{HST} imaging to depth $F110W\gtrsim 27.2$ mag unveils a candidate host offset by only $\sim$\,$0.7\arcsec$ from the sGRB's optical localization (magenta circle). The size of the circle corresponds to the uncertainty on the GRB position. The images are oriented such that North is up and East is to the left.
} \label{fig: 130912A_HST_vs_Keck}
\end{figure*}

\begin{figure*}
\begin{tabular}{ccc}
\subfloat{\includegraphics[width = 2.2in]{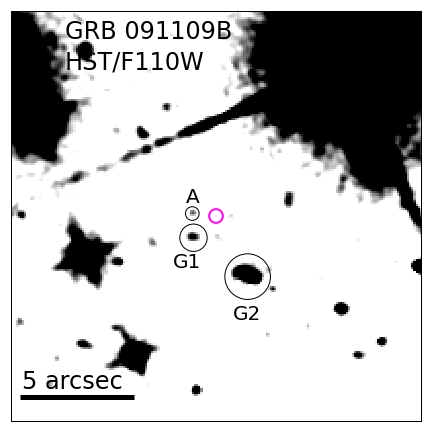}} \hspace{-0.5cm} &
\subfloat{\includegraphics[width = 2.2in]{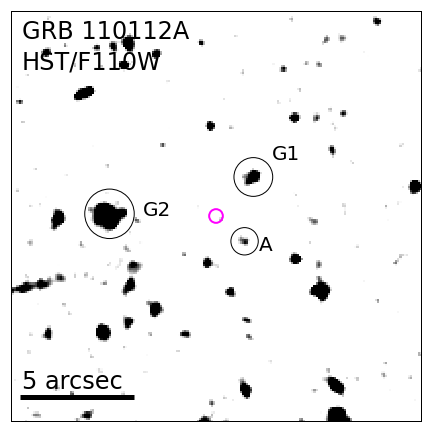}} \hspace{-0.5cm} &
\subfloat{\includegraphics[width = 2.2in]{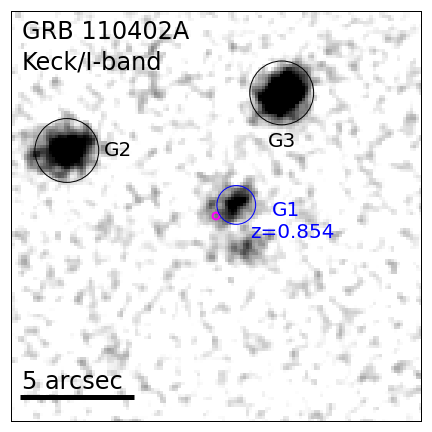}} \vspace{-0.5cm} \\ \vspace{-0.5cm}
\subfloat{\includegraphics[width = 2.2in]{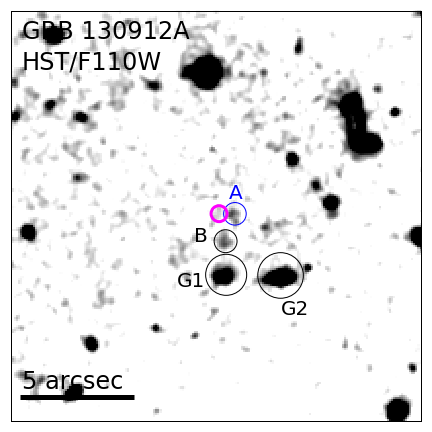}} \hspace{-0.5cm} &
\subfloat{\includegraphics[width = 2.2in]{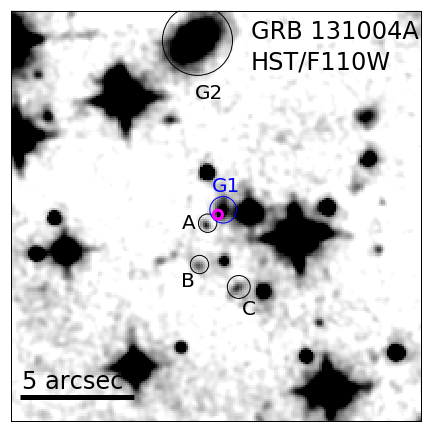}} \hspace{-0.5cm} &
\subfloat{\includegraphics[width = 2.2in]{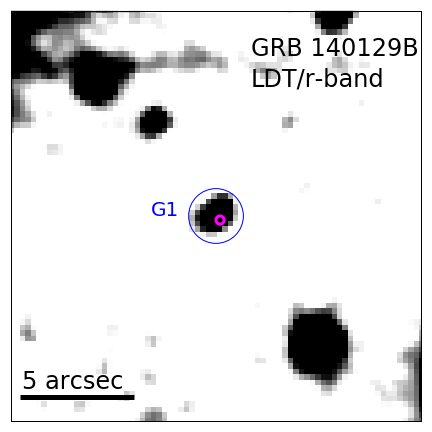}} \\ \vspace{-0.5cm}
\subfloat{\includegraphics[width = 2.2in]{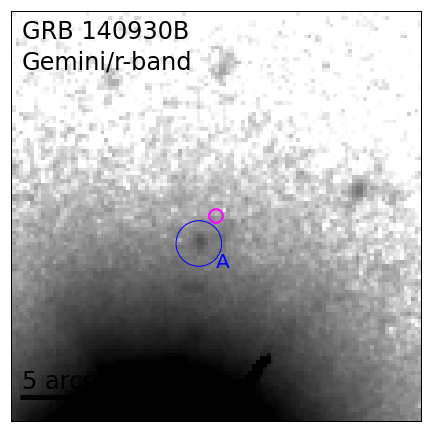}} \hspace{-0.5cm} &
\subfloat{\includegraphics[width = 2.2in]{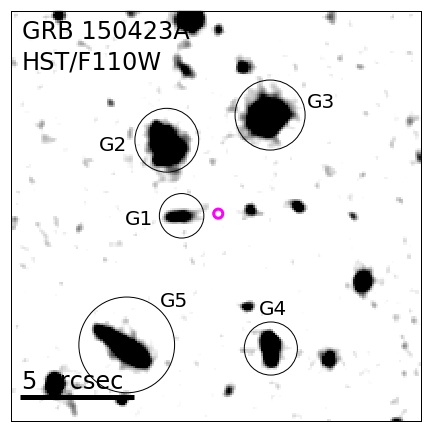}} \hspace{-0.5cm} &
\subfloat{\includegraphics[width = 2.2in]{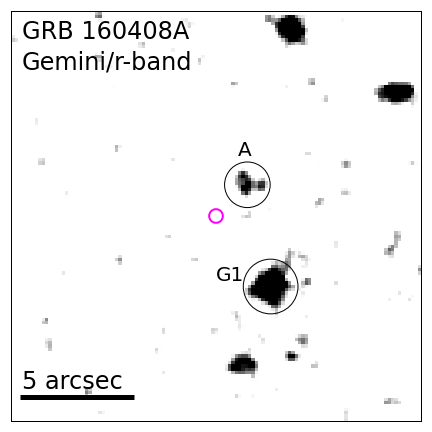}} \\
\subfloat{\includegraphics[width = 2.2in]{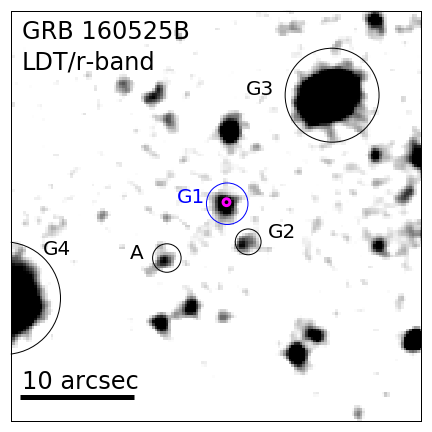}} \hspace{-0.5cm} &
\subfloat{\includegraphics[width = 2.2in]{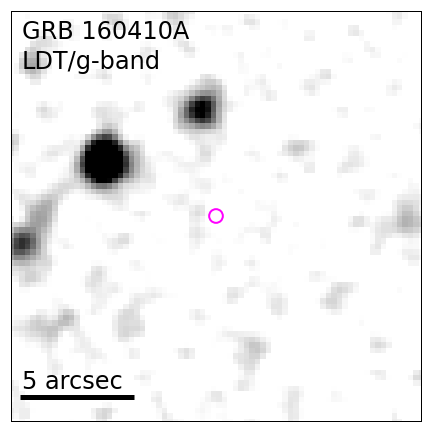}} \hspace{-0.5cm} &
\subfloat{\includegraphics[width = 2.2in]{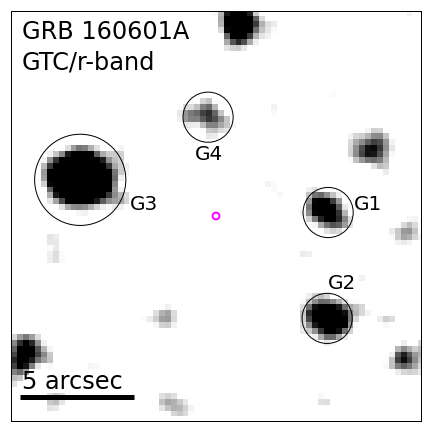}} \\
\end{tabular}
\caption{Host galaxy finding charts for optically localized sGRBs. The magenta circle represents the sGRB localization (with the size corresponding to the error in arcseconds), and the putative host galaxy is designated by a blue circle (those lacking a blue circle are  observationally hostless). Other candidate hosts are marked by black circles and labeled by G1, G2, G3, etc., with increasing offset from the sGRB's localization. Nearby objects that are too faint for star-galaxy classification (\S \ref{sec: source detection}) are labeled as A, B, C, etc. 
The size of each field is represented by the scalebar. 
In each figure, North is up and East is to the left. The figures have been smoothed for display purposes. 
} \label{fig: GalOpt}
\end{figure*}

\begin{figure*}
\begin{tabular}{ccc}
\subfloat{\includegraphics[width = 2.2in]{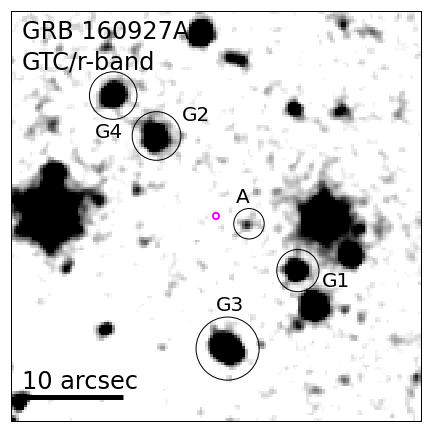}} \hspace{-0.5cm} &
\subfloat{\includegraphics[width = 2.2in]{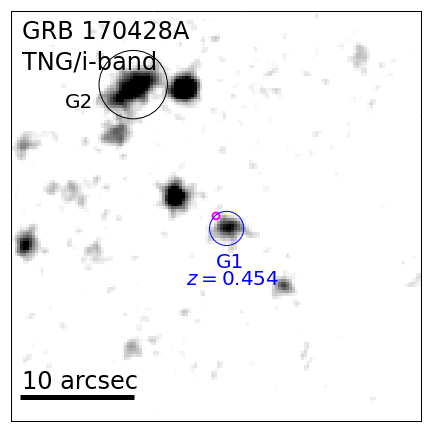}} \hspace{-0.5cm} &
\subfloat{\includegraphics[width = 2.2in]{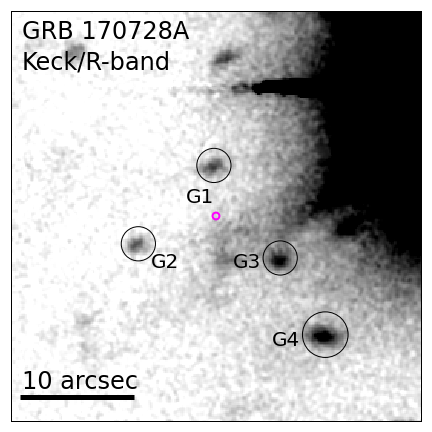}} \vspace{-0.5cm} \\ 
\subfloat{\includegraphics[width = 2.2in]{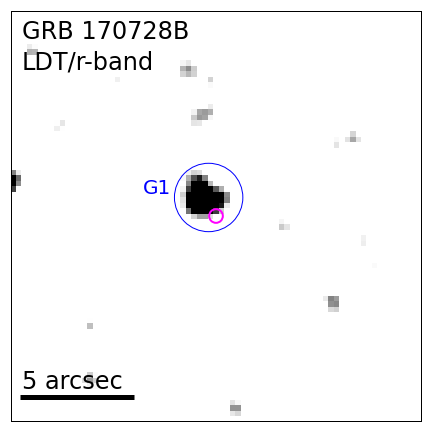}} \hspace{-0.5cm} &
\subfloat{\includegraphics[width = 2.2in]{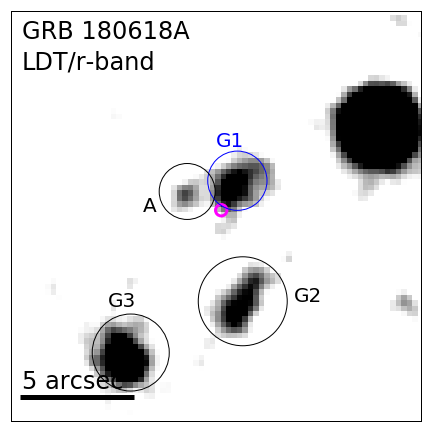}} \hspace{-0.5cm} &
\end{tabular}
\contcaption{
} 
\end{figure*}

\begin{figure*}
\begin{tabular}{ccc}
\subfloat{\includegraphics[width = 2.2in]{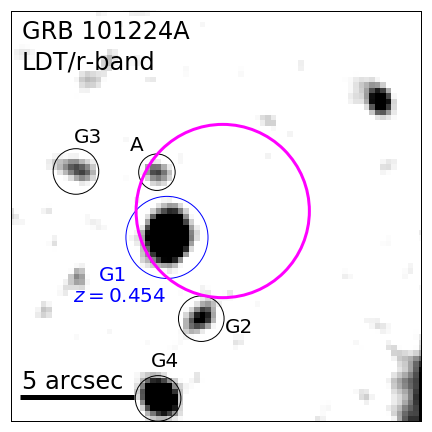}} \hspace{-0.5cm} &
\subfloat{\includegraphics[width = 2.2in]{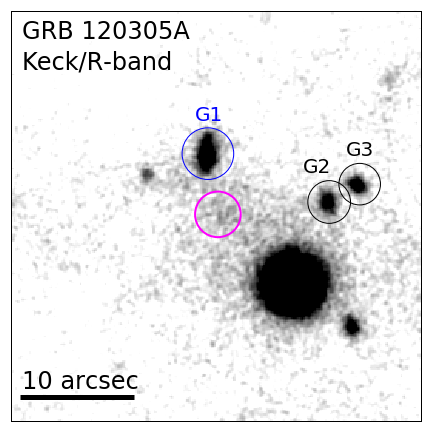}} \hspace{-0.5cm} &
\subfloat{\includegraphics[width = 2.2in]{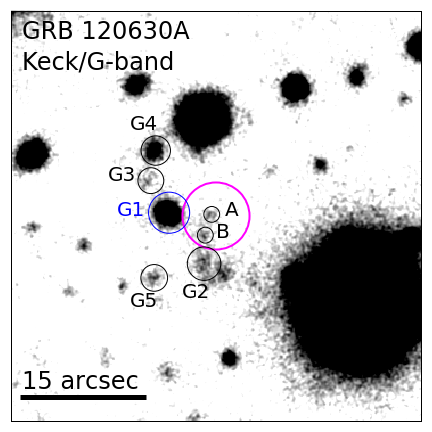}} \vspace{-0.5cm} \\ \vspace{-0.5cm}
\subfloat{\includegraphics[width = 2.2in]{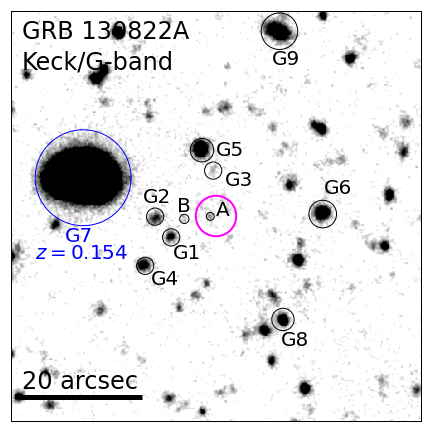}} \hspace{-0.5cm} &
\subfloat{\includegraphics[width = 2.2in]{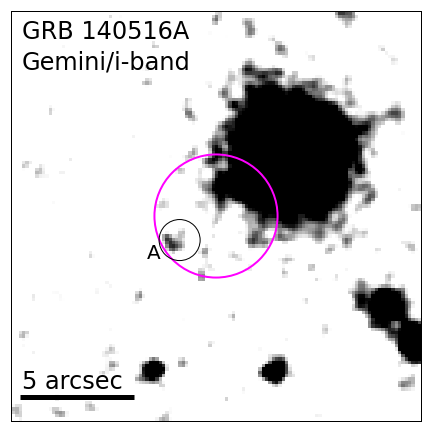}} \hspace{-0.5cm} &
\subfloat{\includegraphics[width = 2.2in]{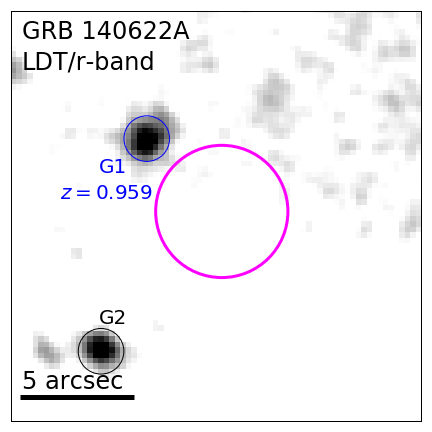}}  \\ \vspace{-0.5cm}
\subfloat{\includegraphics[width = 2.2in]{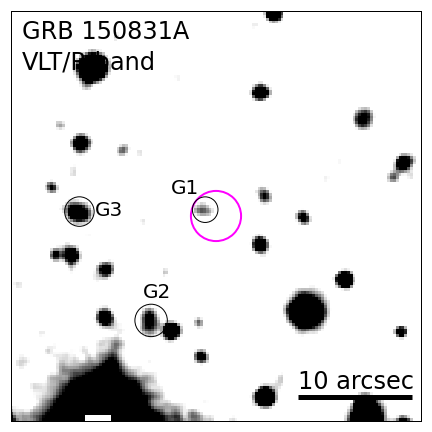}} \hspace{-0.5cm} &
\subfloat{\includegraphics[width = 2.2in]{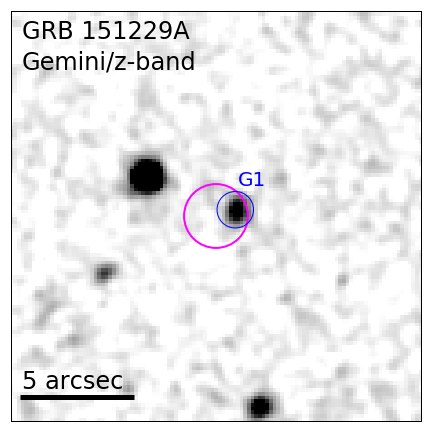}} \hspace{-0.5cm} &
\subfloat{\includegraphics[width = 2.2in]{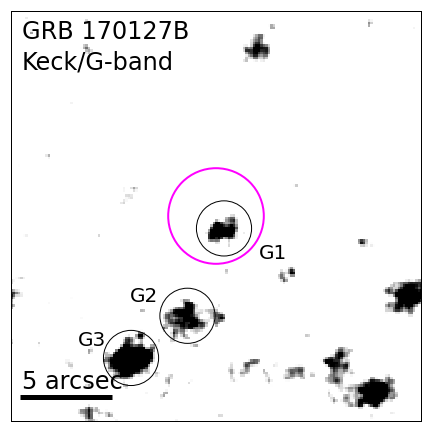}} \vspace{0.25cm}
\end{tabular}
\caption{Same as Figure \ref{fig: GalOpt} but for X-ray localized sGRBs.
}
\label{fig: GalXray}
\end{figure*}

\begin{figure*}
\begin{tabular}{ccc}
 \vspace{-0.5cm}
\subfloat{\includegraphics[width = 2.2in]{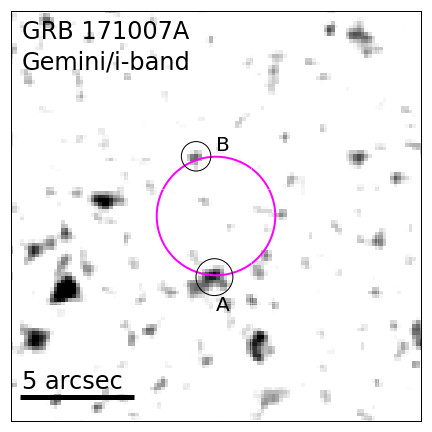}} \hspace{-0.5cm} &
\subfloat{\includegraphics[width = 2.2in]{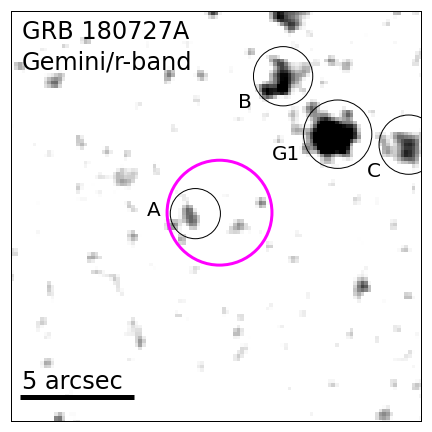}} \hspace{-0.5cm} &
\subfloat{\includegraphics[width = 2.2in]{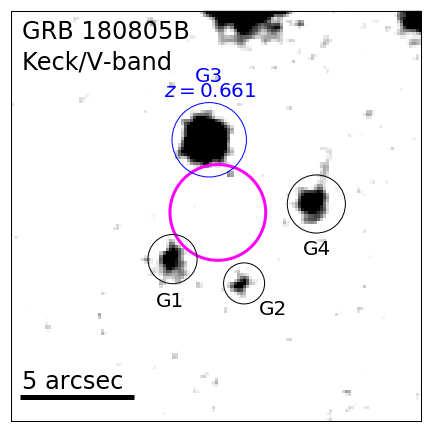}}  \\
\subfloat{\includegraphics[width = 2.2in]{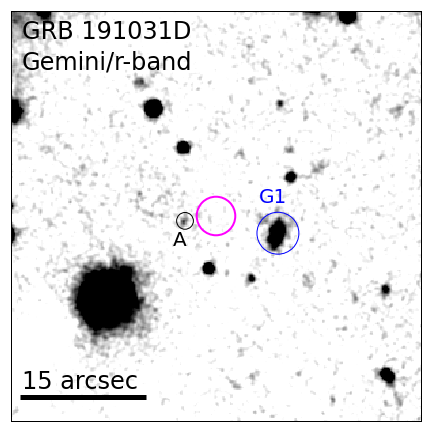}} \hspace{-0.5cm} &
\subfloat{\includegraphics[width = 2.2in]{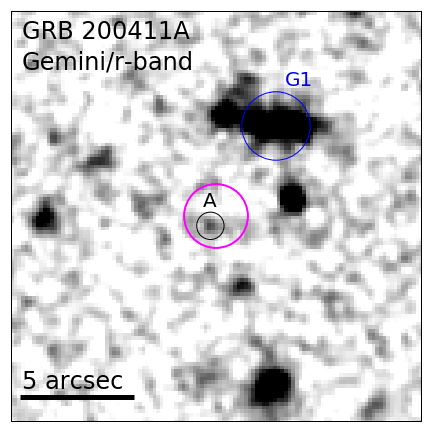}} \hspace{-0.5cm} &
\end{tabular}
\contcaption{}
\end{figure*}

\section{Methods}
\label{sec: methods}



In order to determine the putative host galaxy for each GRB, we  began by identifying all galaxies near the GRB position in our late-time imaging. 
The source detection and classification (star-galaxy separation) procedure is outlined in Section \ref{sec: source detection}. The late-time images were aligned with respect to the afterglow discovery images to precisely determine the host offset from the GRB position, as outlined in \S \ref{sec: Offset Measurements}. The host association was then determined through probabilistic arguments based on the observed sky density of galaxies in Section \ref{sec: Pcc}. The results of our analysis for each GRB are presented in \S \ref{sec: Results}.

\subsection{Source Detection and Classification}
\label{sec: source detection}

Source detection was performed using the \texttt{SExtractor} package after applying a Gaussian filter with a FWHM of 3 pixels\footnote{This value has been utilized in past studies of GRB host galaxies \citep{OConnor2021,Lyman2017}.}. We required that a source consist of a minimum area of 5 pixels at $>$\,$1\sigma$ above the background (\texttt{DET\_THRESH}\,$=$\,$1$). 
The source detection was visually inspected to prevent erroneous blending of adjacent sources. 

Source photometry was computed using the \texttt{SExtractor} \texttt{MAG\_AUTO} parameter, which utilizes Kron apertures. 
In the case of faint sources, the magnitude was computed using seeing matched aperture photometry with the aperture (\texttt{MAG\_APER}) diameter set to the FWHM of the image's point-spread function (PSF). 
The photometry was calibrated for each instrument as outlined in \S \ref{sec:imaging description}. 
The candidate host galaxy photometry for each GRB is presented in Table \ref{tab: host properties}.


In order to determine whether a detected source could be identified as a galaxy we utilized the \texttt{SExtractor} \texttt{SPREAD\_MODEL} parameter. First, we ran \texttt{SExtractor} to identify bright, unsaturated and isolated point-like objects.  We selected them based on their SNR, FWHM, \texttt{CLASS\_STAR} parameter ($>$\,$0.8$), and ellipticity ($<$\,$0.2$). We further imposed  \texttt{FLAGS}\,$<$\,$1$, which excludes sources that are saturated, blended, or too close to the image boundary.
These point-like sources were then passed to \texttt{PSFEx} \citep{Bertin2011,Bertin2013} to estimate the image PSF.
This was then fed to \texttt{SExtractor} to estimate the \texttt{SPREAD\_MODEL} parameter which, for each detected source, measures the deviation of the source profile from the local normalized image PSF.
Point-like sources are characterized by \texttt{SPREAD\_MODEL} $\approx$\,$0$,  
whereas extended objects deviate significantly from the local PSF and have \texttt{SPREAD\_MODEL} $>$\,$0$. For sources smaller than the image PSF (e.g., cosmic rays or spurious detections), \texttt{SPREAD\_MODEL} $<$\,$0$. These star-galaxy classifiers become more uncertain for fainter sources, and we considered
the classification as inconclusive for sources with SNR $\lesssim$\,$5$.


\subsection{Offset Measurements}
\label{sec: Offset Measurements}

In order to precisely localize the GRB with respect to a candidate host galaxy, we utilized relative astrometry to align our late-time images with the afterglow discovery image.
In our sample, 14 sGRBs (45\%) do not have an optical localization, and we relied on the \textit{Swift}/XRT enhanced positions \citep{Goad2007,Evans2009}.
The associated errors are assumed to follow Rayleigh statistics \citep{Evans2014,Evans2020}, 
and in our work are computed at the 68\% level of the Rayleigh distribution. The afterglow positional uncertainty $\sigma_{AG}$ from XRT is therefore derived as 
$\sigma_{AG}\approx$ err$_{90}/1.42$ \citep{Pineau2017}, where err$_{90}$ is the 90\% error typically reported by the \textit{Swift} team\footnote{\url{https://www.swift.ac.uk/xrt\_positions/}}.

The remaining 17 sGRBs (55\% of the total sample) have an optical counterpart, and for these bursts
we obtained publicly available discovery images
from the Ultra-Violet Optical Telescope \citep[UVOT;][]{Roming2005} on-board \textit{Swift}, the 8.1-m Gemini North Telescope, the GTC, the VLT, the 4.2-m William Herschel Telescope (WHT), the 3.6m Telescopio Nazionale Galileo (TNG), and the 2-m Liverpool Telescope.

We applied standard procedures for reduction and calibration of these ground-based images, and used \texttt{SExtractor} for afterglow localization. 
For the \textit{Swift}/UVOT data (GRBs 110402A, 131004A, and 170728A) we used the \texttt{uvotimsum} task within \texttt{HEASoft v6.27.2} to co-add multiple exposures. This produces a higher signal-to-noise afterglow detection. The afterglow localization error (statistical) was then determined using the \texttt{uvotdetect} task.

We used \texttt{SExtractor} to identify common point sources in both the late-time and discovery images, and then \texttt{SCAMP} to compute the astrometric solution. The rms uncertainty $\sigma_\textrm{tie}$ in the offset of astrometric matches between the late-time and afterglow images provides the uncertainty in the sGRBs localization on the late-time image frame, and is included within the determination of the host offset error \citep{Bloom2002}. 

The projected offset $R_o$ is then determined by measuring the distance between the afterglow centroid and the host galaxy's center.
The latter is determined as the barycenter of the pixel distribution using the parameters \texttt{XWIN\_IMAGE} and \texttt{YWIN\_IMAGE}
and its uncertainty $\sigma_\textrm{host}$ 
is derived by adding in quadrature the positional error in both directions. The parameters \texttt{XWIN\_IMAGE} and \texttt{YWIN\_IMAGE} are calculated within a circular Gaussian window instead of the isophotal footprint of each object. The Gaussian window function is determined separately for each object based on the circular diameter containing half the object's flux. Therefore, \texttt{XWIN\_IMAGE} and \texttt{YWIN\_IMAGE} are not affected by detection threshold or irregularities in the background, whereas isophotal centroid measurements take into account only pixels with values higher than the detection threshold. 
The afterglow centroid and its associated uncertainty $\sigma_\textrm{AG}$ are determined  with \texttt{SExtractor} using the same methodology. 
The uncertainty in the sGRB offset is computed as $\sigma_R = \sqrt{\sigma_\textrm{tie}^2+\sigma_\textrm{AG}^2+\sigma_\textrm{host}^2}$ \citep{Bloom2002,FongBerger2013}.

The offset and uncertainty for each GRB is recorded in Table \ref{tab: host properties}. 
For each candidate host galaxy, we also determine the half-light radius ($R_e$) as measured by \texttt{SExtractor} (with \texttt{FLUX\_RADIUS} = 0.5). This allows us to compute a host-normalized offset (see the discussion in \S \ref{sec: offset distribution}).

\subsection{Host Galaxy Assignment}
\label{sec: Pcc}

The association of a GRB to a host galaxy relies on probabilistic arguments based on the likelihood of finding a random galaxy near the GRB localization. This is estimated by computing the probability to detect a galaxy of equal magnitude or brighter within a given region on the sky \citep[e.g.,][]{Bloom2002,Bloom2007,Berger2010a}. If the probability is too high or equivalent for multiple galaxies in the field (see Figure \ref{fig: 130912A_HST_vs_Keck}), the GRB is considered observationally hostless. 
Using the methods outlined by \citet{Bloom2002}, the probability of chance coincidence is 
\begin{align}
    P_{cc}=1-e^{-\pi R^2 \sigma(\lesssim m)}
    \label{eqn: Pcc}
\end{align}
where $R$ is the effective angular offset of the galaxy from the GRB position. For XRT localized GRBs, or those where a galaxy is not detected coincident to the GRB position, the effective angular offset is given by $R=\max\Big(3\sigma_R,\sqrt{R_o^2+4R_e^2}\Big)$, where $3\sigma_R$\,$\approx$\,$1.59\times$ err$_{90}$ \citep[see, e.g., Section 4.2 of][]{Pineau2017}. 
If the GRB has a precise (sub-arcsecond) localization, and lies within the visible light of a galaxy, we adopt $R=2R_e$ \citep{Bloom2002}.

The quantity $\sigma(\lesssim$\,$m)$ in Equation \ref{eqn: Pcc} denotes the number density of galaxies brighter than magnitude $m$ based on deep optical and infrared surveys \citep[e.g., the Hubble Deep Field;][]{Metcalfe2001}. For our optical observations, we utilize $\sigma(\lesssim$\,$m)$ based on $r$-band number counts from \citet{Hogg1997}. For infrared observations, we use the \textit{H}-band (\textit{HST}/$F160W$ filter) number counts presented by \citet{Metcalfe2006,Galametz2013}. 
The magnitude for each galaxy is corrected for Galactic extinction \citep{Schlafly2011} prior to computing the probability. This is done because the galaxy number counts used in this work \citep{Hogg1997,Metcalfe2006,Galametz2013} were derived from observations of high Galactic latitude fields, where the extinction is negligible.

For each sGRB, we computed the probability of chance coincidence for all galaxies identified within $1\arcmin$ of the sGRB position. 
We require that the putative host galaxy for each sGRB has $P_{cc}\lesssim 0.1$ to be considered a robust association, otherwise we deem the sGRB to be observationally hostless. At offsets $>$\,$1\arcmin$, a $P_{cc}$\,$\lesssim$\,$0.1$ requires an extremely bright galaxy $r$\,$\lesssim$\,$16$ mag, which would not be missed in our imaging. We also note that the largest angular offset reported for a sGRB is $\sim$\,$16\arcsec$ for GRB 061201 \citep[][]{Stratta2007}, which we consider to be observationally hostless based on $P_{cc}$\,$>$\,$0.1$. All events with confident host associations are located at smaller angular offsets. In many cases there are a number of faint extended objects ($r$\,$\gtrsim$\,$23$ mag) at $\gtrsim$\,$10\arcsec$ which we remove from our analysis due to their high probability of chance coincidence $P_{cc}$\,$\gtrsim$\,$0.5$. The remaining galaxies in the field are then considered candidate hosts; see Figure \ref{fig: 130912A_HST_vs_Keck} for an example finding chart for sGRB 130912A based on deep Keck and \textit{HST} imaging. We report the results of our search for each sGRB in Appendix \ref{sec: appendixsampleanalysis}, and their finding charts are displayed in Figures \ref{fig: GalOpt} and \ref{fig: GalXray}. Sources classified as a galaxy are denoted by G1, G2, G3, etc., by increasing offset from the GRB position, whereas sources which could not be classified are labeled as A, B, C, etc., in the same manner.

The probability of chance coincidence reported for each sGRB (Table \ref{tab: host properties}) is based on $r$-band number counts when possible, but if the galaxy is only detected in redder filters we include this probability instead using the number counts presented by \citet{Capak2007} for the $i$-band and \citet{Capak2004} for the $z$-band.

\subsection{Galaxy SED Modeling}
\label{sec:prospector}

For those events with well-sampled galaxy SEDs but lacking a spectroscopic redshift, 
we obtained a photometric redshift 
by modeling the SED using \texttt{prospector} \citep{Johnson2019} with the methods previously utilized by \citet{OConnor2021,Dichiara2021,Piro2021}. We note that these photometric redshifts were determined based on the assumption that the photometric jump between two filters is due to the 4000 \AA\, break. A large break is indicative of an older stellar population.

We adopted a \citet{Chabrier2003} initial mass function (IMF) with integration limits of 0.08 and 120 $M_{\odot}$ (\texttt{imf\_type = 1}), an intrinsic dust attenuation $A_V$ using the extinction law of  \citet[][\texttt{dust\_type = 2}]{{Calzetti2000}}, and a delayed-$\tau$ star formation history (\texttt{sfh=4}). Furthermore, we include nebular emission lines using the photoionization code \texttt{Cloudy} \citep{Ferland2013}. In the cases of sGRBs 151229A,
180618A, and 191031D we turned off nebular emission lines as their spectra (Table \ref{tab: SpecObs}) did not display bright or obvious emission features. The synthetic SEDs derived from these model parameters were calculated using the flexible stellar population synthesis (FSPS) code \citep{Conroy2009} using WMAP9 cosmology \citep{Hinshaw2013}.

\begin{figure*}
\begin{tabular}{cc}
\includegraphics[width = 2\columnwidth]{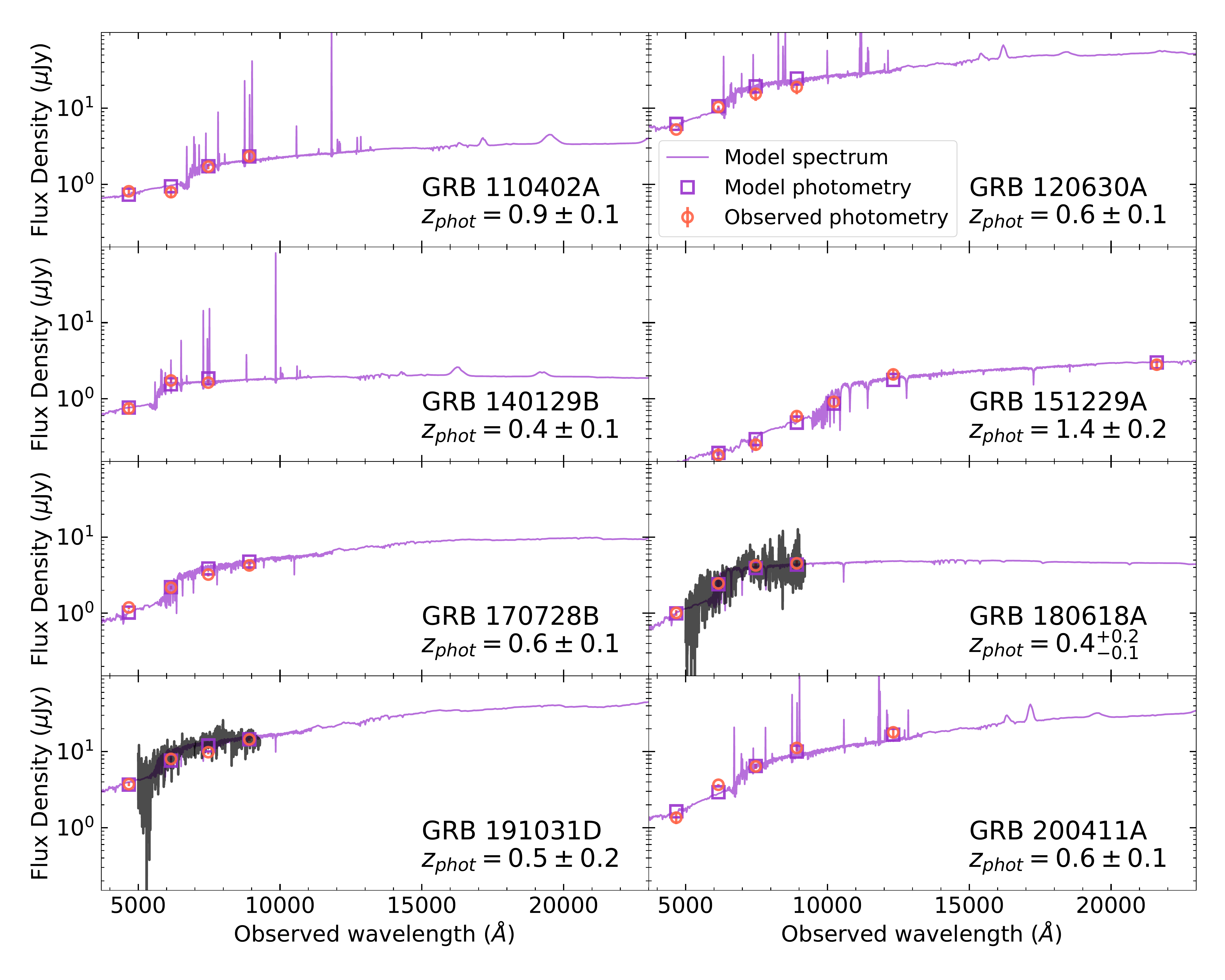}
\end{tabular}
\caption{Spectral energy distributions of sGRB host galaxies with photometric redshifts determined in this work. The best fit model spectrum (solid line) and model photometry (squares) describing the galaxy SED is compared to the extinction-corrected photometry (circles). The observed Gemini spectrum, smoothed with a Savitzky-Golay filter, for the host galaxies of GRBs 180618A and 191031D is shown by a solid black line (see Table \ref{tab: SpecObs}).
} \label{fig: SED_fits}
\end{figure*}

The free model parameters are: the redshift $z$, the total stellar mass formed $M$, the age $t_{\rm age}$ of the galaxy, the e-folding timescale $\tau$, the intrinsic reddening $A_V$, and the metallicity $Z$. These parameters are further used to compute the stellar mass $M_*$. We adopt uniform priors in log $t_{\rm age}$, log $\tau$, log $Z$, $A_V$ as in \citet{Mendel2014}. The prior on the photometric redshift is uniform between $z_\textrm{phot}=0$\,$-$\,$3$. However, only for sGRBs with a UV detection of their afterglow (e.g., sGRBs 110402A and 140129B; see Appendix \ref{sec: appendixsampleanalysis}) from \textit{Swift}, we adopt $z_\textrm{phot}=0$\,$-$\,$1.5$. The fits were performed using the dynamic nested sampling method implemented in the \texttt{DYNESTY} package \citep{dynesty}. 
The best fit model SEDs and the resulting photometric redshift estimates are displayed in Figure \ref{fig: SED_fits}. The photometric redshifts for these sGRBs are recorded in Table \ref{tab: host properties}, and the stellar mass is reported in their individual sections in Appendix \ref{sec: appendixsampleanalysis} as well as Table \ref{tab: prospectorfit}. In Table \ref{tab: prospectorfit} we likewise record the star formation rate (SFR), which is computed as outlined in \citet{OConnor2021}.

\begin{table}
    \centering
    \caption{
    Results of our \texttt{prospector} SED modeling. We present the photometric redshift, stellar mass, and star formation rate. The SED fits are displayed in Figure \ref{fig: SED_fits}.
    }
    \label{tab: prospectorfit}
    \begin{tabular}{lccccc}
    \hline
    \hline
\textbf{Source} & \textbf{$z_\textrm{phot}$} & \textbf{$\log(M_*/M_\odot)$} & \textbf{SFR ($M_\odot$ yr$^{-1}$)} \\
    \hline
   110402A$^{a,b}$  & $0.9\pm0.1$ & $9.5^{+0.4}_{-0.2}$  & $2^{+5}_{-1}$ \\[0.9mm]
   120630A & $0.6\pm0.1$ & $9.1\pm0.1$  & $30\pm15$ \\[0.9mm] 
   140129B & $0.4\pm0.1 $ &  $9.1\pm0.1$ & $0.4^{+0.7}_{-0.2}$ \\[0.9mm] 
   151229A & $1.4\pm0.2$ & $10.3\pm0.2$  & $0.4^{+1.6}_{-0.2}$ \\[0.9mm] 
   170728B$^a$ & $0.6\pm0.1$ & $9.7\pm0.2$ & $2^{+3}_{-1}$ \\[0.9mm] 
  180618A$^a$ & $0.4^{+0.2}_{-0.1}$ & $9.6\pm0.3$  & $0.1^{+0.3}_{-0.1}$\\[0.9mm] 
   191031D & $0.5\pm0.2$ & $10.2\pm0.2$ & $8\pm6$ \\[0.9mm] 
   200411A & $0.6\pm0.1$ & $10.4\pm 0.1$  & $3^{+6}_{-2}$ \\[0.9mm] 
    \hline
    \hline
    \end{tabular}
    \begin{flushleft}
      \quad \footnotesize{$^a$ Short GRB with extended emission.} \\
      \quad \footnotesize{$^a$ This GRB also has a spectroscopic redshift $z=0.854$ determined in this work.} \\
   \end{flushleft}
\end{table}

\section{Results}
\label{sec: Results}

In this work, we have analyzed the host galaxies and environments of 31 sGRBs; 17 with a sub-arcsecond position from optical observations and 14 with only an XRT localization (Figure \ref{fig: localization}).  In Figures \ref{fig: GalOpt} and \ref{fig: GalXray}, we display a finding chart for each sGRB in our sample. We find that 18 events (see Table \ref{tab: host properties}) are associated to a host galaxy ($P_{cc}$\,$<$\,$0.1$), while 13 events are deemed observationally hostless. With respect to previous work, we have adopted the $P_{cc}$ threshold previously used by \citet{Bloom2002} and \cite{Berger2010a}, whereas other authors have utilized lower thresholds, such as 0.01 \citep{Tunnicliffe2014} or 0.05 \citep{FongBerger2013}. 
We demonstrate below that our choice is robust and ensures a low number of spurious associations. 

Based on our host galaxy assignments, we identify a spectroscopic redshift for 5 sGRBs in our sample (sGRBEEs 110402A, 160410A, and 180805B, and GRBs 101224A and 140622A; see Tables \ref{tab: host properties} and \ref{tab: SpecObs}). In addition, we derive a photometric redshift for 8 events (sGRBEEs 110402A and 170728B, and GRBs 120630A, 140129B, 151229A, 180618A, 191031D and 200411A; Figure \ref{fig: SED_fits} and Table \ref{tab: host properties}). The detailed analysis for each sGRB is reported in Appendix \ref{sec: appendixsampleanalysis}, and the magnitudes and offsets for the putative host galaxies are presented in Table \ref{tab: host properties}.

We estimate the number of spurious galaxy associations in our sample following \citet{Bloom2002}. The probability that all sGRB host galaxies discovered in this work are a chance alignment with the GRB localization is given by 
\begin{equation}
    P_\textrm{false}=\prod^{m}_{k=1} P_k=4.8\times 10^{-25}
\end{equation}
where $m=18$ (the number of host galaxies we associate to sGRBs in this work) and $P_k$ is the probability of chance coincidence for each sGRB computed using Equation \ref{eqn: Pcc} based on $r$-band number counts (\S \ref{sec: Pcc}). If we compute $P_\textrm{false}$ for the optical and X-ray localized samples separately, we obtain $P_\textrm{false}=3.4\times 10^{-15}$ and $1.4\times 10^{-10}$, respectively. 
Moreover, the probability that every galaxy has a real, physical association to these GRBs can be estimated using 
\begin{equation}
    P_\textrm{real}=\prod^{m}_{k=1} (1-P_k)=0.36.
\end{equation}
If we consider again the optical and X-ray localized samples individually we find $P_\textrm{real}=0.76$ and $0.48$, respectively. As expected, the galaxy associations for the optically localized sample (Figure \ref{fig: GalOpt}) are more robust, but even the XRT only sample yields a similar result to the value ($P_\textrm{real}=0.48$) presented by \citet{Bloom2002} for their sample of long GRBs. Furthermore, we estimate $\sim2-3$ spurious associations out of our sample of 31 events \citep{Bloom2002}. The spurious associations are likely dominated by the XRT localized events. Based on these probabilistic arguments, we consider the host associations determined in this work to be robust, with minimal contamination due to chance alignment.



\begin{figure} 
\centering
\includegraphics[width=\columnwidth]{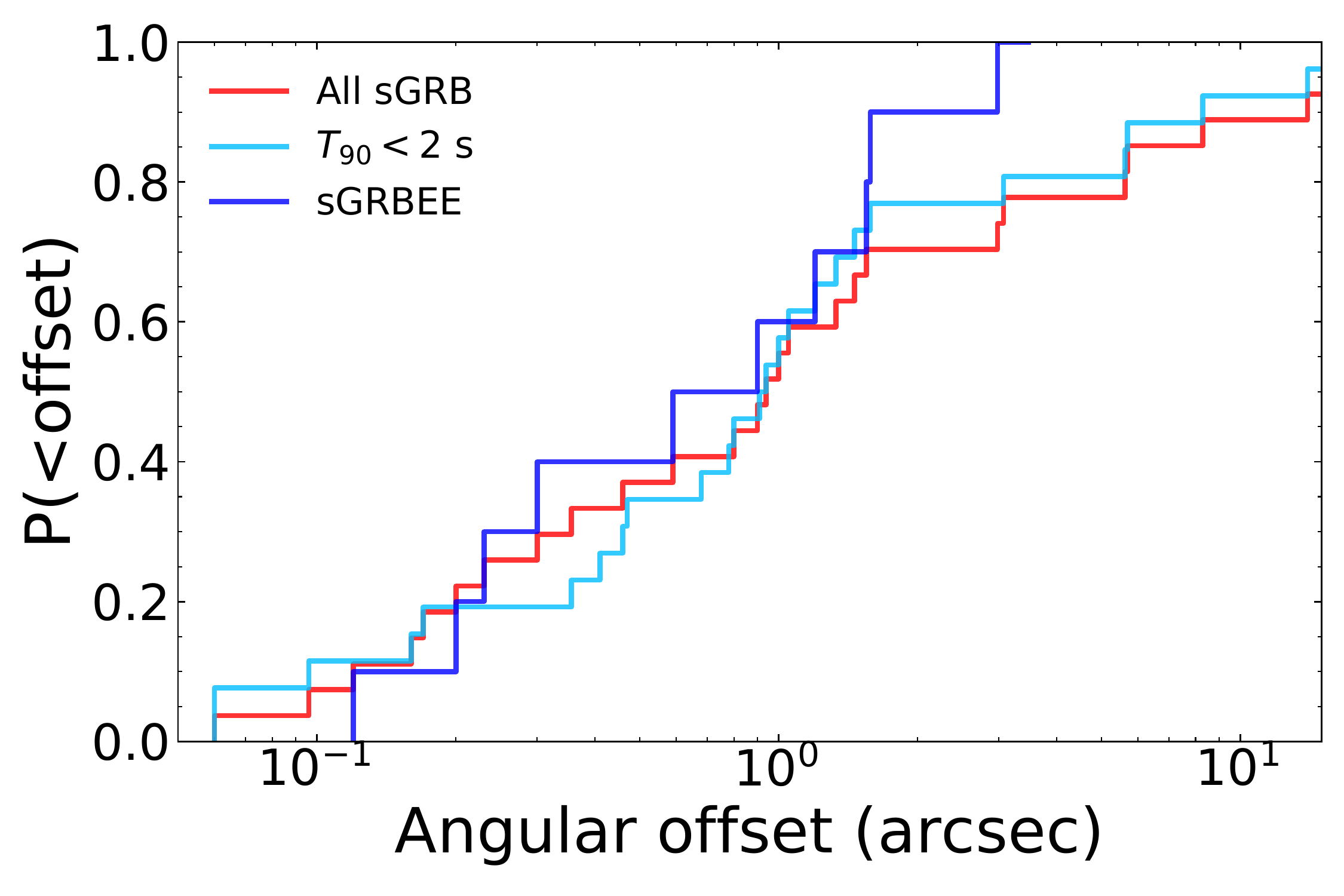}
\includegraphics[width=\columnwidth]{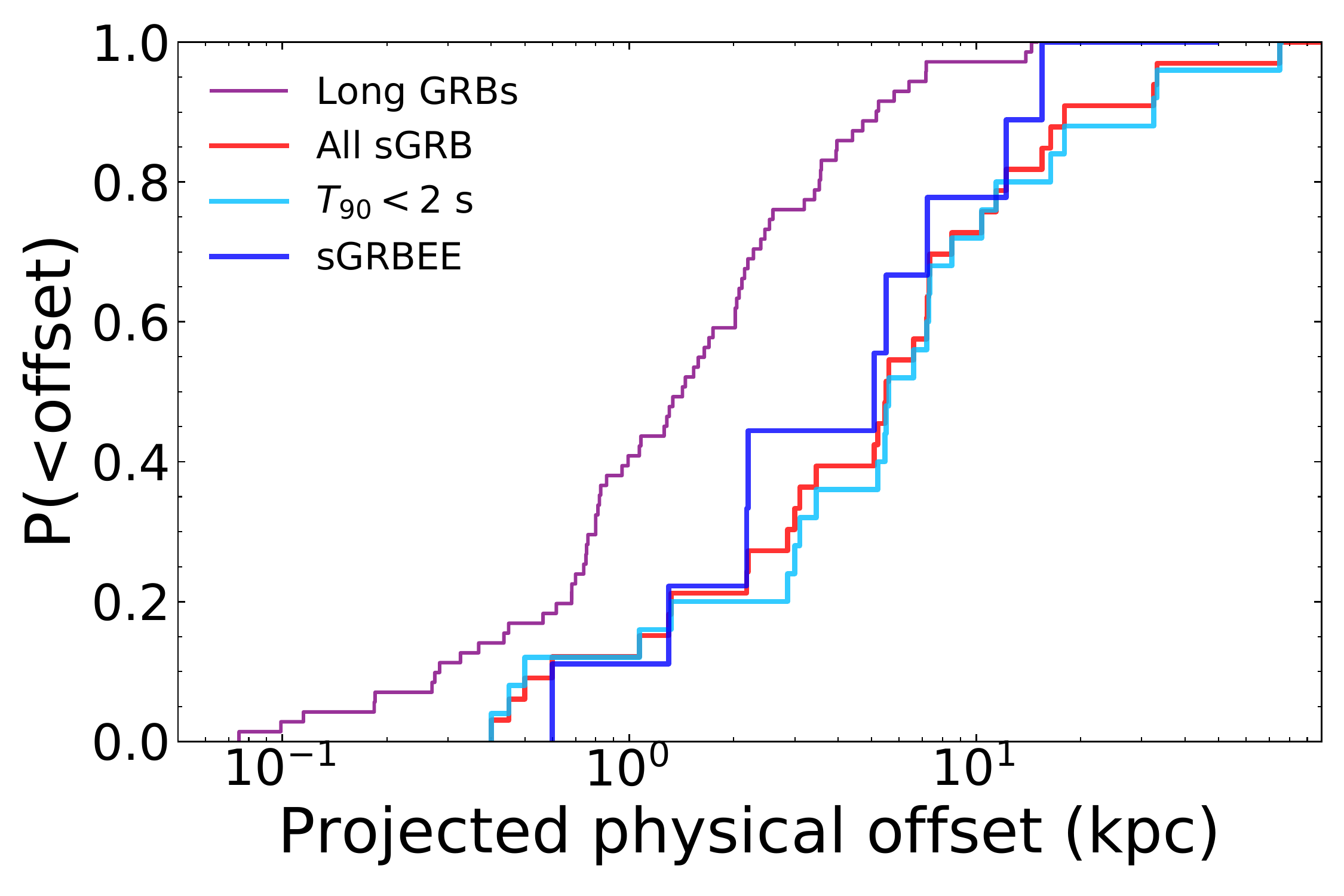}
\includegraphics[width=\columnwidth]{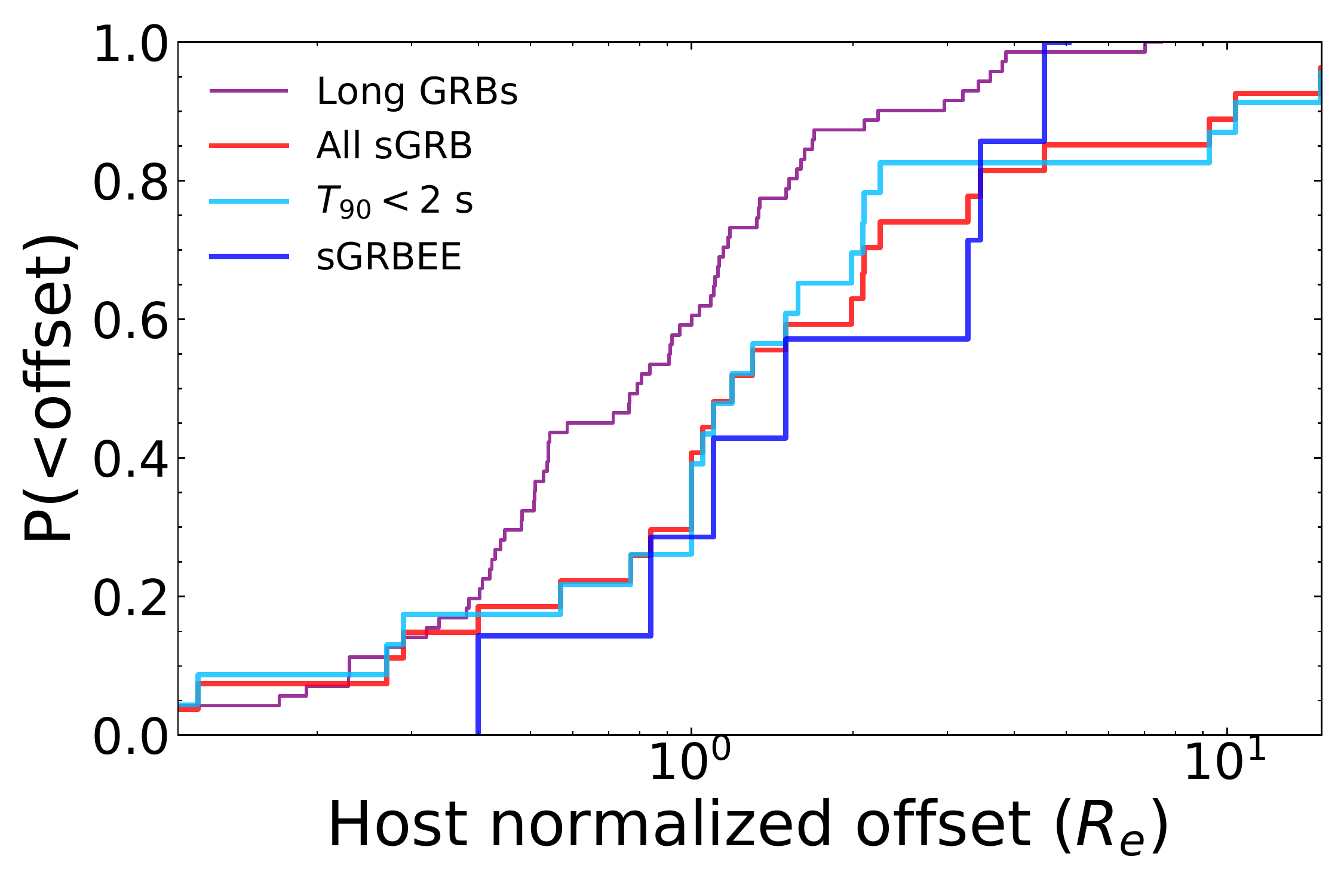}
\caption{\textit{\textbf{Top}}: 
Cumulative distribution of angular offsets for all sub-arcsecond localized sGRBs in our sample (red). We split the sample of all sGRBs into two sub-samples: the sample of sGRBs with $T_{90}$\,$<$\,$2$ s (cyan) and the remaining 10 events displaying EE (blue).
\textit{\textbf{Middle}}: Cumulative distribution of projected physical offsets for 33 sGRBs with sub-arcsecond localization (red). The offsets of long GRBs (purple) are displayed for comparison \citep{Blanchard2016}. 
\textit{\textbf{Bottom}}: Same as middle panel but for host-normalized offsets.
}
\label{fig: offset_dist}
\end{figure}


We now compare the properties of the host galaxies determined in this work to other large samples previously presented within the literature \citep[e.g.,][]{Fong2013,Tunnicliffe2014}. 
To do so, we supplement the 31 sGRBs that we analyzed with 41 events (29 sub-arcsecond) from the literature with deep host galaxy searches. Out of these 72 well-studied events, we find that 37 have a spectroscopic redshift, 11 have a photometric redshift, 20 are observationally hostless, and 15 display extended emission.

In order to perform a one-to-one comparison with our homogeneously selected sample, we excluded events from the literature which did not satisfy our selection criteria (specified in \S \ref{sec: sampleselection} and Table \ref{fig: outline}): including $A_V$\,$<$\,$1.5$ mag, $\sigma_\textrm{AG}$\,$<$\,$4\arcsec$, and a \textit{Swift}/BAT detection of the prompt emission. 
These criteria exclude a number of sGRBs typically included in other samples: sGRBs 050509B, 060502B, 090621B
, 100206A, 161104A, and sGRBEE 061210 
are excluded due to the large error ($>$\,$4\arcsec$) of their XRT localization, sGRBEE 050724 does not satisfy $A_V$\,$<$\,$1.5$ mag, and sGRBEE 050709 (\textit{HETE}), sGRBEE 060121 (\textit{HETE}), and sGRB 070707 (\textit{INTEGRAL}) are excluded as they were not detected with \textit{Swift}/BAT. 

The probabilities of chance coincidence for X-ray localized sGRBs were recalculated with the XRT enhanced positions derived using \texttt{HEASOFT} v6.28.
Different versions of the XRT calibration database and analysis software may change the error radius by up to 50\% of its value, and this step ensures that all the X-ray positions are based on the same calibration database (\texttt{HEASOFT} v6.28).
The resulting probabilities uniformly adopt the $3\sigma$ positional error (see \S \ref{sec: Pcc}), while in the literature different conventions (e.g., 68\% or 90\% CL) were sometimes adopted.


Based on this re-analysis, 3 XRT localized events (sGRBs~050813, 061217, and 070729) are found to have candidate hosts with $P_{cc}$\,$>$\,$0.1$, and are hereafter considered observationally hostless. 
This leaves us with only 9 sGRBs in the literature sample 
with both an XRT localization and a putative host galaxy 
(sGRBs 051210, 060801, 080123, 100625A, 101219A, 121226A, 141212A, 150120A, and 160624A). Including the events in this work, this sample doubles to 18 XRT localized events with a putative host. The impact of these XRT events is discussed in \S \ref{sec: XRT_offset}.

\subsection{Offset Distribution}
\label{sec: offset distribution}


\subsubsection{Sub-arcsecond Localized}
\label{sec: subarcsec offset}


We begin by studying the angular offset distribution (Figure \ref{fig: offset_dist}; top panel) for 34 sGRBs with sub-arcsecond positions. With a few exceptions, 
this sample coincides with the 
sample of optically-localized bursts, which have a typical uncertainty of $\sim$\,$0.2\arcsec$ on their offset. 
The measured angular offsets range between $0.06\arcsec$ (GRB 090426; \citealt{Antonelli2009,Levesque2010}) to $16\arcsec$ (GRB 061201; \citealt{Stratta2007}), with 70\% of the bursts lying $<$\,$2\arcsec$ from their putative host galaxy's center. For comparison, GRB 170817A was located at $10.6\arcsec$ (2 kpc) from its galaxy's center \citep{Levan2017,Im2017}.




We convert angular offsets into projected physical offsets by using the sGRB distance scale, typically derived from the putative host galaxy. 
For sGRBs without a measured redshift (8 events; $\sim 20\%$ of the sub-arcsecond localized sample), we adopt the median redshift (\S \ref{sec: redshift dist}), $z$\,$\approx$\,$0.5$, for sGRBs in our sample\footnote{
We note that the subset of events without a measured redshift are very unlikely to reside at $z$\,$<$\,$0.5$, and are more likely between $z$\,$\sim$\,$0.5$\,$-$\,$1$, where the difference in angular scale is $D_\theta(z=1.0)/D_\theta(z=0.5)$\,$\approx$\,$1.3$. We find that varying the redshift of these events does not significantly affect our results.}. 
We find that the physical offsets of sGRBs 
range from 0.4~kpc to 75~kpc with a median of $5.6$ kpc (Figure \ref{fig: offset_dist}; middle panel, red line). This is slightly larger than the median of 4.5 kpc from \citet{FongBerger2013} and a factor of $4\times$ larger than the median value for long GRBs \citep{Bloom2002,Lyman2017}. This result is consistent with the $<$\,$10$ kpc median sGRB offset derived by \citet{OConnor2020}, and with the expectations from binary population synthesis of BNS mergers \citep[see, e.g.,][]{Fryer1999,Bloom1999,Belczynski2006,Church2011,Mandhai2021,Perna2021}, although some modeling efforts predict larger median offsets \citep{Zemp2009,Wiggins2018}.

The last quantity to explore is the host-normalized offset, which provides the most uniform comparison between the location of sGRBs with respect to their galaxies (Figure \ref{fig: offset_dist}; bottom panel). 
We find that the median host normalized offset of the entire sGRB sample (sub-arcsecond localized) is  $R_o/R_e$\,$\sim$\,$1.2$ (Figure \ref{fig: offset_dist}; bottom panel). However, our dataset includes both high-resolution \textit{HST} imaging and seeing-limited ground-based observations, and the latter might bias the inferred half-light radii of faint unresolved galaxies to larger values. By performing a homogeneous analysis of the \textit{HST} dataset only, we derive $R_o/R_e$\,$\sim$\,$2$, consistent with the value from the literature \citep{FongBerger2013}. For comparison, the median host normalized offset for long GRBs  is $R_o/R_e$\,$\sim$\,$0.6$ (\citealt[][]{Blanchard2016,Lyman2017}).

Furthermore, based on Figure \ref{fig: offset_dist}, we find that the offset distribution of this sample of sGRBEEs (dark blue lines) is a factor of $3-4\times$ further extended than long GRBs (purple lines). A KS test between the two samples yields $p_\textrm{KS}\approx0.04$ (in both host normalized and physical offset), rejecting the null hypothesis that they are drawn from the same distribution at the $\sim$\,$2\sigma$ level. This provides additional and independent support to the hypothesis that their progenitors are different from those of long GRBs \citep{Norris2006,Gehrels2006,Gal-Yam2006}. 

Moreover, we find that the offset distributions (angular, physical, and host normalized) for classical sGRBs with $T_{90}$\,$<$\,$2$ s (Figure \ref{fig: offset_dist}; cyan lines) and those displaying EE (Figure \ref{fig: offset_dist}; dark blue lines) are consistent with being drawn from the same distributions. The comparison in Figure \ref{fig: offset_dist} is made for 24 classical sGRBs and 10 sGRBEEs, all of which have a sub-arcsecond localization. If we include the offsets to the lowest $P_{cc}$ candidate hosts for hostless events (see \S \ref{sec: including hostless}), increasing the sample sizes to 34 sGRBs and 11 sGRBEEs, we find the same result. 
This suggests that regardless of whether classical sGRBs and sGRBEEs are created by different progenitor systems, their merger environments are indistinguishable based on these limited number of events.

\begin{figure} 
\centering
\includegraphics[width=\columnwidth]{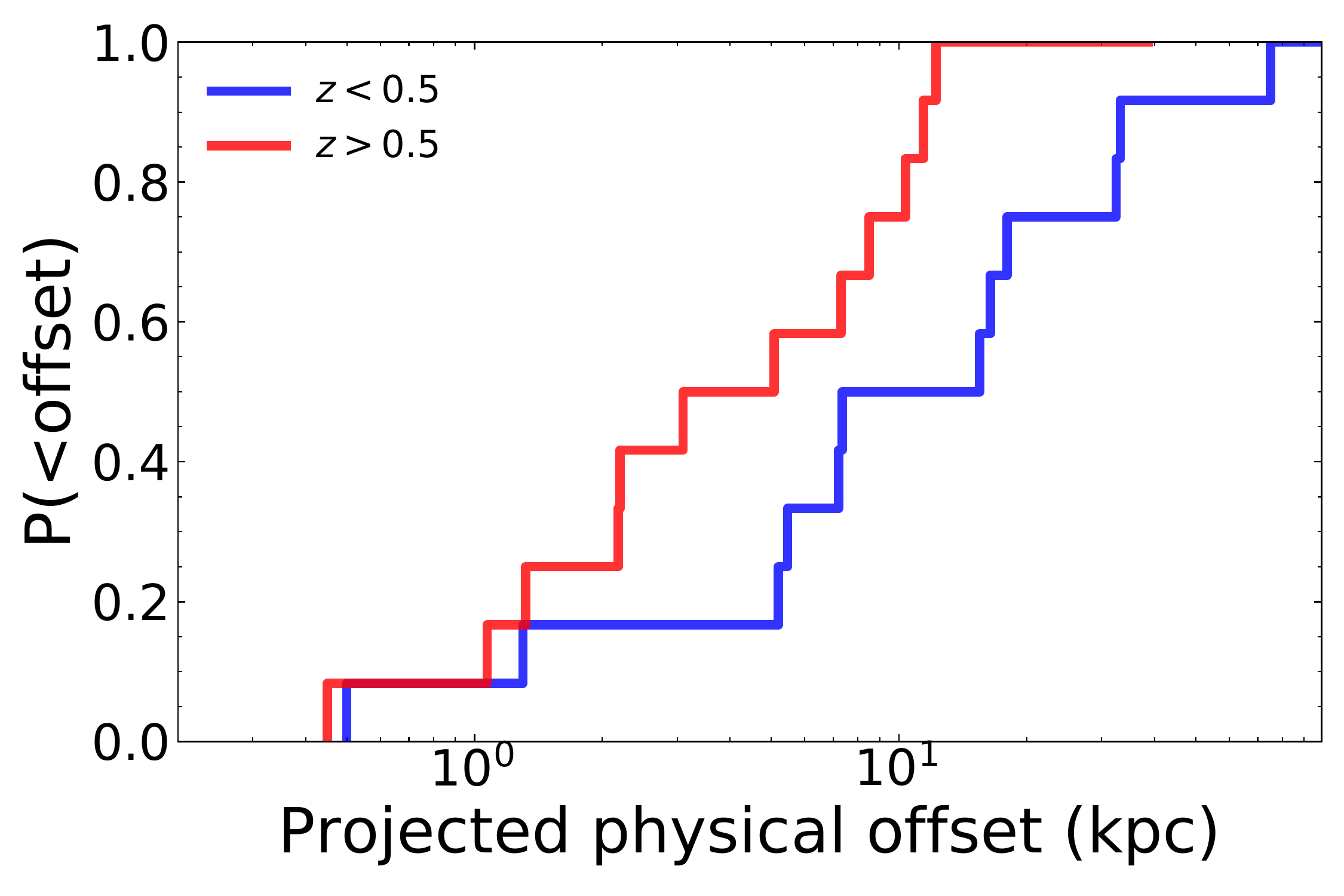}
\includegraphics[width=\columnwidth]{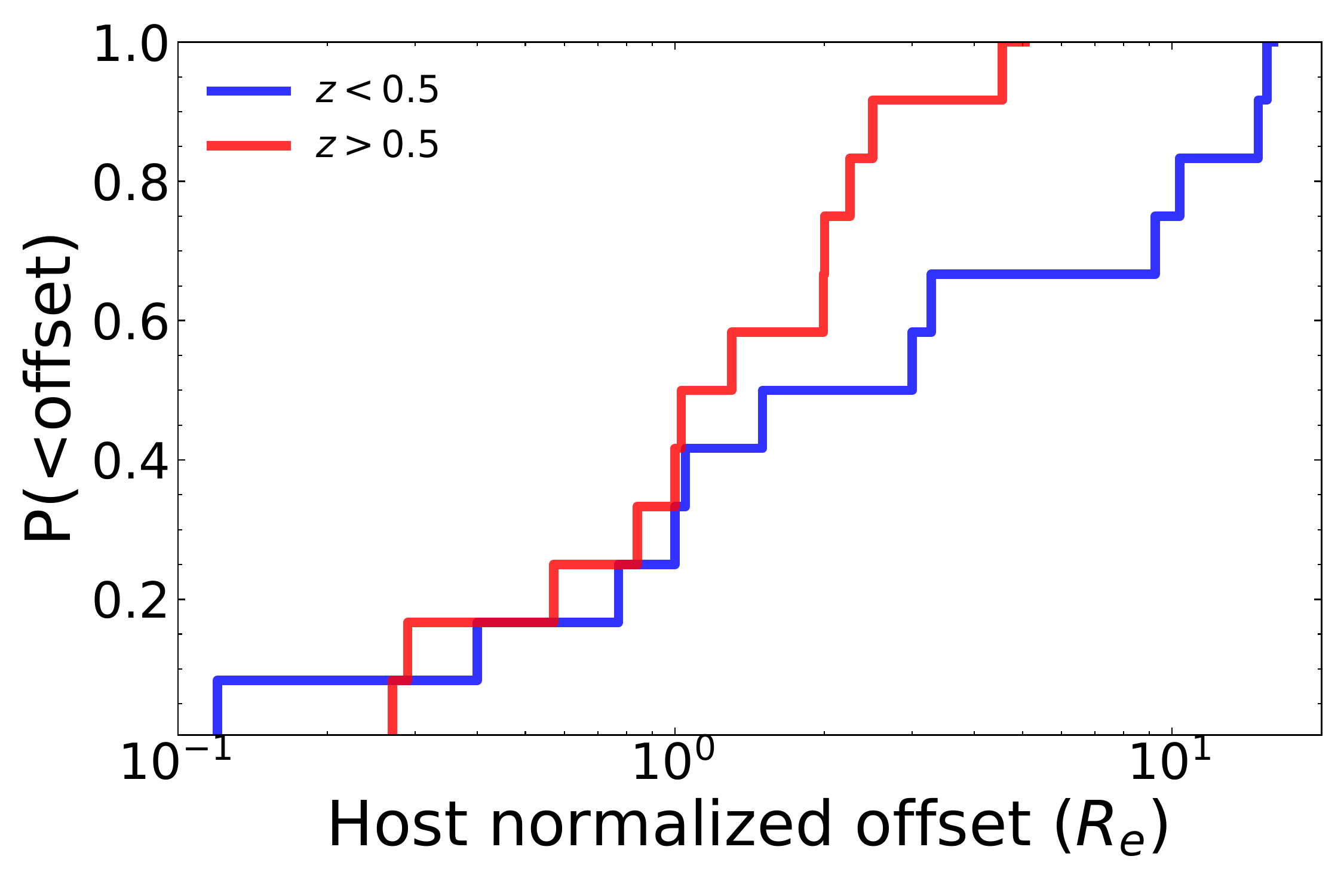}
\caption{
\textit{\textbf{Top}}: Cumulative distribution of projected physical offsets for sGRBs with both a sub-arcsecond localization and spectroscopic redshift at $z$\,$<$\,$0.5$ (blue) and $z$\,$>$\,$0.5$ (red). 
\textit{\textbf{Bottom}}: Same as the top panel but for host-normalized offsets.}
\label{fig: offset_redshift_normed}
\end{figure}

We also explored whether there was an evolution of the observed offset distribution with redshift. In this analysis, we focus only on events with a measured and secure spectroscopic redshift. In Figure \ref{fig: offset_redshift_normed}, we separate the physical offsets for sub-arcsecond localized GRBs into two distributions with $z$\,$<$\,$0.5$ and $z$\,$>$\,$0.5$. 
The median offset for sGRBs at $z$\,$<$\,$0.5$ (7.5 kpc) is a factor of $\sim$\,$2\times$ higher than those at $z$\,$>$\,$0.5$ (3.2 kpc), despite a KS test supporting that they are drawn from the same distribution ($p_\textrm{KS}$\,$=$\,$0.09$).  
In addition, no sGRBs at $z$\,$>$\,$0.5$ have a projected physical offset $>$\,$15$ kpc, compared to 50\% of those at $z$\,$<$\,$0.5$. 
If we perform the same comparison for the host normalized offset distribution (Figure \ref{fig: offset_redshift_normed}), we find that the two samples are again consistent with being drawn from the same distribution with $p_\textrm{KS}$=$0.25$, despite all events at $>$\,$5R_e$ being located at low redshifts. 
Although the distributions are similar statistically, the lack of large offsets at $z$\,$>$\,$0.5$ is suggestive of a redshift evolution effect. The physical implications of this possible redshift evolution are discussed in \S \ref{sec: discussion_bias}.

\subsubsection{Including XRT Localized sGRBs}
\label{sec: XRT_offset}

The previous section focused on sub-arcsecond localized events,
however, the majority of sGRBs have only an XRT localization.
For the sample of 99 events satisfying our selection criteria (\S \ref{sec: sampleselection}), 
the median error on the XRT enhanced position 
is $\sim$\,$1.8\arcsec$.
Due to this large uncertainty,
often comparable to the measured angular offset, 
XRT localized events are difficult to include in the offset distribution. Here, we adopt a Bayesian formalism to identify the true distribution of offsets for XRT localized GRBs. Following \citet{Bloom2002}, we assume that the probability density distribution of the GRB's offset from its host galaxy follows a Rice distribution \citep{Wax1954}, denoted by $\mathcal{R}(x,\mu,\sigma)$ where $\mu$ and $\sigma$ are the shape parameters. 

Applying Bayes' theorem, the posterior distribution for the true offset, $R_\textrm{true}$, of the GRB from its host galaxy's center given the observed offset, $R_\textrm{obs}$, and its uncertainty, $\sigma_R$, is 
\begin{equation}
    P(R_\textrm{true}|R_\textrm{obs})=\frac{P(R_\textrm{obs}|R_\textrm{true})P(R_\textrm{true})}{P(R_\textrm{obs})},
\end{equation}
where the probability density for the likelihood $P(R_\textrm{obs}|R_\textrm{true})$ is given by the Rice distribution $\mathcal{R}(R_\textrm{obs},R_\textrm{true},\sigma_R)$. 

\begin{figure} 
\centering
\includegraphics[width=\columnwidth]{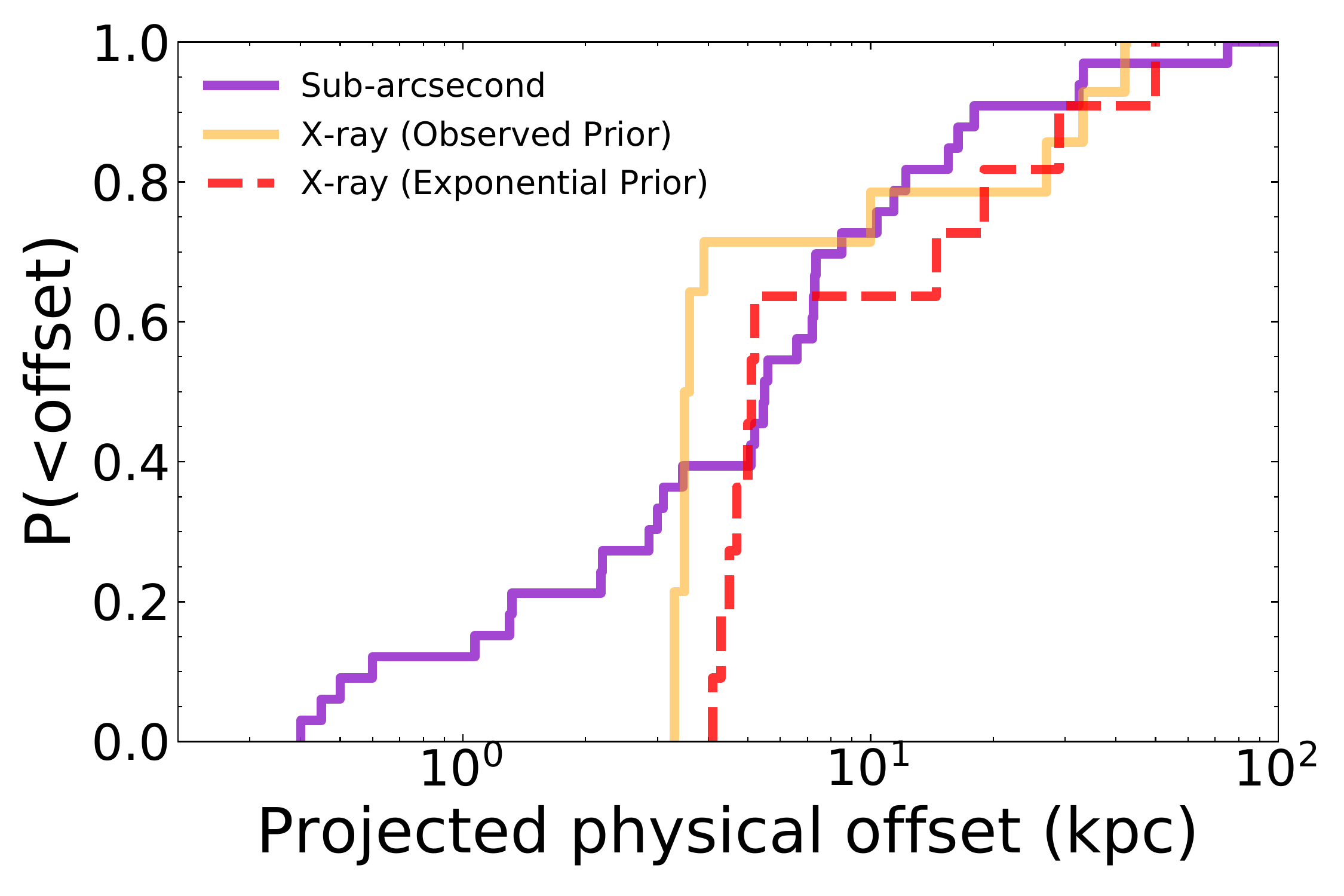}
\caption{Cumulative distribution of sGRB offsets for the sample of sub-arcsecond localized events (purple) compared to X-ray localized events for two different priors (\S \ref{sec: XRT_offset}):  \textit{i}) the ``observed'' prior (yellow) and \textit{ii}) the ``exponential'' prior (red).}
\label{fig: offset_dist_withxray}
\end{figure}

The choice of prior distribution, $P(R_\textrm{true})$, can have a significant impact on the unknown posterior. 
While simple priors may appear to minimize our assumptions on the underlying distribution, we note that they are generally unrealistic. For example, assuming that the GRB has an equal probability of occurring anywhere in a circle surrounding the galaxy's centroid (i.e., uniform probability in area), such that $P(R_\textrm{true})$\,$\propto$\,$R_\textrm{true}$, preferentially favors larger radii. Whereas both observations of sGRBs (Figure \ref{fig: offset_dist}) and models of BNS systems \citep{Bloom1999} find that the significant majority of systems form at $<$\,$10$ kpc. 
Therefore, we consider two different prior distributions:
\textit{i}) following the observed distribution of physical offsets for sub-arcsecond localized sGRBs (Figure \ref{fig: offset_dist}),
and \textit{ii}) assuming that GRBs form following an exponential profile $P(R_\textrm{true})\propto \exp(-R_\textrm{true}/R_*)$ where $R_*$ is taken to be the half-light radius of each galaxy. In Figure \ref{fig: offset_dist_withxray}, we refer to these priors as ``observed'' and ``exponential''.

We choose to adopt the median value of the posterior distribution $P(R_\textrm{true}|R_\textrm{obs})$ for each GRB's offset, and include these XRT localized GRBs within the cumulative distribution of sGRB offsets. In Figure \ref{fig: offset_dist_withxray} we demonstrate how the X-ray localized events impact the offset distribution for the two prior distributions. 
The ``observed'' and ``exponential'' priors only cause a marginal deviation from the sub-arcsecond only distribution. Therefore, based on this analysis, the offsets of X-ray localized events are not inherently different from those with an optical localization. 

\subsubsection{Including Hostless sGRBs}
\label{sec: including hostless}

\begin{figure}
\includegraphics[width =\columnwidth]{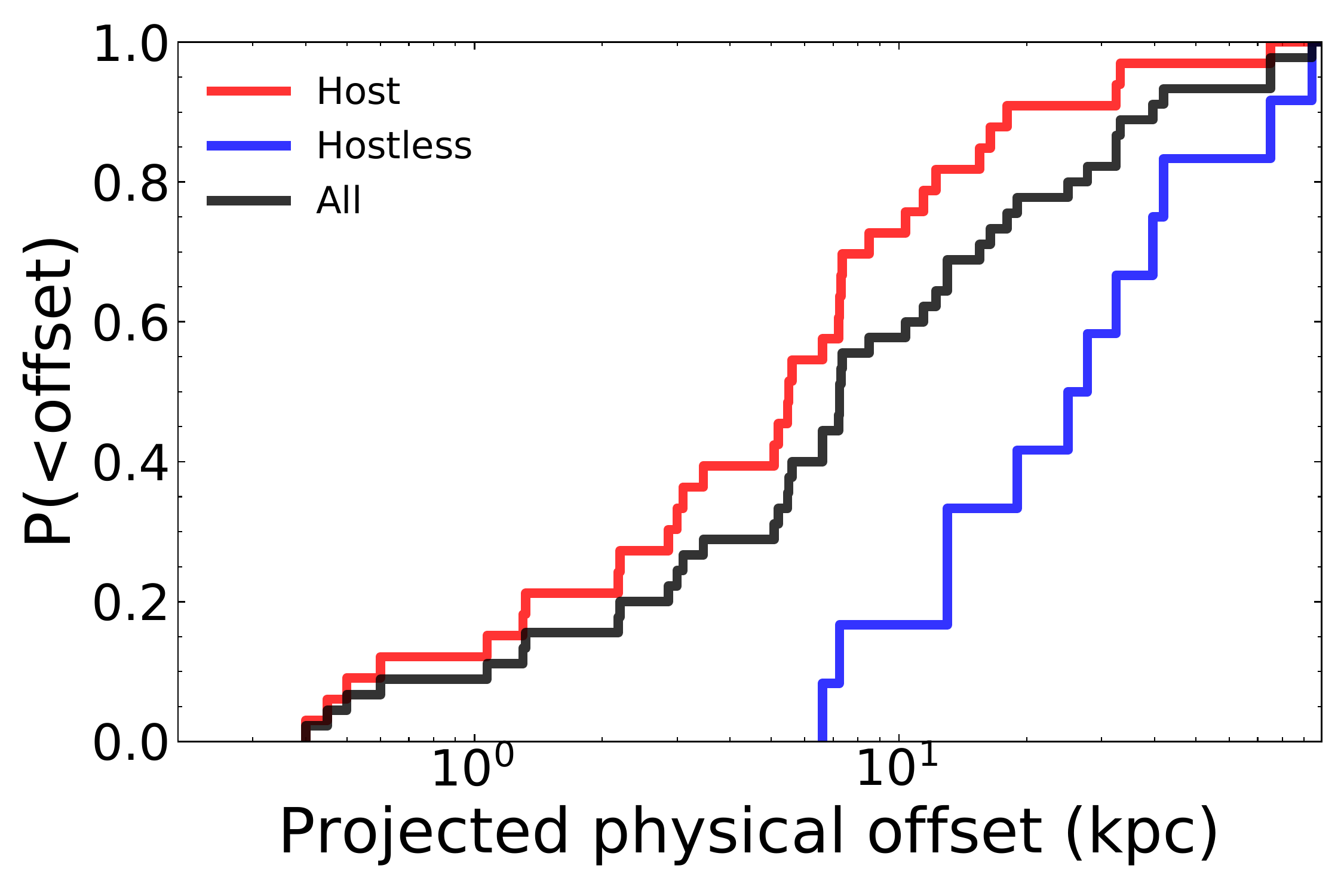}
\caption{Cumulative distribution of projected physical offsets for sub-arcsecond localized sGRB with a putative host (red) and for those which are hostless (blue); the total population is shown in black.
}
\label{fig: host_vs_hostless_offset}
\end{figure}

Up to this point, we have focused on the offset distribution of sGRBs with a confident host galaxy association ($P_{cc}$\,$<$\,$0.1$). 
Here, we include in our study 12 sub-arcsecond localized observationally hostless events. 
For these bursts, we identify the galaxy with the lowest
chance probability $P_{cc}$ and measure the offset between the burst position and the galaxy's centroid (Appendix \ref{sec: appendixsampleanalysis}). Only 2 of these events are located within 10~kpc of their most likely host and, as a result, the median offset for the sample is 26.4~kpc, $5\times$ larger than the value derived in \S \ref{sec: subarcsec offset} (see also Figure \ref{fig: host_vs_hostless_offset}). We further examine the implications of these hostless events in \S \ref{sec: discussion_hostless}. 



\begin{figure*} 
\centering
\includegraphics[width=1.5\columnwidth]{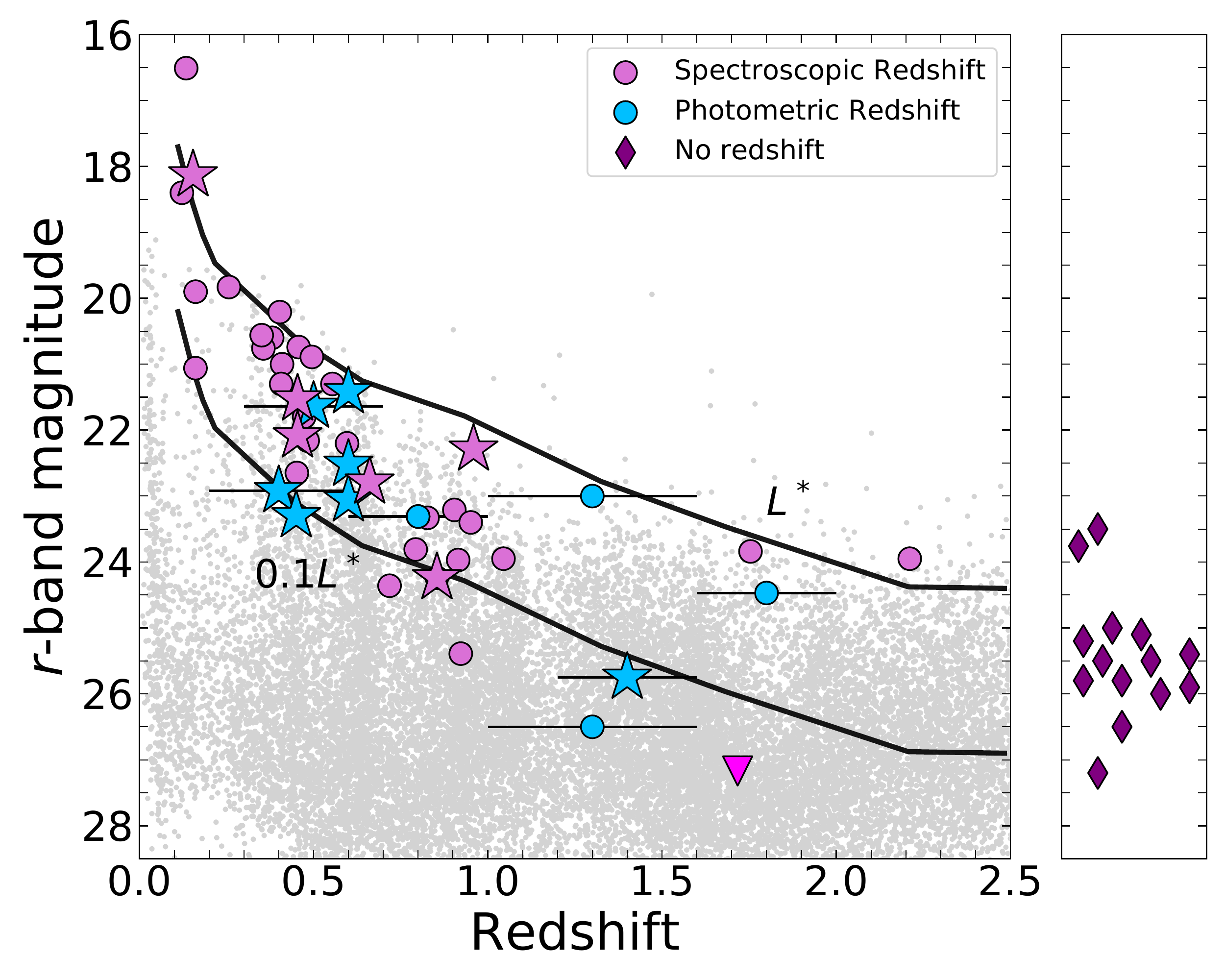}
\caption{Host galaxy $r$-band magnitude versus redshift for the sample of sGRBs included in this work. Spectroscopic (light purple) and photometric (blue) redshift measurements from the literature are represented by circles, and those determined in this work by stars.
The small light gray circles represent galaxies in the CANDELS UDS. The black lines demonstrate the range of $0.1$\,$-$\,$1.0L^*$ galaxies as a function of redshift \citep{Brown2001,Ilbert2005,Willmer2006,Reddy2009,Finkelstein2015}. 
The deep constraint on the host galaxy of GRB 160410A \citep{AguiFernandez2021} is marked by a downward magenta triangle. 
In the right panel, we show the $r$-band magnitude for the host galaxies of sGRBs without a known redshift (dark purple diamonds), including the lowest $P_{cc}$ candidate host of observationally hostless events (see \S \ref{sec: including hostless}). 
Magnitudes have been corrected for Galactic extinction \citep{Schlafly2011}.
}
\label{fig: rband_vs_z}
\end{figure*}

\begin{figure} 
\centering
\includegraphics[width=\columnwidth]{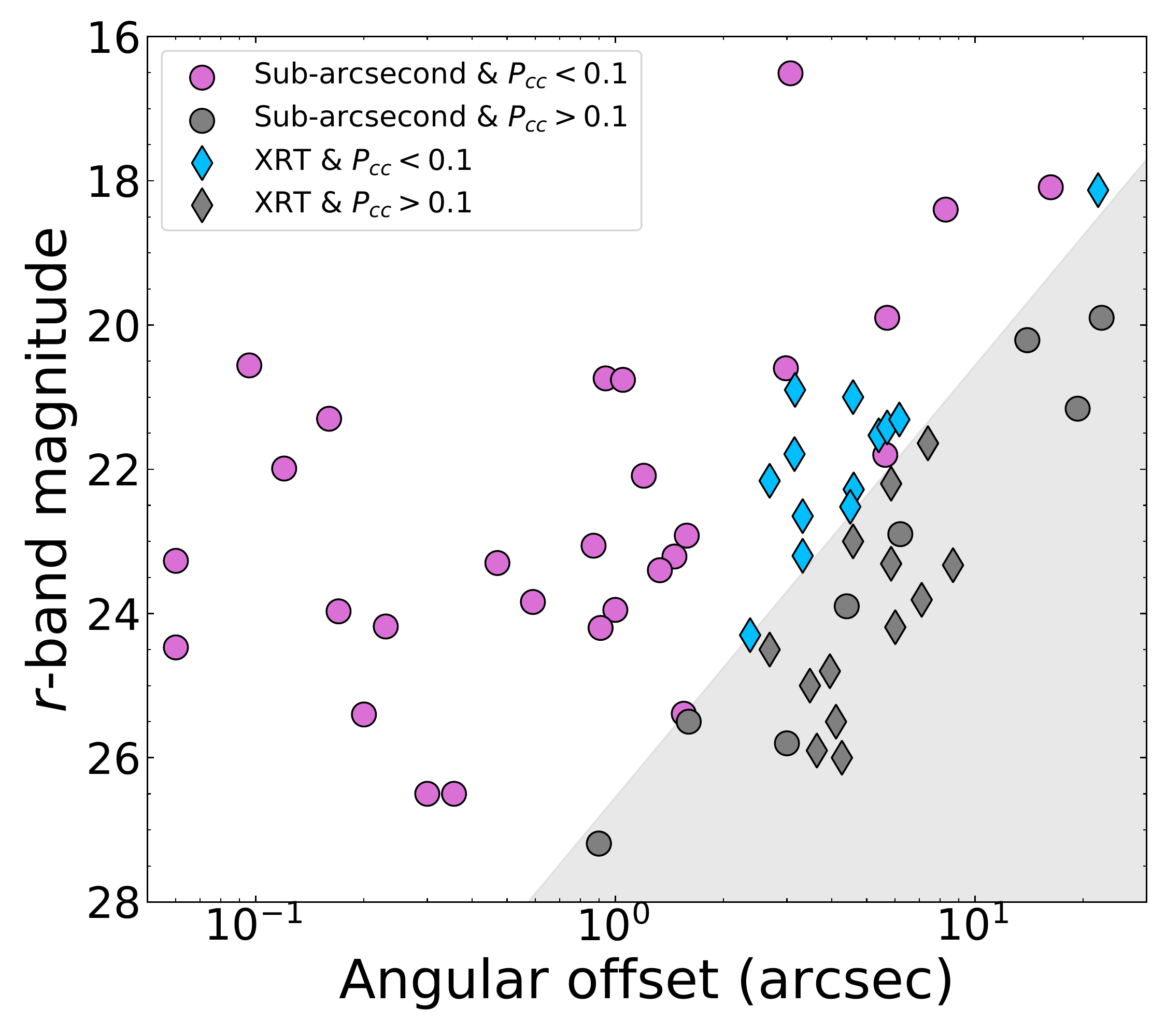}
\caption{Host galaxy $r$-band magnitude versus angular offset for the sample of sGRBs included in this work. 
We also include GRBs where the galaxy with the lowest probability of chance coincidence has $P_{cc}$\,$>$\,$0.1$ (gray). 
The shaded gray region marks where $P_{cc}$\,$>$\,$0.1$.
}
\label{fig: rband_vs_offset}
\end{figure}

\subsection{Host Luminosities}
\label{sec: hostlum}

In Figure \ref{fig: rband_vs_z}, we display the apparent $r$-band magnitude (corrected for Galactic extinction) of sGRB host galaxies plotted against their redshift. By comparing the brightness of these galaxies to a sample of $\sim$\,$30,000$ galaxies from the Cosmic Assembly Near-infrared Deep Extragalactic Legacy Survey project \citep[CANDELS,][]{Koekemoer2011,Grogin2011} Ultra Deep Survey \citep[UDS;][]{Galametz2013}, we confirm that the host galaxies of sGRBs trace the brightest galaxies ($0.1$\,$-$\,$1.0L^*$) at each redshift. In the right panel of Figure \ref{fig: rband_vs_z}, we report the $r$-band magnitude of candidate host galaxies without a known redshift, including the lowest $P_{cc}$ candidate host galaxies of observationally hostless events. 

We have identified that 4 sub-arcsecond localized observationally hostless events within our sample (e.g., GRBs 150423A, 160408A, 160601A, 160927A) have lowest $P_{cc}$ candidates (see \S \ref{sec: including hostless}) with faint $r$-band magnitudes ($r$\,$\gtrsim$\,$24.5$ mag; corrected for Galactic extinction). 
When compared to typical sGRB host galaxies (Figure \ref{fig: rband_vs_z}) this is suggestive of either \textit{i}) an origin at $z$\,$>$\,$1$ or \textit{ii}) a population of under-luminous sGRB galaxies ($<$\,$0.1L^*$). Even if under-luminous, these galaxies would have to occur at $z$\,$>$\,$0.5$ in order to avoid an unexplained gap in luminosity (Figure \ref{fig: rband_vs_z}) between faint galaxies and the known bright hosts at low-$z$. We note that there are only a handful of examples of low luminosity ($<$\,$0.1L^*$) sGRB host galaxies in GRBs 070714B \citep{Cenko2008}, 101219A \citep{Fong2013}, 120804A \citep{Berger2013,Dichiara2021}, and 151229A (this work), all of which reside at $z$\,$>$\,$0.5$.

We observe the same trend in the observationally hostless sample of XRT localized sGRBs (e.g., GRB 140516A, 150831A, 170127B, 171007A, 180727A); there are faint $r$\,$\gtrsim$\,$24.5$ mag candidates detected within their XRT localization's, which range from $2.2$\,$-$\,$2.7\arcsec$ (90\% CL). 

We emphasize that none of these events are located near bright, low-$z$ galaxies (none within $60\arcsec$) from which they could have been kicked. 
This is in contrast to other observationally hostless events, such as sGRBs 061201, 090515, and 091109B, where the most likely host galaxy is a bright, low-$z$ galaxy at a significant offset. We discuss this further in \S \ref{sec: discussion_hostless}. 

In Figure \ref{fig: rband_vs_offset}, we show the $r$-band magnitude of sGRB host galaxies versus the angular offset of the sGRB from its host for both X-ray (diamonds) and optically localized GRBs (circles). The gray shaded region represents the region precluded from a strong host association, due to $P_{cc}$\,$>$\,$0.1$. Based on the distribution of XRT localized events we find that it is difficult to associate a galaxy fainter than $r$\,$>$\,$23.5$ to a GRB lacking a precise, sub-arcsecond localization. 
While the brightest sGRBs may have an X-ray localization (from \textit{Swift}/XRT) of $\sim$\,$1.4$\,$-$\,$1.5\arcsec$ (90\% CL), the majority are less precisely localized to $>2\arcsec$. 
As such, the majority of X-ray localized sGRBs are limited to associations with galaxies brighter than $r$\,$<$\,$23.5$ mag, decreasing the likelihood of association with galaxies at $z$\,$>$\,$1$ (see \S \ref{sec: discussion_bias}).

\subsection{Redshift Distribution}
\label{sec: redshift dist}

\begin{figure} 
\centering
\includegraphics[width=\columnwidth]{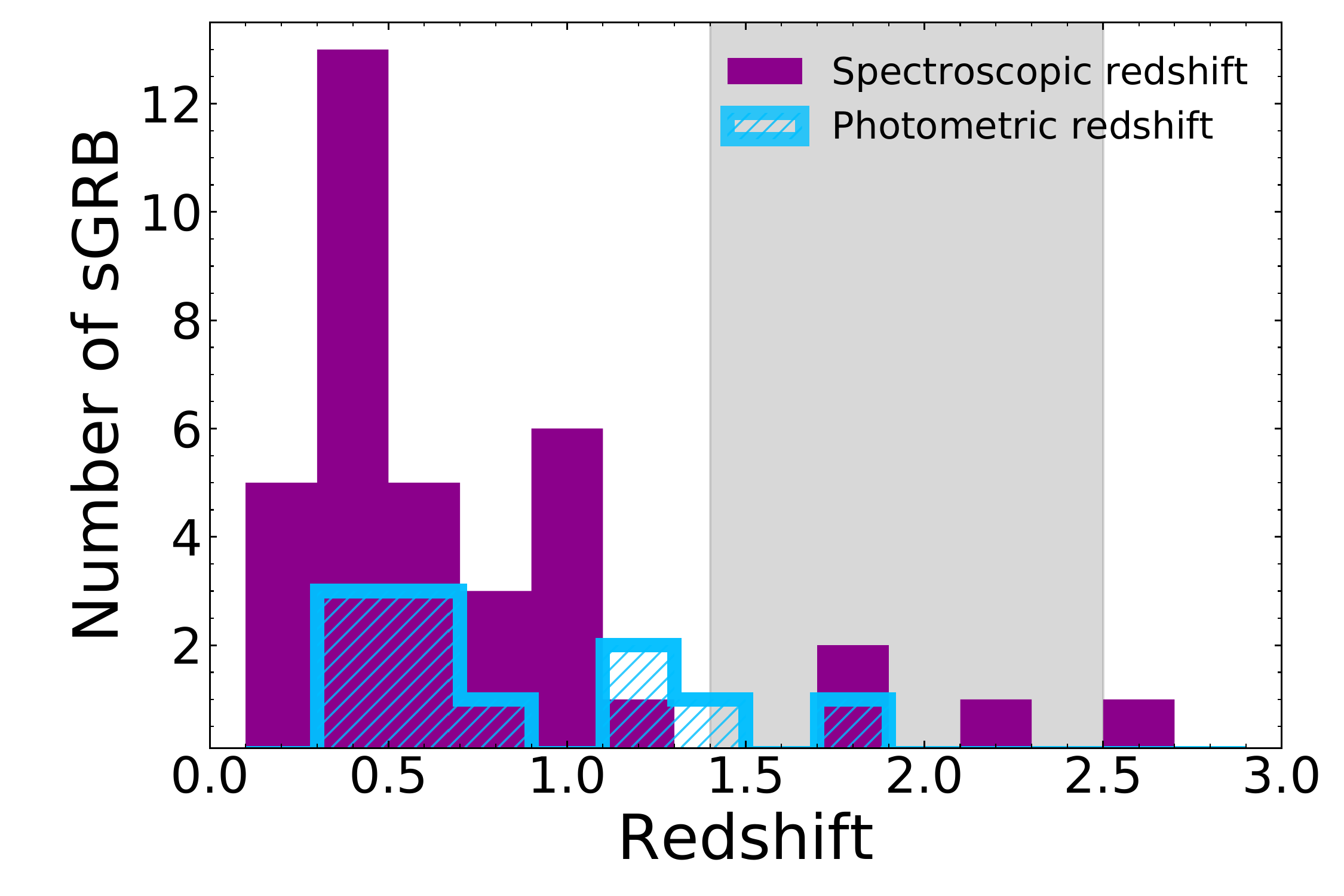}
\includegraphics[width=\columnwidth]{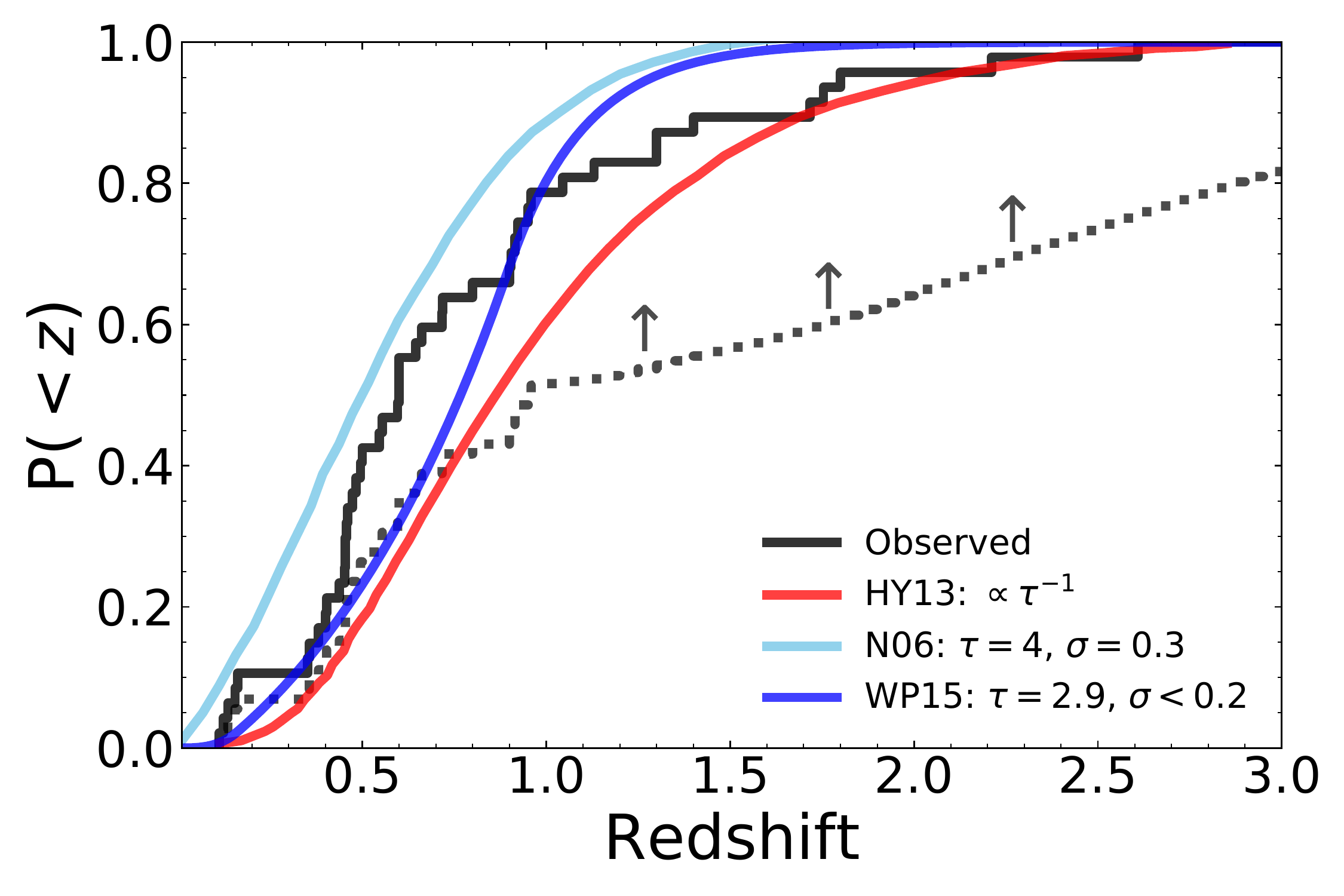}
\caption{\textit{\textbf{Top}}: Histogram of the observed spectroscopic redshifts (purple) for 36 sGRBs matching our selection criteria. We also show a sample of photometric redshifts (blue) for 12 additional events. The gray solid region marks the ``redshift desert'' between $1.4$\,$<$\,$z$\,$<2.5$. 
\textit{\textbf{Bottom}}: Cumulative distribution of sGRB redshifts (black) compared to the expected distribution for several different DTDs \citep{Nakar2006,HaoYuan2013,Wanderman2015}. In these models, $\tau$ represents the delay time. For log-normal distributions, the width of the distribution is given $\sigma$ \citep{Nakar2006,Wanderman2015}. 
The dashed black line represents a lower limit to $P(<z)$ assuming $\sim$\,$50\%$ of the population occurs at $z$\,$>$\,$1$ with a negligible delay time. 
}
\label{fig: redshift_dist}
\end{figure}

Our sample consists of 72 well-localized sGRBs (including the sub-class of sGRBEEs) observed in homogeneous conditions. Of these, 37 (51\%) have a spectroscopic redshift, 11 (16\%) a photometric redshift, and 24 (33\%) lack a distance measurement. 
Only three of these redshift measurements come from direct afterglow spectroscopy, whereas the large majority are determined from the putative host galaxy. 
In Figure \ref{fig: redshift_dist} (top panel), we display a histogram of the observed redshift distribution. 
The median value is $z$\,$\approx$\,$0.5$ for the sample of spectroscopic redshifts, and $z$\,$\approx$\,$0.6$ for the combined sample of photometric and spectroscopic redshifts. 
By adding $4$ spectroscopic redshifts at $z$\,$>$0.5 and 
$7$ photometric redshifts at $z$\,$>$0.4, our work mainly populates the upper tail of the distribution. This shows the importance of deep imaging and spectroscopy, using large aperture $8$\,$-$\,$10$m telescopes, in probing the most distant sGRBs and their faint host galaxies. 
However, only 1 of our events lies at $z>1$ (Table \ref{tab: host properties}). This is not surprising as our survey is optically-driven and affected by complex selection effects,  such as the so-called ``redshift desert'' ($1.4$\,$<$\,$z$\,$<2.5$; also marked in Figure \ref{fig: redshift_dist}) where common nebular emission lines are shifted towards infrared wavelengths. 
A similar systematic survey of sGRBs at nIR wavelengths would be essential to complement our study and extend the redshift distribution of sGRBs.  

The number of distant sGRBs is an important constraint
for progenitor models and their delay time distribution (DTD).
In Figure \ref{fig: redshift_dist} (bottom panel), we show the cumulative distribution of sGRB redshifts (including photometric redshifts) compared to predictions based on different DTD models. 
The two models commonly adopted in the literature are:  \textit{i}) a log-normal distribution \citep{Nakar2006,Wanderman2015} and \textit{ii}) a power-law with decay index between $\sim$\,$-1$ to $-1.5$ \citep{HaoYuan2013}. 

%


A KS test between our distribution and the \citet{Nakar2006} model yields $p_{KS}$\,$=$\,$10^{-2}$, rejecting the null hypothesis that the observed redshift distribution is drawn from their model.
The observed distribution appears instead consistent with the power-law DTD models with slope $\sim$\,$-1$ to $-1.5$\footnote{We note that the redshift distribution also depends on the assumptions as to the SFH, gamma-ray luminosity function, detector sensitivity, and minimum delay time, and can therefore be different even for the same DTD.}. 
However, a significant population of bursts with no known redshift exists. Our survey identifies that their likely host galaxies are much fainter than the rest of the sample (Figure \ref{fig: rband_vs_z}), and a likely explanation is that these bursts represent a missing population of high-$z$ sGRBs. A larger number of $z$\,$>$\,$0.5$ events increases the tension with the log-normal DTD models.

In the most extreme case, these would be prompt mergers with a negligible delay time between formation and merger. 
In Figure \ref{fig: redshift_dist} we show the implications of this scenario. 
The dotted black line represents the hypothetical redshift distribution derived assuming that all the bursts with no known redshift follow the SFH of the Universe \citep{Moster2013}. 
This sets a lower limit to the true redshift distribution and helps constrain the parameter space allowed by observations.
By assuming that sGRB progenitors are described by a single DTD function, 
the \citet{HaoYuan2013} curve is consistent with all the observing constraints.

\subsection{Circumburst Environment}
\label{sec: environment}

\begin{figure} 
\centering
\includegraphics[width=\columnwidth]{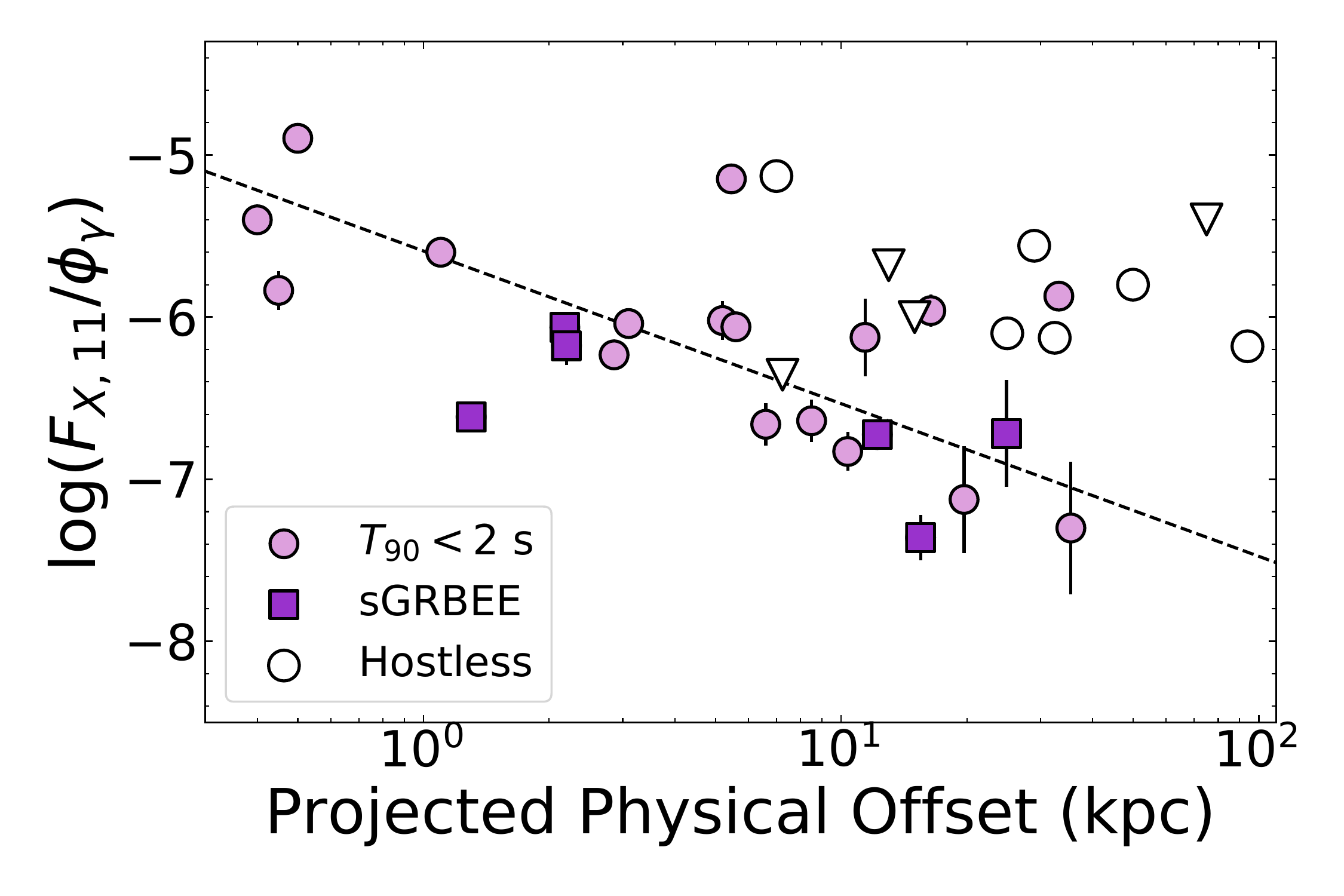}
\caption{Ratio of $0.3$\,$-$\,$10$ keV X-ray flux at 11-hours, $F_{X,11}$, to the $15$\,$-$\,$150$ keV gamma-ray fluence, $\phi_\gamma$, versus the projected physical offset from the sGRB host galaxy. sGRBs with $T_{90}$\,$<$\,$2$ s are represented by light purple circles, sGRBEE by dark purple squares and observationally hostless events (adopting the offset to their lowest $P_{cc}$ candidate host) are displayed by light gray circles. Events with upper limits on $F_{X,11}$ are shown by downward triangles. 
The sample of events is compiled from \citet{Nysewander2009,Berger2014,OConnor2020,OConnor2021}.}
\label{fig: fluxfluence_vs_offset}
\end{figure}


In this section, we explore the consistency between the observed offsets of sGRBs around their galaxies and their inferred circumburst environment based on observations of their afterglows in X-rays. 
First, we use the onset of the X-ray afterglow from \textit{Swift}/XRT to set a lower limit to the circumburst density for each of the 31 bursts in our sample (see \citealt{OConnor2020} and our Appendix \ref{appendix: densitycalc}). Of these 31 bursts we find that $<33\%$ have a circumburst density consistent with $n_\textrm{min}$\,$<$\,$10^{-4}$ cm$^{-3}$, setting an upper limit to the fraction of sGRBs in this sample occurring in a IGM-like environment (physically hostless; see Appendix \ref{appendix: densitycalc}). Of these potentially low-density events, 5 are observationally hostless (Table \ref{tab: XrayAGprop}).  



Moreover, we searched for a correlation between the GRB offsets and their high-energy properties. In particular, the ratio of the X-ray flux at 11-hours, $F_{X,11}$, to the prompt gamma-ray fluence, $\phi_\gamma$, is known to probe the circumburst density such that $F_{X,11}/\phi_\gamma$\,$\propto$\,$n^{1/2}$ \citep{Sari1998,Wijers1999,Granot2002}. This is valid only in the synchrotron slow cooling regime when the cooling frequency lies above the X-ray band, and does not accounts for energy injection from the central engine. Moreover, this quantity $F_{X,11}/\phi_\gamma$ is independent of distance. 
In Figure \ref{fig: fluxfluence_vs_offset}, we observe that there is a large scatter in the correlation (see also \citealt{OConnor2020}). Although GRBs with small offsets tend to occupy the upper part of the plot, and those with larger offsets the lower part, no trend can be conclusively established.



We find no evidence for a population of bursts in a rarefied environment (i.e., a low ratio of X-ray flux to gamma-ray fluence in comparison to other events at a similar offset. For example, see GRB 211211A, \citealt{Troja2022}). 
Instead, we find that observationally hostless sGRBs (e.g., sGRBs 061201, 091109B, 110112A, 111020A, 160601A, 160927A) are not X-ray faint when compared to the overall population, as they all lie above $\log(F_{X,11}/\phi_\gamma)$\,$\gtrsim$\,$-6.1$. 
While these events have no secure host association, we paired them with their most likely host galaxy to calculate their offsets in Figure~\ref{fig: fluxfluence_vs_offset}. 
However, the X-ray brightness of their afterglows does not support the large offset/low density scenario implied by these galaxy's associations and may suggest that they reside in faint hosts at $z$\,$>$\,$1$.

\section{Discussion}
\label{sec: discussion}



\subsection{A Redshift Evolution of sGRB Locations}
\label{sec: discussion_bias}

By exploring the distribution of sGRB offsets at $z$\,$<$\,$0.5$ and $z$\,$>$\,$0.5$ (Figure \ref{fig: offset_redshift_normed}; top panel), we identified a redshift evolution in the locations of sGRBs around their galaxies. Based on our analysis, there are no events with $z$\,$>$\,$0.5$ at physical offsets $>$\,$15$ kpc, compared to 50\% at $z$\,$<$\,$0.5$. We examine three possible factors which could be at the origin of the observed trend: \textit{i}) an evolution of the host galaxy size,  \textit{ii}) an intrinsic property of their progenitors, or \textit{iii}) an observational bias against dim high-z galaxies. 

The increased size of sGRB host galaxies over cosmic time possibly leads to a larger birth radius of the progenitor, and therefore a larger offset. 
This is consistent with observations of galaxy size evolution following the relation $R_e$\,$\propto$\,$(1+z)^{-\alpha}$ with $\alpha$\,$\approx$\,$0.6$\,$-$\,$1.3$ \citep[see, e.g.,][]{Dahlen2007,vanderWel2008,Papovich2012,Ribeiro2016,Allen2017,Paulino-Afonso2017} leading to growth by a factor of $\sim$\,$2\times$ between $z$\,$=$\,$1$ and the present. It is not clear if this growth is completely due to a true galaxy evolution effect or an observational bias due to surface brightness dimming with distance. Nonetheless, we show that, when normalized by the host galaxy's size, the two distributions at $z$\,$<$\,$0.5$ and $z$\,$>$\,$0.5$ move closer to each other (Figure \ref{fig: offset_redshift_normed}). In particular, for offsets $<$\,$R_e$ they seem to track each other well. However, we find that all events with offsets $>$\,$5R_e$ reside only in low-$z$ galaxies.



\begin{figure} 
\centering
\includegraphics[width=\columnwidth]{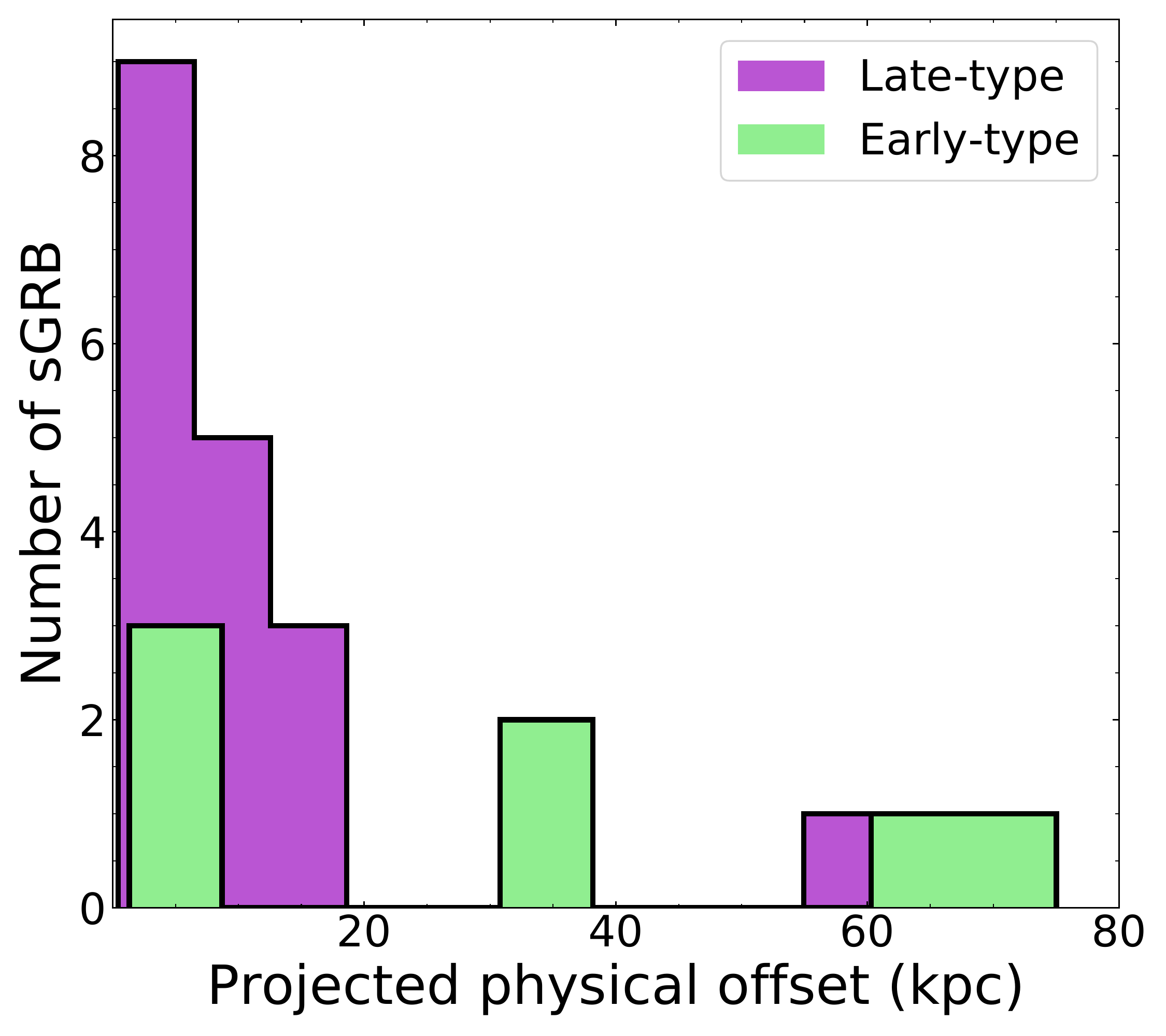}
\caption{Histogram of projected physical offset of sGRBs from their host galaxies. The distribution for late-type galaxies is shown in purple, and early-type hosts in green \citep{Gompertz2020,Paterson2020,OConnor2021}. We have limited the sample to those with classified galaxy type and an error on their offset of $<$\,$20\%$. 
}
\label{fig: offset_vs_type}
\end{figure}

By correlating the physical offset with host galaxy type (see Figure \ref{fig: offset_vs_type}), we find that low-$z$ early-type galaxies preferentially host these sGRBs with large spatial offsets. These events are commonly interpreted as highly kicked BNS systems \citep{Behroozi2014,Zevin2020} or BNS mergers dynamically formed in globular clusters \citep{Salvaterra2010,Church2011}. 
However, we note that an alternative possibility is that the sGRB progenitors were formed in the extended stellar halo of their galaxy \citep{PeretsBeniamini2021}, and as such do not require large natal kicks. Thus, the large host normalized offsets may be due to the fact that $R_e$ is not a good tracer of the extended stellar halo in early-type galaxies \citep{DSouza2014,Huang2018}. 

Another physical explanation for this evolution is that systems merging at low redshifts had a longer delay time between formation and merger of the binary, allowing them to travel further distances than those merging at higher redshifts. However, through population synthesis, \citet{Perna2021} found the opposite trend: simulated BNS at high redshift reach a larger distance from their host galaxies. In fact, they found that $\sim$\,$20\%$ of BNS systems in simulated galaxies at $z$\,$=$\,$1$ reach offsets $>$\,$15$ kpc, whereas none have been identified observationally. Future population synthesis modeling, specifically using inferences from observations of Galactic BNS systems \citep{Beniamini2016,Beniamini2016p2,Tauris2017,Abbott2017kick,Kruckow2018,VignaGomez2018,Andrews2019a,BeniaminiPiran2019}, is required to discern whether these results are expected under different assumptions for the delay time and natal kick distributions.

Nevertheless, we bear in mind that an alternative scenario to explain the redshift evolution is an observational bias against faint high-$z$ galaxies. This bias can most easily be understood based on Figure \ref{fig: rband_vs_z}, where the decrease in host galaxy apparent magnitude as a function of redshift is displayed. For instance, above $z$\,$>$\,$1$ the majority of galaxies in the universe are fainter than $r$\,$>$\,$23.5$ mag, with a significant fraction dimmer than $r$\,$>$\,$25$ mag. In order to associate a GRB to such faint galaxies (Figure \ref{fig: rband_vs_offset}) requires an offset of  $\lesssim$\,$3\arcsec$ (corresponding to $\lesssim$\,$25$ kpc, assuming $z$\,$\approx$\,$1$). This condition becomes more stringent if the probability of chance coincidence cutoff threshold is decreased from the 10\% value used in this work (\S \ref{sec: Pcc}). For example, adopting a cutoff value of 5\%, as used in previous studies \citep{FongBerger2013}, requires an offset $\lesssim 2.2\arcsec$ or, equivalently, $\lesssim$\,$18$ kpc, even for sub-arcsecond localized sGRBs. Surprisingly, even a Milky Way-like spiral galaxy at $z$\,$\approx$\,$1$ ($r$\,$\approx$\,$23$ mag) will have a probability of chance alignment larger than 5\% (10\%) if the projected physical offset is $>$\,$20$ (30) kpc \citep{Tunnicliffe2014}. Therefore, we find that it is unlikely, based on probabilistic grounds, to associate high-$z$ sGRBs to galaxies at large physical offsets. This bias may explain, at least in part, the observed redshift evolution of sGRB offsets and should be taken into account when comparing the observed offset distribution to progenitor models.

\subsection{Hostless Short GRBs}
\label{sec: discussion_hostless}

\subsubsection{Observationally Hostless Fraction}

\begin{figure} 
\centering \includegraphics[width=\columnwidth]{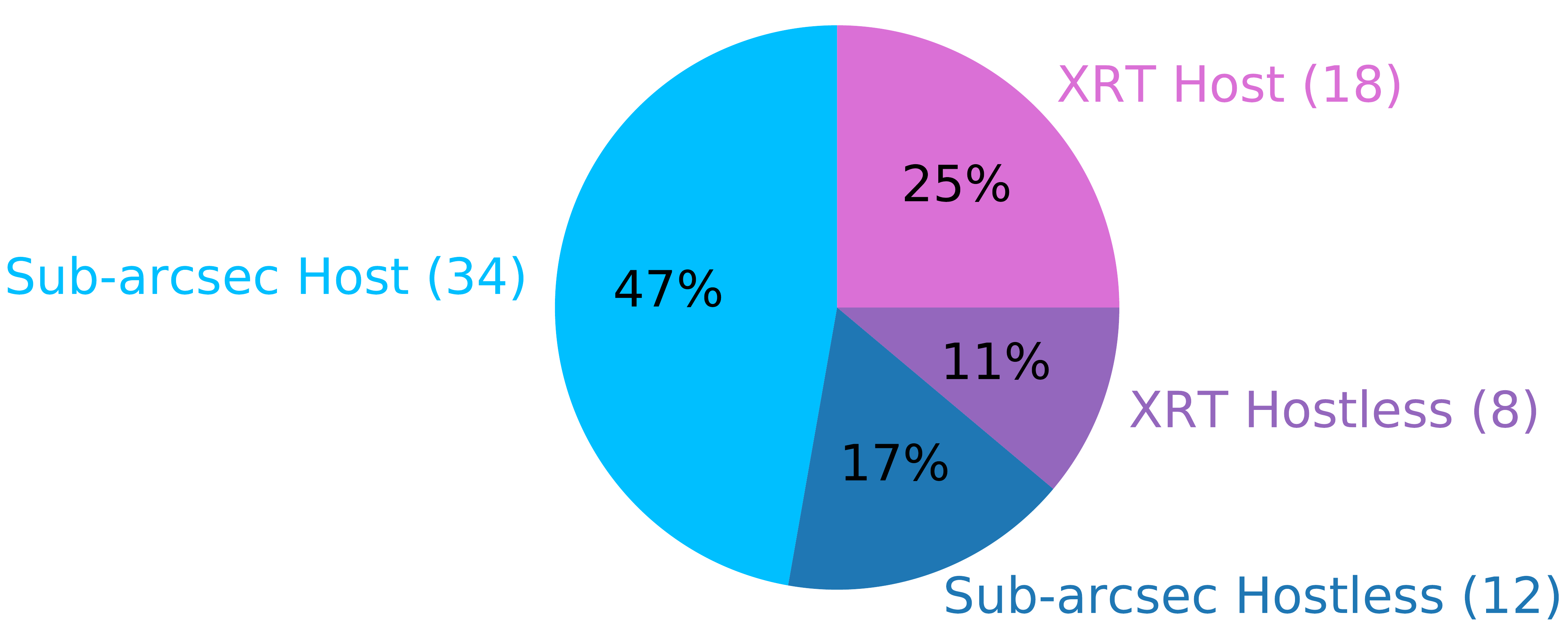}
\caption{Breakdown of the fraction of 72 events considered in this work into those with a putative host galaxy and those that are considered hostless. We have separated these events further based on their localization either with XRT (purple) or to a sub-arcsecond position (blue). The total fraction of hostless events is 28\% (11\% XRT and 17\% sub-arcsecond localized).  The total number of hostless events is 20, with 12 of them having a sub-arcsecond localization. 
}
\label{fig: new_pie}
\end{figure}

We have selected a homogenous sample (\S \ref{sec: sampleselection}) of short GRBs detected by \textit{Swift}/BAT of which 72 have a sensitive search for their host galaxy. We identify that $\sim$\,$28\%$ (20 events) of these 72 events are observationally hostless (see Figure \ref{fig: new_pie} for a breakdown of the fraction of events with and without a host separated by their localization). This fraction is higher than the value of $17\%$ reported by \citet{Fong2013}. We find that this difference is mainly driven by the larger sample of X-ray localized events studied in our work. Considering only the sample with sub-arcsecond positions, 
the hostless fraction is $26\%$, consistent between the two works. 

As the fraction of hostless sub-arcsecond localized events is consistent with the full population, we find that our result is not driven by the lower accuracy of X-ray localized events. In fact, in \S \ref{sec: XRT_offset}, we demonstrated that the offsets of X-ray localized events are consistent with the locations of sub-arcsecond localized sGRBs (Figure \ref{fig: offset_dist_withxray}). This suggests that any selection bias against large offsets or low-density environments acts on both samples in the same way.

\subsubsection{Interpretation of Hostless Events}

We emphasize that there is a lingering ambiguity as to the origin of hostless short GRBs. The main scenarios are that \textit{i}) the GRB was kicked to a substantial distance from its birth galaxy, such that the probability of chance alignment is large, or \textit{ii}) the GRB merged in a faint, undetected galaxy at a smaller angular distance. However, the diagnosis for individual events is complicated, and it is difficult to distinguish between these two scenarios. 
For instance, the hostless sGRBs presented by \citet{Berger2010a} are located at a significant offset ($30$\,$-$\,$75$ kpc) from bright low-$z$ galaxies ($z<0.5$). However, despite their brightness, the probability of chance coincidence is $\gtrsim 10\%$. Therefore, it is not clear whether these sGRBs are truly associated to these low-$z$ galaxies, or whether they reside in faint, undetected hosts ($H$\,$>$\,$26$ mag). The interpretation has a direct impact on the energetics, redshift (\S \ref{sec: redshift dist}), and delay time distributions of sGRBs.

In this work, we have tripled the number of observationally hostless sGRBs (from 7 to 20 events). We find that half of the observationally hostless sGRBs lack any nearby (low-$z$) candidate host. These events are more likely to have exploded in faint $r\gtrsim 24.5$ mag galaxies (see \S \ref{sec: hostlum}) that are consistent with $0.1-1.0L^*$ galaxies at $z>1$. We note, however, that an alternative explanation is that these represent a population of low luminosity ($<$\,$0.1 L^*$) galaxies hosting sGRBs at $z$\,$<$\,$1$, although this is at tension with the population of well-determined sGRB hosts ($0.1$\,$-$\,$1 L^*$; \citealt{Berger2010a}) and with predictions from population synthesis modeling, which find that BNS systems preferentially form in the most massive (brightest) galaxies \citep{Behroozi2014,Mapelli2018,Artale2019,Artale2020,Adhikari2020,Mandhai2021,Chu2022}. 

Previous work in the literature \citep[see, e.g., ][]{Berger2010a,Tunnicliffe2014} has focused on the likelihood to detect faint galaxies at high-$z$, as opposed to the large probability of chance coincidence even in the event that a galaxy is detected. We find that despite detecting these faint galaxies, they are difficult to confidently associate to the GRB using the standard probability of chance coincidence methodology \citep{Bloom2002}. This is indicative of an observational bias against faint galaxies (see also \S \ref{sec: discussion_bias}).

We note that a larger population of sGRBs at $z$\,$>$\,$1$ implies a steep DTD with an increased fraction of events with short delay times, as deduced based on Galactic BNS systems \citep[][]{BeniaminiPiran2019}. This would further disfavor log-normal DTD models (\S \ref{sec: redshift dist}), and support a primordial formation channel for these events.

or the
We further explored the sample of observationally hostless events that lie close to low-$z$ galaxies. We exploited their high-energy properties to probe their environments (\S \ref{sec: environment}), as their circumburst density can be used to constrain their allowed physical offset \citep{OConnor2020}. Figure \ref{fig: fluxfluence_vs_offset} shows a weak correlation between X-ray afterglow brightness with the sGRB location, such that a larger offset leads to fainter X-ray emission. The X-ray constraints for hostless events are either too shallow or inconsistent with the observed trend. Although this does not conclusively rule out that these hostless sGRBs could be mergers kicked out into the IGM (physically hostless), it does not offer observational support and leaves their nature undetermined. Rapid and deep X-ray observations with next-generation instruments (e.g., the \textit{Athena X-ray observatory}; \citealt{athena}) will be capable of probing X-ray fluxes of $\sim$\,$10^{-16}$ erg cm$^{-2}$ s$^{-1}$ within 12 hr of the GRB trigger, and, therefore, will be able to detect the low flux regime of physically hostless sGRBs. \textit{Athena} will require input from dedicated GRB and gamma-ray missions \citep{Piro2021Athena}, such as \textit{SVOM} \citep{SVOM2011,SVOM2015}, \textit{THESEUS} \citep{THESEUS}, the \textit{Gamow Explorer} \citep{White2021}, \textit{SIBEX} \citep{RomingSIBEX}, \textit{STROBE-X} \citep{Ray2019}, and \textit{AXIS} \citep{AXIS}, among other proposed and future missions, in order to rapidly locate and target sGRBs.

We note that the main factor preserving the ambiguity in interpreting these events is that the distance scale to the sGRB is not known.
Therefore, in order to disentangle between faint hosts and large offsets we require better constraints as to the distance to short GRBs. The most critical observational tests are \textit{i}) rapid afterglow spectroscopy to determine redshift independent of the galaxy association (e.g., GRB 160410A; this work and \citealt{AguiFernandez2021}), \textit{ii}) the conclusive identification of a kilonova, providing indirect evidence of the GRB distance scale \citep{Troja2019, Chase2022}, or \textit{iii}) the advent of next generation GW detectors capable of detecting compact binaries at cosmological distances  \citep{Punturo2010,Dwyer2015}.

\section{Conclusions}
\label{sec: conclusions}

We carried out a systematic study of the host galaxies of 31 short GRBs. This analysis effectively doubles the sample of well-studied sGRB host galaxies, leading to a total of 72 events fitting our selection criteria with sensitive searches for their host. We assign a spectroscopic redshift to 5 of these events, and derive a photometric redshift 
for 7 others. Based on the results of this study, we present the subsequent findings:

\begin{enumerate}[leftmargin=\parindent]
\item The sub-arcsecond localized population of sGRBs has a median projected physical offset of $5.6$ kpc ($4\times$ larger than for long GRBs; \citealt{Blanchard2016,Lyman2017}), with $70\%$ of events occurring at $<$\,$10$ kpc from their host's nucleus.
\item We find that 28\% of sGRBs (20 out of 72) lack a putative host galaxy to depth $r$\,$>$\,$26$ mag. For half of these hostless bursts, the most likely host is a faint ($r$\,$>$\,$24.5$ mag) galaxy consistent with a high redshift origin ($z$\,$>$\,$1$). 
\item Based on this evidence and the larger sample of $48$ redshifts, we have presented improved constraints on the redshift distribution of sGRBs. We find that 20\% of sGRBs with known redshift lie above $z$\,$>$\,$1$, although this number could be as high as 50\% when including the population of events with no known host. The data is inconsistent with log-normal DTDs for their progenitors, and instead favors power-law models with index $-1$ or steeper.
\item By correlating the high-energy properties of sGRBs with their locations, we find evidence of a possible trend linking the X-ray brightness to the distance from the host galaxy. We point out that hostless events, if associated to their most likely nearby galaxy, do not follow this trend. Hence, their X-ray brightness does not lend support to their interpretation as mergers in a rarefied medium.
\item We find that sGRBEEs are inconsistent with the offset distribution of long GRBs in both projected physical offset and host normalized offset. This conclusion is reached independently of classical sGRBs.
\item 
Lastly, we uncover that the low redshift population of sGRBs is further offset by a factor of $2\times$ from their hosts compared to the sample at $z$\,$>$\,$0.5$ with the median value increasing from 3.2 to 7.5 kpc. This redshift evolution can be explained either by a physical evolution in their progenitors or the larger size of low-$z$ galaxies. Another possibility is that the apparent redshift evolution is due to a selection bias against faint galaxies that reside at higher redshifts.
\end{enumerate}


We emphasize that while late-time observations alone cannot allow for concrete host associations for events at $>$\,$50$ ($25$) kpc past $z$\,$\gtrsim$\,$0.1$ ($1.0$), rapid optical spectroscopy can determine the GRB's distance scale and yield a confident host galaxy assignment. Moreover, rapid and deep optical and infrared observations can lead to the identification of a kilonova, providing an indication of the GRB's distance. These transient are expected to be detectable out to $z$\,$\sim$\,$1$ with both current (\textit{James Webb Space Telescope}; \textit{JWST}) and future observatories (e.g., the 39-m Extremely Large Telescope; \citealt{ELT}).

In addition, the combination of next generation GW detectors (i.e., Einstein Telescope and Cosmic Explorer; \citealt{Punturo2010,Dwyer2015}) with EM observations can allow for confident associations (out to $z$\,$\sim$\,$4$\,$-$\,$10$; \citealt{Hall2019,Singh2021}) as the distance of the GW event can be compared to nearby galaxies. 
This will allow us to unambiguously distinguish between the large offset scenario and a high-$z$ explanation for observationally hostless sGRBs. 

Lastly, future infrared observations with \textit{HST} and \textit{JWST} will probe lower stellar mass galaxies as a function of redshift (Figure \ref{fig: 130912A_HST_vs_Keck}), allowing for more robust limits on the possible faint (high-$z$) galaxies these sGRBs. High resolution observations would also allow for an accurate morphological analysis of the detected hosts, leading to a better understanding of the ratio of early- to late-type galaxies, which yields important information as to the age and formation channels of sGRB progenitors and can illuminate whether events at large offsets are due to kicks or formation in their galaxy's halo.

\section*{Acknowledgements}

The authors would like to thank the reviewer for their thoughtful feedback on the manuscript. B.~O. acknowledges useful discussions with Phil Evans and Geoffrey Ryan. B.~O. thanks Amy Gottlieb for assistance in obtaining LDT observations.

B.~O. was partially supported by the National Aeronautics and Space Administration through grants NNX16AB66G, NNX17AB18G, and 80NSSC20K0389, through \textit{Chandra} Award Numbers GO021065A, GO021062A, and GO122068X issued by the \textit{Chandra} X-ray Center, which is operated by the Smithsonian Astrophysical Observatory for and on behalf of the National Aeronautics Space Administration under contract NAS8-03060, and by the National Science Foundation through grant no. 12850. P.~B.'s research was supported by a grant (no. 2020747) from the United States-Israel Binational Science Foundation (BSF), Jerusalem, Israel. J.~B.~G. acknowledges financial support from the Spanish Ministry of Science and Innovation (MICINN) through the Spanish State Research Agency, under Severo Ochoa Program 2020-2023 (CEX2019-000920-S). This project has received funding from the European Research Council (ERC) under the European Union’s Horizon 2020 research and innovation programme, grant  101002761 (BHianca; PI: Troja). RS.~R. acknowledges support under the CSIC-MURALES project with reference 20215AT009 and from the State Agency for Research of the Spanish MCIU through the Center of Excellence Severo Ochoa award to the Instituto de Astrof\'isica de Andaluc\'ia (SEV-2017-0709).

This work made use of data supplied by the UK \textit{Swift} Science Data Centre at the University of Leicester. 
This research has made use of the Keck Observatory Archive (KOA), which is operated by the W. M. Keck Observatory and the NASA Exoplanet Science Institute (NExScI), under contract with the National Aeronautics and Space Administration.
Based on observations obtained at the international Gemini Observatory, a program of NSF's OIR Lab, which is managed by the Association of Universities for Research in Astronomy (AURA) under a cooperative agreement with the National Science Foundation on behalf of the Gemini Observatory partnership: the National Science Foundation (United States), National Research Council (Canada), Agencia Nacional de Investigaci\'{o}n y Desarrollo (Chile), Ministerio de Ciencia, Tecnolog\'{i}a e Innovaci\'{o}n (Argentina), Minist\'{e}rio da Ci\^{e}ncia, Tecnologia, Inova\c{c}\~{o}es e Comunica\c{c}\~{o}es (Brazil), and Korea Astronomy and Space Science Institute (Republic of Korea). 
The \textit{HST} data (ObsID: 14685) used in this work was obtained from the Mikulski Archive for Space Telescopes (MAST). STScI is operated by the Association of Universities for Research in Astronomy, Inc., under NASA contract NAS5-26555. Support for MAST for non-\textit{HST} data is provided by the NASA Office of Space Science via grant NNX09AF08G and by other grants and contracts. 
These results also made use of Lowell Observatory's Lowell Discovery Telescope (LDT), formerly the Discovery Channel Telescope. Lowell operates the LDT in partnership with Boston University, Northern Arizona University, the University of Maryland, and the University of Toledo. Partial support of the LDT was provided by Discovery Communications. LMI was built by Lowell Observatory using funds from the National Science Foundation (AST-1005313). 
This paper makes use of data obtained from the Isaac Newton Group of Telescopes Archive which is maintained as part of the CASU Astronomical Data Centre at the Institute of Astronomy, Cambridge. 
This work is based on data from the GTC Public Archive at CAB (INTA-CSIC), developed in the framework of the Spanish Virtual Observatory project supported by the Spanish MINECO through grants AYA 2011-24052 and AYA 2014-55216. The system is maintained by the Data Archive Unit of the CAB (INTA-CSIC). 
Based on observations made with the Liverpool Telescope operated on the island of La Palma by Liverpool John Moores University in the Spanish Observatorio del Roque de los Muchachos of the Instituto de Astrofisica de Canarias with financial support from the UK Science and Technology Facilities Council.
Additionally, this work is based on data obtained from the ESO Science Archive Facility.
We additionally made use of Astropy, a community-developed core Python package for Astronomy \citep{Astropy2018}. 

\section*{Data Availability}

The data underlying this article will be shared on reasonable request to the corresponding author.

\begin{table*}
 	\centering
 	\caption{Log of imaging observations of sGRB host galaxies.
 	}
 	\label{tab: observations}
 	\begin{tabular}{lccccccccccc}
    \hline
       \hline 
       \\[-2.5mm]
   \textbf{GRB} & \textbf{$T_{90}^c$} & \textbf{RA} & \textbf{Dec} &\textbf{Obs. Date}  &   \textbf{Telescope} & \textbf{Instrument} & \textbf{Filter} & \textbf{Exp.} &  \textbf{AG Image$^{b}$} & \textbf{AB Mag$^{d}$} &\textbf{$A_\lambda$}  \\
   & \textbf{(s)} &  \textbf{(J2000)} & \textbf{(J2000)} & \textbf{(UT)} & & & & \textbf{(s)} & &   & \textbf{(mag)} \\
    \hline
   091109B & 0.3 & 07:30:56.61 & -54:05:22.85 & 11-10-2009   & VLT & FORS2  & \textit{R}  & 3600 & Y & ... &   ... \\
   ... & ... & ... & ... & 11-01-2016  & \textit{HST} & WFC3 & \textit{F110W} & 5600 & ... & $>27.3$ &   0.13 \\
           \hline
  101224A & 0.2 & 19:03:41.72 & 45:42:49.5 & 06-11-2020 & LDT & LMI & \textit{g} & 750 & ... &$22.71\pm0.06$  & 0.17 \\
   ... & ... & ... & ... & 06-11-2020 & LDT & LMI & \textit{r} & 750 & ... & $22.11\pm0.06$ & 0.12 \\
      ... & ... & ... & ... & 06-11-2020 & LDT & LMI & \textit{i} & 750 & ... & $21.91\pm0.05$ & 0.09\\
         ... & ... & ... & ... & 06-11-2020 & LDT & LMI & \textit{z} & 800 & ... &  $21.84\pm0.05$& 0.06 \\
   \hline
   110112A & 0.5 & 21:59:43.85 & 26:27:23.9 &  01-12-2011  & WHT & ACAM &  \textit{i} & 900 & Y & ... &...\\
      ... & ... & ... & ... & 10-13-2016   & \textit{HST} & WFC3 & \textit{F110W} & 5200 & ... & $>27.3$ &   0.05\\
   \hline
  110402A$^{a}$  & 56 & 13:09:36.53  & 61:15:09.9 & 04-02-2011 & \textit{Swift} & UVOT & $\textit{wh}$ & 1630 &  Y &... & ... \\
      ... & ... & ... & ... & 05-27-2014  & Keck & LRIS & \textit{B} & 180 & ...& $24.19\pm0.11$ &  0.06 \\
    ... & ... & ... & ... & 05-27-2014  & Keck & LRIS & \textit{I} & 570 & ...& $23.35\pm0.10$ &  0.03  \\
   ... & ... & ... & ... & 08-03-2020  & Gemini & GMOS-N & \textit{r} & 900 & ...& $24.24\pm0.20$  &  0.04  \\
     ... & ... & ... & ... & 05-05-2021  & LDT & LMI & \textit{i} & 1500 & ...& $23.35\pm0.09$ &  0.03  \\
  ... & ... & ... & ... & 05-06-2021  & LDT & LMI & \textit{z} & 2100 & ...& $23.0\pm0.16$  &  0.02  \\
    \hline
  120305A  & 0.1 & 03:10:08.68 & 28:29:31.0 & 03-13-2012	  & Gemini  & GMOS-N & \textit{i} & 2340 & ... & $21.56\pm0.08$ & 0.71 \\
 ... & ... & ... & ... & 03-06-2014	  & LDT  & LMI & \textit{r} & 2700 & ... &   $22.32\pm0.09$ & 0.81 \\
  ... & ... & ... & ... & 10-25-2014	&   Keck& LRIS&\textit{G}& 3000 & ...  & $23.00\pm0.06$  & 1.30 \\
    ... & ... & ... & ... & 10-25-2014	&  Keck& LRIS&\textit{R} & 2750 & ...& $22.28\pm0.04$  & 0.75 \\
      ... & ... & ... & ... & 11-09-2021  & LDT  & LMI & \textit{y} & 1980 & ... &   $<20.6$ & 0.38  \\
  \hline
  120630A  & 0.6 & 23:29:11.07 & 42:33:20.3 & 07-01-2012   & Gemini  & GMOS-N & \textit{r} & 500 & ...& $21.60\pm0.06$ &  0.21 \\
    ... & ... & ... & ... &07-01-2012& Gemini & GMOS-N & \textit{i} & 500 & ...& $21.25\pm0.07$  &  0.19 \\
      ... & ... & ... & ... &07-01-2012 & Gemini & GMOS-N & \textit{z} & 500 & ...& $21.08\pm0.05$  &  0.14 \\
  ... & ... & ... & ... & 09-05-2014  & LDT & LMI & \textit{r} & 700 & ...&  $21.56\pm0.05$ &  0.21 \\
 ... & ... & ... & ... &09-05-2014 & LDT & LMI & \textit{i} & 400 & ...& $21.16\pm0.06$ &  0.16 \\
  ... & ... & ... & ... &2014-09-05	 & LDT & LMI & \textit{z} & 800 & ...& $21.0\pm0.2$ &  0.11 \\
  ... & ... & ... & ... &10-25-2014 & Keck & LRIS & \textit{R} & 3300 & ...& $21.63\pm0.04$ &  0.20 \\
    ... & ... & ... & ... &10-25-2014 & Keck & LRIS & \textit{G} & 3600 & ...& $22.45\pm0.03$ &  0.34 \\
   ... & ... & ... & ... & 11-09-2021	  & LDT  & LMI & \textit{y} & 1980 & ... &   $<20.2$ & 0.09  \\
      ... & ... & ... & ... & -- & WISE & --  & \textit{W1} & -- & ...& $19.48\pm0.05$  & 0.02  \\
    ... & ... & ... & ... & -- & WISE & --  & \textit{W2} & -- & ...& $19.61\pm0.08$  & 0.016 \\

  \hline
  130822A  & 0.04 & 01:51:41.27 & -03:12:31.7 & 08-23-2013	 & Gemini & GMOS-N & \textit{i} & 600 & ...& $17.79\pm0.03$ &  0.05 \\
  ...  & ... & ... & ... & 10-25-2014 & Keck & LRIS & \textit{G} & 3000 & ...& $18.84\pm0.03$ & 0.08 \\
   ...  & ... & ... & ... & 10-25-2014	 & Keck & LRIS & \textit{R} & 2750 & ...& $18.18\pm0.03$ & 0.05 \\
  \hline
  130912A  & 0.3 & 03:10:22.23 & 13:59:48.7 & 09-13-2013  & WHT & ACAM & \textit{i} & 900 & Y & ... &  ... \\
  ...  & ... & ... & ... & 02-25-2014  & LDT & LMI & \textit{r} & 2700 & ...& $>24.9$ & 0.56 \\
  ...  & ... & ... & ... &10-25-2014   & Keck & LRIS &  \textit{G} & 2400 & ...& $>26.3$ & 0.90 \\
 ...  & ... & ... & ... & 10-25-2014  & Keck & LRIS & \textit{R}  & 2750  & ...& $>26.2$ & 0.52 \\
  ...  & ... & ... & ... & 01-09-2017  & \textit{HST} & WFC3 & \textit{F110W} & 5200 & ...& $>27.2$ & 0.22 \\
   \hline
    131004A   & 1.5 & 19:44:27.08& -02:57:30.2 & 10-04-2013 & \textit{Swift} & UVOT & \textit{wh} & 520 & Y & ... &  ... \\
      ...& ... &...& ...& 10-07-2013  & Keck & MOSFIRE & \textit{$K_s$} & 290  & ...& $>22.3$  & 0.08 \\
     ...  & ... & ... & ... & 10-11-2016 & \textit{HST} & WFC3 & \textit{F110W} & 5212 & ...& $25.80\pm0.05$  & 0.22 \\
          \hline
   140129B & 1.35 & 21:47:01.66 & +26:12:23.0 & 01-29-2014 & \textit{Swift}  & UVOT   & \textit{wh}  & 150 & Y & ... & ... \\
   ... & ... & ... & ... & 06-10-2014 &  LDT & LMI & \textit{r} & 1500 & ... & $23.55\pm0.10$ & 0.20 \\
     ... & ... & ... & ... & 11-03-2019 & LDT  & LMI  & \textit{r} & 1200  & ...& $23.50\pm0.09$  & 0.20 \\
   ... & ... & ... & ... & 08-06-2021 & LDT  & LMI  & \textit{g} & 1200  & ...& $24.52\pm0.18$  & 0.30 \\
   ... & ... & ... & ... & 08-06-2021 & LDT  & LMI  & \textit{i} & 1200  & ...& $23.52\pm0.10$  & 0.15 \\
   ... & ... & ... & ... & 08-06-2021 & LDT  & LMI  & \textit{z} & 1000  & ...& $<23.0$  & 0.11 \\
             \hline
140516A & 0.2 & 16:51:57.40 & 39:57:46.3 & 05-16-2014 & Gemini & GMOS-N & \textit{i} & 1800 & ... & $>26.1$ & 0.02 \\
   ... & ... & ... & ... & 09-04-2014 & LDT  & LMI  &  \textit{r} & 4200 & ...& $>25.0$ & 0.03 \\
     ... & ... & ... & ... & 10-15-2019 & Keck  & MOSFIRE  &  $K_s$ & 1800 & ...& $>23.6$ & 0.005 \\
            \hline
  140622A & 0.13 & 21:08:41.53 & -14:25:9.5 & 08-05-2021 & LDT & LMI    & \textit{g}   & 1200 &...  & $22.75\pm0.07$ &  0.22\\
   ... & ... & ... & ... & 08-05-2021 & LDT  & LMI  & \textit{r} &1200   & ...& $22.43\pm0.07$   & 0.15 \\
   ... & ... & ... & ... & 08-05-2021 & LDT  & LMI  & \textit{i} & 750  & ...& $21.95\pm0.06$  & 0.11 \\
   ... & ... & ... & ... & 08-05-2021 & LDT  & LMI  & \textit{z} & 800  & ...&  $22.0\pm0.2$ &0.08 \\
  \hline
  140930B  & 0.8 &  00:25:23.4 &24:17:41.7 & 10-01-2014	& Gemini & GMOS-N & \textit{r} & 1350 & Y & ... &... \\ 
    ...   & ... & ... & ... & 10-02-2014 & Gemini & GMOS-N & \textit{r} & 1350 & Y & ... & ... \\
   ...  & ... & ... & ... & 08-01-2020 & Gemini & GMOS-N &  \textit{r} & 1650 & ...& $23.8\pm0.2$ & 0.06 \\
  \hline
  150423A  & 0.08 & 14:46:18.86 & 12:17:00.70 &  04-23-2015 & VLT & FORS2 & \textit{R} & 300 & Y & ... & ... \\
    ...  & ... & ... & ... & 02-03-2017  & \textit{HST} & WFC3 & \textit{F110W} & 5200 & ...& $>27.2$ & 0.02 \\
          \hline
  150831A & 1.15 & 14:44:05.84 & -25:38:06.4 & 09-01-2016 & VLT & FORS2 & \textit{R}  & 2400 & ...& $>25.8$ & 0.22 \\
     ... & ... & ... & ... & 03-07-2017 & VLT & FORS2  &\textit{I}  &2400 & ...& $>24.5$  & 0.16 \\
... & ... & ... & ... &  07-29-2020 & Gemini & GMOS-S & \textit{i} & 2040 & ...  & $>25.7$ & 0.16 \\

          \hline
    \end{tabular}
 \end{table*}

 \begin{table*}
 	\centering
 	\contcaption{}
 	\begin{tabular}{lccccccccccc}
    \hline
       \hline 
       \\[-2.5mm]
   \textbf{GRB} & \textbf{$T_{90}^c$} & \textbf{RA} & \textbf{Dec} &\textbf{Obs. Date}  &   \textbf{Telescope} & \textbf{Instrument} & \textbf{Filter} & \textbf{Exp.} &  \textbf{AG Image$^{b}$} & \textbf{AB Mag$^{d}$} &\textbf{$A_\lambda$}  \\
   & \textbf{(s)} &  \textbf{(J2000)} & \textbf{(J2000)} & \textbf{(UT)} & & & & \textbf{(s)} & &   & \textbf{(mag)} \\
     \hline
  151229A  & 1.4 &21:57:28.78& 	-20:43:55.2 & 03-08-2019 & LDT & LMI & \textit{r}&1200 & ... & $>24.5$ & 0.05 \\
  ...& ... &...& ...& 07-30-2019   & Gemini &GMOS-S & \textit{z}&1920 & ...& $24.47\pm0.10$ & 0.03 \\
  ...& ... &...& ...& 10-15-2019   & Keck & MOSFIRE & \textit{Y} & 1340 & ...& $24.0\pm0.2$ & 0.03 \\
  ...& ... &...& ...& 08-11-2020   & Gemini & GMOS-N & \textit{r}&2250 & ...& $25.75\pm0.20$  & 0.05  \\
  ...& ... &...& ...& 06-16-2021  & LDT & LMI & \textit{i}& 900 & ... &  $>23.8$  & 0.04   \\
    ...& ... &...& ...& 07-22-2021   & Gemini & F2 & \textit{J}& 1680 & ...& $23.10\pm0.18$ & 0.02 \\
      ...& ... &...& ...& 07-22-2021  & Gemini & F2 & $K_s$ & 1680 & ...& $22.78\pm0.19$  & 0.01  \\
  ...& ... &...& ...& 07-30-2021   & Gemini & GMOS-S & \textit{i}& 1680 & ...&  $25.41\pm0.20$  & 0.04  \\
    \hline 
    160408A & 0.3 & 08:10:29.81 & 71:07:43.7 & 04-08-2016 & Gemini & GMOS-N & \textit{r} & 900 & Y & ... & ... \\
   ... & ... & ... & ... & 04-09-2016 & Gemini & GMOS-N  & \textit{r}  & 900 & ...& $>25.8$ & 0.06 \\
    ... & ... & ... & ... & 03-29-2020 & LDT & LMI  & \textit{g}  & 1500 & ... & $>24.6$ & 0.08 \\
      ... & ... & ... & ... & 03-29-2020 & LDT & LMI  & \textit{r}  & 1500 & ...& $>24.5$ & 0.06 \\
        ... & ... & ... & ... & 03-29-2020 & LDT & LMI  & \textit{i}  & 1500 & ...& $>24.2$ & 0.04 \\
          ... & ... & ... & ... & 03-29-2020 & LDT & LMI  & \textit{z}  & 1500 & ...& $>23.7$ & 0.03 \\
          \hline
      160410A$^a$ & 96 & 10:02:44.37 & 03:28:42.4  & 04-10-2016 & \textit{Swift} & UVOT & \textit{wh} & 540 & Y & ...   & ... \\
         ... & ... & ... & ... & 04-28-2016 & Keck  & DEIMOS   & \textit{R}  &330 & ... & $>25.0$ & 0.05  \\
   ... & ... & ... & ... & 04-28-2016 & Keck  & DEIMOS & \textit{I}  & 330 & ... & $>24.2$ & 0.03 \\
  ... & ... & ... & ... & 12-15-2020 & LDT & LMI & \textit{r} &2100 & ...& $>24.5$ & 0.05 \\
 ... & ... & ... & ... & 02-06-2021 & LDT & LMI  & \textit{g} & 1950 & ...& $>24.9$ & 0.07 \\
           \hline
  160525B & 0.3  &09:57:32.23	 & 51:12:24.9 & 05-25-2016 & \textit{Swift}& UVOT &\textit{wh} & 150 & Y & ... & ... \\
   ... & ... & ... & ... &  01-29-2020 &  LDT & LMI& \textit{g} & 1200  &  ...&  $23.30\pm0.15$ & 0.03 \\
      ... & ... & ... & ... & 01-29-2020  & LDT & LMI  &\textit{r}& 1200  & ...& $23.29\pm0.09$ & 0.02 \\
           ... & ... & ... & ... &02-29-2020 & LDT & LMI & \textit{i}  &1500 & ...& $23.29\pm0.18$ & 0.016 \\
       ... & ... & ... & ... & 12-15-2020 & LDT & LMI  & \textit{z}  & 2000 & ...& $23.4\pm0.3$  & 0.012 \\
  \hline
  160601A  & 0.12 & 15:39:43.97 & 64:32:30.5 & 06-02-2016  &  Gemini & GMOS-N & \textit{r} & 900 & Y & ... & ... \\
   ... & ... & ... & ... & 06-03-2016  & LDT  & LMI & \textit{r} & 720 & ...& $>24.6$ & 0.05 \\
   ... & ... & ... & ... & 09-08-2016  & GTC  & OSIRIS & \textit{r} & 1680 & ...& $>25.9$ & 0.05 \\
     ... & ... & ... & ... & 03-25-2019 & Keck & MOSFIRE  & $K_s$ & $2400$ & ... & $>23.5$ & 0.01 \\
 ... & ... & ... & ... & 08-01-2020	 & Gemini & GMOS-N &  \textit{r} & 1800 & ... & $>25.6$  & 0.05 \\
        ... & ... & ... & ... & 02-05-2021 & LDT & LMI  & \textit{g}  & 800& ...& $>22.5$ & 0.07 \\
       ... & ... & ... & ... & 02-05-2021 & LDT & LMI  & \textit{i}  & 1200 & ...& $>22.5$ & 0.04 \\
            ... & ... & ... & ... & 02-05-2021 & LDT & LMI  & \textit{z}  & 1500 & ...& $>22.0$  & 0.03 \\
 
  \hline
  160927A  & 0.48 & 17:04:58.22 & 17:19:54.9 & 09-28-2016  & GTC & OSIRIS & \textit{r} & 1915 & Y & ... &  ... \\
   ... & ... & ... & ... & 02-23-2017   & GTC   & OSIRIS  &\textit{r}  & 1200 & ...& $>26.1$ & 0.15 \\
   ... & ... & ... & ... & 05-20-2018    & LDT  &  LMI & \textit{r} & 300 & ...& $>24.3$ &0.15 \\
     ... & ... & ... & ... & 10-06-2018    & Keck  &  LRIS & \textit{G} & 2760 & ...& $>25.9$ & 0.25 \\
         ... & ... & ... & ... & 10-06-2018    & Keck  &  LRIS & \textit{R} & 600 & ...& $>25.2$ & 0.14 \\
     ... & ... & ... & ... & 09-04-2019    & Keck  &  LRIS & \textit{Z} & 800 & ...& $>24.8$  & 0.10 \\
   ... & ... & ... & ... & 08-01-2020    & Gemini & GMOS-N & \textit{i} & 720 & ...& $>26.0$ & 0.13 \\
  \hline
   170127B & 0.5  &01:19:54.47 &-30:21:28.6 & 2018-01-27 & Gemini & GMOS-S & \textit{g} & 1800 &...& $>24.2$  & 0.06 \\
   ... & ... & ... & ... & 2018-10-06 & Keck & LRIS   &\textit{G}  & 2520 & ...& $>26.1$  & 0.07 \\
    ... & ... & ... & ... & 2018-10-06  & Keck & LRIS &\textit{R}  & 1720 & ...& $>26.0$ & 0.04 \\
    ... & ... & ... & ... & 2019-09-04  & Keck & LRIS &\textit{G}  & 1920 & ...& $>26.0$  & 0.07 \\
    ... & ... & ... & ... & 2019-09-04 & Keck & LRIS &\textit{I}  & 1600 & ...& $>25.9$ & 0.04 \\
    ... & ... & ... & ... & 2019-10-15  & Keck & MOSFIRE &\textit{J}  & 2010 & ...& $>24.1$ & 0.01 \\
    ... & ... & ... & ... & 01-30-2021 & Gemini & GMOS-S &\textit{z}  & 1440 & ...& $>23.9$ & 0.03 \\
    \hline
    170428A& 0.2 & 22:00:18.78 &  26:54:57.0 & 04-29-2017 & LDT  & LMI  & \textit{i}  & 1200 & ... & $22.2\pm0.2$  & 0.09 \\ 
  ... & ... & ... & ... &  05-01-2017 & TNG & LRS & \textit{i} & 1470 & ... & $22.05\pm0.15$ & 0.09 \\
    ... & ... & ... & ... & 05-01-2017 & TNG & LRS & \textit{z} & 1620 & ... & $21.94\pm0.15$ &   0.06 \\
   ... & ... & ... & ... & 05-21-2018 & LDT & LMI  & \textit{g} & 100 & ...& $>23.5$ & 0.17 \\
... & ... & ... & ... & 05-21-2018 & LDT & LMI  & \textit{r} & 200 & ...& $22.21\pm0.10$ & 0.12 \\
... & ... & ... & ... & 05-21-2018 & LDT & LMI  & \textit{i} & 200 & ...& $21.93\pm0.15$  & 0.09 \\
... & ... & ... & ... & 05-21-2018 & LDT & LMI  & \textit{z} & 100 & ...& $22.1\pm0.3$ & 0.06 \\
      \hline
 170728A  & 1.3  &03:55:33.17 & 12:10:54.7 & 07-28-2017 & \textit{Swift} & UVOT & \textit{wh} & 150 & Y &... & ... \\
    ... & ... & ... & ... & 01-14-2018  & Keck & LRIS  & \textit{G}  & 1380 & ...& $>25.3$ & 0.76 \\
    ... & ... & ... & ... & 01-14-2018 & Keck & LRIS  & \textit{R}  & 1380 & ...& $>25.1$ & 0.44 \\
    ... & ... & ... & ... &  01-08-2019 & LDT & LMI  &   \textit{r} & 900 & ...& $>24.6$ & 0.47 \\
        \hline
 170728B$^a$ & 48 & 15:51:55.47 &  70:07:21.1 & 07-28-2017 & \textit{Swift} & UVOT & \textit{wh} & 900 & Y &... &... \\ 
   ... & ... & ... & ... & 11-03-2019 & LDT & LMI  & \textit{r} & 1200 & ...& $23.13\pm0.06$ & 0.06 \\
     ... & ... & ... & ... & 12-07-2019 & LDT & LMI  & \textit{g} & 900 & ...& $23.82\pm0.06$ & 0.09 \\
    ... & ... & ... & ... & 12-07-2019 & LDT & LMI & \textit{i}  & 1200 & ...& $22.67\pm0.05$ & 0.04 \\
     ... & ... & ... & ... & 12-07-2019 & LDT  & LMI  & \textit{z} & 1200 & ... & $22.36\pm0.15$ & 0.03  \\
          \hline
   171007A$^a$ & 68 &09:02:24.14 & 42:49:08.8 & 01-09-2020  & LDT & LMI & \textit{r}  & 1200 & ... & $>24.9$ &  0.04\\
    ... & ... & ... & ... & 02-01-2021  & Gemini  & GMOS-N  & \textit{i}  & 1440 & ...& $>26.1$ & 0.03 \\
     \hline
    \end{tabular}
\end{table*}

 \begin{table*}
 	\centering
 	\contcaption{}
 	\begin{tabular}{lccccccccccc}
    \hline
       \hline 
       \\[-2.5mm]
   \textbf{GRB} & \textbf{$T_{90}^c$} & \textbf{RA} & \textbf{Dec} &\textbf{Obs. Date}  &   \textbf{Telescope} & \textbf{Instrument} & \textbf{Filter} & \textbf{Exp.} &  \textbf{AG Image$^{b}$} & \textbf{AB Mag$^{d}$} &\textbf{$A_\lambda$}  \\
   & \textbf{(s)} &  \textbf{(J2000)} & \textbf{(J2000)} & \textbf{(UT)} & & & & \textbf{(s)} & &   & \textbf{(mag)} \\
           \hline
  180618A$^a$ & 47  & 11:19:45.87  &  73:50:13.5 & 06-18-2018  & Liverpool & IO:I & \textit{r} & 60 & Y  & ... & ... \\
   ... & ... & ... & ... & 04-07-2019 & LDT & LMI  & \textit{r} & 1200 & ...& $23.08\pm0.08$  & 0.16 \\
   ... & ... & ... & ... & 12-07-2019 & LDT & LMI  & \textit{g} & 1200 & ...& $24.11\pm0.12$ & 0.22 \\
   ... & ... & ... & ... & 12-07-2019 & LDT & LMI  & \textit{i} & 1200 & ...& $22.45\pm0.10$ & 0.12 \\
   ... & ... & ... & ... & 05-05-2021 & LDT & LMI  & \textit{z} & 1800 & ...& $22.34\pm0.12$ & 0.09 \\
   ... & ... & ... & ... & 05-05-2021 & LDT & LMI  & \textit{y} & 1400 & ...& $>21.5$  & 0.06 \\
            \hline
  180727A & 1.1 & 23:06:39.68 & -63:03:06.7  & 10-14-2018 & Gemini  & GMOS-S  & \textit{i}  & 2520 & ...& $>26.0$ & 0.03 \\
      ... & ... & ... & ... & 07-28-2019 & Gemini  & GMOS-S  & \textit{r}  & 1560 & ...& $>26.1$  & 0.04 \\
         ... & ... & ... & ... & 07-30-2019 & Gemini  & GMOS-S  & \textit{g}  & 1800 & ...& $>26.3$ & 0.06 \\
          ... & ... & ... & ... & 07-30-2019 & Gemini  & GMOS-S  & \textit{z}  & 1800 & ...& $>26.0$ & 0.02 \\
          \hline
  180805B$^a$ & 122 & 01:43:07.59 & -17:29:36.4 & 09-10-2018 & Keck & LRIS  & \textit{G} & 1920  & ...& $23.52\pm0.07$ & 0.06 \\
     ... & ... & ... & ... & 09-10-2018 &Keck & LRIS  & \textit{I} & 1600 & ...& $22.34\pm0.12$ & 0.03 \\
 ... & ... & ... & ... & 09-04-2019 &Keck & LRIS  & \textit{V} & 1680  & ...& $22.83\pm0.09$ & 0.04 \\
     ... & ... & ... & ... & 09-04-2019 & Keck & LRIS  & \textit{Z} & 1400 & ... & $22.01\pm0.14$ & 0.02 \\
       ... & ... & ... & ... & 10-15-2019 &Keck & MOSFIRE & $K_s$ & 1800  & ...& $21.23\pm0.15$  & 0.005 \\
      ... & ... & ... & ... & 01-16-2021 & LDT & LMI & \textit{z} & 2000 & ...& $21.98\pm0.09$ & 0.02 \\
         \hline
   191031D & 0.3 &18:53:09.57 &47:38:38.8 & 11-02-2019  & Gemini & GMOS-N & \textit{r} & 720 & ... & $21.78\pm0.05$ & 0.14 \\
    ... & ... & ... & ... & 11-03-2019 & LDT  & LMI  & \textit{g}  &1200& ...& $22.89\pm0.07$  & 0.21 \\
    ... & ... & ... & ... &04-18-2021 & LDT & LMI  &   \textit{i} & 600 & ...& $21.3\pm0.2$  & 0.11 \\
    ... & ... & ... & ... &04-18-2021 & LDT & LMI  &  \textit{z} & 700 & ...& $21.3\pm0.3$ & 0.08 \\
    ... & ... & ... & ... &04-18-2021 & LDT & LMI  &  \textit{y} & 700 & ...& $21.1\pm0.3$  & 0.07 \\
    ... & ... & ... & ... & -- & PS1 & --  & \textit{i} & -- & ...& $21.53\pm0.06$  & 0.11 \\
    ... & ... & ... & ... & -- & PS1 & --  & \textit{z} & -- & ...& $21.03\pm0.03$  & 0.08  \\
    ... & ... & ... & ... & -- & WISE & --  & \textit{W1} & -- & ...& $19.6\pm0.15$  & 0.014  \\
    ... & ... & ... & ... & -- & WISE & --  & \textit{W2} & -- & ...& $20.16\pm0.30$  & 0.01 \\

\hline
  200411A  & 0.3  &03:10:39.39	 & -52:19:03.4 &01-25-21 & Gemini & GMOS-S & \textit{r} & 1800 & ...& $22.55\pm0.03$ & 0.03\\
        ... & ... & ... & ... & -- & DES & --  &\textit{g}  & -- & ...& $23.6\pm0.2$  & 0.06  \\
         ... & ... & ... & ... & -- & DES & --  &\textit{r}  & -- & ...& $22.6\pm0.1$  & 0.04  \\
          ... & ... & ... & ... & -- & DES & --  &\textit{i}  & -- & ...& $21.9\pm0.1$  & 0.03  \\
           ... & ... & ... & ... & -- & DES & --  &\textit{z}  & -- & ...& $21.3\pm0.1$  &  0.02 \\
     ... & ... & ... & ... & -- & VISTA & --  &\textit{J}  & -- & ...& $20.9\pm0.2$  & 0.01   \\
     ... & ... & ... & ... & -- & VISTA & --  &\textit{K}  & -- & ...& $20.0\pm0.2$  & 0.005   \\
      ... & ... & ... & ... & -- & WISE & --  & \textit{W1} & -- & ...& $20.0\pm0.1$  & 0.003  \\
    ... & ... & ... & ... & -- & WISE & --  & \textit{W2} & -- & ...& $20.2\pm0.3$  & 0.003 \\
     \hline
    \end{tabular}
    \begin{flushleft}
    \quad \footnotesize{$^a$ sGRBEE.} \\
    \quad \footnotesize{$^b$ Afterglow image used for relative alignment.}\\
    \quad \footnotesize {$^c$ $T_{90}$ values were retrieved from the \textit{Swift} BAT GRB catalog \citep{Lien2016}.}\\
   \quad \footnotesize {$^d$ Host galaxy magnitudes not corrected for Galactic extinction $A_\lambda$ \citep{Schlafly2011}.}\\
\end{flushleft}
\end{table*}

\begin{table*}
 	\centering
 	\caption{Log of spectroscopic observations of sGRB host galaxies. The redshift and the emission or absorption lines of the spectroscopic target are also reported.
 	\label{tab: SpecObs}
 	}
 	\begin{tabular}{lccccccccl}
    \hline
       \hline 
       \\[-2.5mm]
   \textbf{GRB} &\textbf{Obs. Date}  &   \textbf{Telescope} & \textbf{Instrument} & \textbf{Grating} & \textbf{$\lambda_\textrm{cen}$} & \textbf{Exp.} & \textbf{Slit Width} & \textbf{Redshift} & \textbf{Lines}   \\
   &\textbf{UT}  &   & & & \textbf{(nm)} & \textbf{(s)} & \textbf{($\arcsec$)} & & \\
     \hline
      060121 &  05-27-2014& Keck & LRIS & 600/4000 & 330 & 2720 & 1.0 & -- & No trace \\
     ...  & ... & ... & ... & 400/8500 & 588 & 2720 & 1.0 & & \\
       \hline
      101224A & 05-27-2014& Keck & LRIS & 600/4000 & 330 & 1570 & 1.0 & $0.4536 \pm 0.0004$ & H$\alpha$,H$\beta$,H$\gamma$ \\
     ...  & ... & ... & ...& 400/8500 & 588 & 1570 & 1.0 & & [OII],[OIII]\\
      \hline 
      110402A$^a$ & 05-27-2014 & Keck  &  LRIS & 400/3400  & 680 & 1800 & 1.0 & $0.854\pm0.001$ & [OII] \\
              ...    & ... &  ... & ... & 400/8500 & 840 & 1800 & 1.0  & & \\
           \hline
      140622A & 05-27-2014& Keck & LRIS & 600/4000 & 330 & 900 & 1.0 & $0.959\pm0.001$ & [OII],[OIII] \\
     ...  & ... & ... & ... & 400/8500 & 588 & 900 & 1.0 & &\\
     \hline
      151229A & 09-10-2018  & Keck & LRIS & 400/3400 & 176 & 5520 & 1.0 & -- & No trace  \\
    ...  & ...  & ... & ...& 400/8500 & 622 & 5520 & 1.0 & &  \\
    ...  & ...  & ... & ... & 400/3400 & 544 & 5320 & 1.0  & & \\
     ... & ...  & ... &...  & 400/8500 & 1021 & 5320  & 1.0  & & \\
        \hline
   160410A$^{a,b}$ & 04-10-2016 & Keck & LRIS & 400/3400 & 176 & 600 &   1.0 & $1.717\pm0.001$ & Ly$\alpha$,[SiII] \\
     ...  & ... & ... &...  & 400/8500 & 622 & 600 & 1.0 & & [AlII] \\
     ...  & ... & ... &  ...& 400/3400  &544 & 600 & 1.0& &  \\
     ...  &...  &...& ... & 400/8500 & 1020 & 600 & 1.0 & & \\
     \hline
      180618A$^a$ & 02-01-2021 & Gemini & GMOS-N & R400 & 710 & 3600 & 1.0  &  $0.4^{+0.2}_{-0.1}$ $^c$ & No lines \\
     \hline
      180805B$^a$ & 09-10-2018 & Keck & LRIS & 400/3400 & 358 & 2440 & 1.0 & $0.6609\pm 0.0004$ &H$\beta$,H$\gamma$  \\
     ...  &... & ... & ... & 400/8500 & 763 & 2440 &  1.0 &  &[OII],[OIII] \\
     \hline
      191031D & 11-03-2019 & Gemini & GMOS-N & R400 & 705 & 3600 & 1.0 & $0.5\pm0.2^c$ & No lines \\
     \hline
    \end{tabular}
\begin{flushleft}
    \quad \footnotesize{$^a$ Short GRB with extended emission.} \\
    \quad \footnotesize{$^b$ Afterglow spectroscopy.} \\
    \quad \footnotesize{$^c$ Photometric redshift $z_\textrm{phot}$ based on \texttt{prospector} \citep{Johnson2019} modeling of the host galaxy SED.} \\
\end{flushleft}
\end{table*}

\begin{table*}
 	\centering
 	\caption{Short GRB host galaxy properties. Magnitudes are corrected for Galactic extinction \citep{Schlafly2011}.
 	}
 	\label{tab: host properties}
 	\begin{tabular}{lcccccccccc}
    \hline
       \hline
       \\[-2.5mm]
   \textbf{GRB} &\textbf{$\sigma_\textrm{tie}$}  &   \textbf{$\sigma_\textrm{AG}$}$^{b}$ & \textbf{$\sigma_\textrm{host}$} & \textbf{$R_o$ ($\arcsec$)} & \textbf{$R_o$ (kpc)} & \textbf{$R_e$ ($\arcsec$)} & \textbf{AB Mag}$^d$ & \textbf{Host?} & \textbf{$P_{cc}^d$}  & $z$  \\
    \hline
    \multicolumn{11}{c}{\textbf{Optical Localization}} \\
    \hline
 091109B & 0.04 & 0.10 & ... & ... & ... & ... & $>27.3^g$ & N & $>0.2^{g}$ & ...      \\[0.5mm]
 110112A & 0.11 & 0.09 & ...  & ... & ... & ...  &  $>27.3^g$  & N &$>0.45^{g}$  & ...      \\[0.5mm]
 110402A$^a$  & 0.15  & 0.07 & 0.05  & $0.91\pm0.17$ & $7.2\pm1.3$ &  0.7  & $24.24\pm0.20$  & Y & 0.03 & 0.854   \\[0.5mm]
  130912A & 0.06 & 0.3  & 0.04 & $0.68\pm0.31$ & $5.6\pm2.6^j$ & $0.32$ & $26.8\pm0.3^{g}$  & Y & $0.08^{g}$  & ...   \\[0.5mm]
   131004A   & 0.16 & 0.05 & 0.01  & $0.41\pm0.17$& $3.1\pm1.3$ & 0.4 & $25.80\pm0.05^{g}$ &Y   & 0.05$^g$ & $0.717$  \\[0.5mm]
 140129B & 0.16 & 0.02 & 0.02  & $0.5\pm 0.2$& $3.0\pm1.0$ & 0.5 & $23.50\pm0.09$& Y  & 0.009 & $0.6\pm0.1^f$   \\[0.5mm]
 140930B & --  & 0.05 & 0.09 & $1.4\pm0.1$ &  $8.8\pm 0.9^i$  & 0.4 & $23.8\pm0.2$ &  Y & 0.02 & ...      \\[0.5mm]
 150423A & 0.06 & 0.04 & ... & ... & ...& ... & $>27.2^{g}$ &N  & $>0.15^{g}$ & ...      \\[0.5mm]
 160408A & -- & 0.02  & ... & ... & ...& ... & $>25.8$ &N  & $>0.13$  & ...      \\[0.5mm]
  160410A$^a$ & 0.16  & 0.08 & ... & ... & ... & ...  & $>25.0$&N  & $>0.5$ & $1.717^e$    \\[0.5mm]
 160525B & 0.21  & 0.11 & 0.07  & $0.06\pm0.25$ & $0.4\pm1.6^i$ & 1.0 & $23.29\pm0.09$  & Y& 0.03 & ...   \\[0.5mm]
  160601A    & 0.02  & 0.02 & ... & ... & ... & ...  & $>25.9$ & Y & $>0.4$  & ...    \\[0.5mm]
 160927A   &  0.04 & 0.08 & ... & ... & ... & ... & $>26.0$ &N   & $>0.5$  & ...   \\[0.5mm]
 170428A & -- & 0.3 & 0.05  & $1.2\pm0.3$ & $7.2\pm1.8$ & 1.2  & $22.09\pm0.10$ &Y & 0.01 & 0.454$^e$  \\[0.5mm]
 170728A & 0.15 & 0.08 & ... & ... & ... & ... & $>24.7$ &N & $>0.2$ & ...     \\[0.5mm]
 170728B$^a$ & 0.22 & 0.07 & 0.06 & $0.78\pm0.24$ & $5.5\pm1.7$ & 0.7 & $23.06\pm0.06$ &Y  &  $0.014$ & $0.6\pm0.1^f$     \\[0.5mm]
 180618A$^a$ & 0.23  & 0.04 & 0.04  & $1.58\pm0.24$ &  $8.8\pm1.3$  & 1.0 & $22.92\pm0.08$ &Y  & 0.03 & $0.4^{+0.2}_{-0.1}$\,$^f$  \\[0.6mm]
     \hline
           \multicolumn{11}{c}{\textbf{XRT Localization}} \\
      \hline
 101224A & ... &3.8  & 0.01 & $2.4\pm2.7^k$& $14\pm17$ & 0.6 & $21.53\pm0.05$ &Y & 0.11/0.10$^h$  & 0.454  \\[0.5mm]
 120305A & ... & 2.0 & 0.05 & $5.4\pm1.4^k$ &  $34\pm9^i$& 1.1 &  $21.53\pm0.04$ & Y  & 0.07 & ...    \\[0.5mm]
  120630A & ... & 4.0 & 0.01 & $5.8\pm2.9^k$ & $40\pm20$ &  0.9 & $21.42\pm0.04$ &Y   & 0.07/0.08$^h$  & $0.6\pm0.1^f$   \\[0.5mm]
  130822A  & ... & 3.3 & 0.003  & $22.0\pm2.3^k$ & $61\pm6$ & 2.7 &  $18.13\pm0.01$  &Y & 0.08/0.06$^h$ & 0.154   \\[0.5mm]
 140516A & ... & 2.7 & ...  & ... & ... & ... & $>26.1$ & N  & $>0.2$  & ...     \\[0.5mm]
 140622A & ... & 2.9 & 0.02 & $4.6\pm2.0^k$ & $38\pm17$ & 1.2 & $22.28\pm0.07$ &Y  & 0.08/0.08$^h$ & 0.959  \\[0.5mm]
 150831A &  ... & 2.2 & ... & ... & ... & ... & $>25.6$ &N & $>0.25$ & ...      \\[0.5mm]
 151229A  & ... & 1.4 & 0.02  & $1.0\pm1.0^k$ & $9\pm9$ & 0.4 &  $25.75\pm0.16$ & Y& 0.25/0.10$^h$ & $1.4\pm0.2^f$  \\[0.5mm]
    170127B & ... & 2.6  &  ...& ... & ...& ... & $>26.0$ &N  & $>0.5$ & ...      \\[0.5mm]
  171007A$^a$ & ... & 2.5 & ... &...  & ... & ...  &  $>26.1$ &N& $>0.5$  & ...   \\[0.5mm]
 180727A & ... &2.3  & ... &...  & ...& ... & $>26.1$ &N & $>0.6$  & ...      \\[0.5mm]
  180805B$^a$ & ... & 2.1  & 0.02 & $3.4\pm1.5^k$ & $25\pm11$ &0.60& $22.79\pm0.09$ &Y & 0.07/0.08$^h$ & 0.661    \\[0.5mm]
  191031D  & ... & 2.3 &  0.02 & $7.4\pm1.7^k$ & $47\pm11$  & 1.1  &  $21.64\pm0.05$ &Y  & 0.12/0.05$^h$  & $0.5\pm0.2^f$    \\[0.5mm]
 200411A & ... & 1.4 & 0.04  & $4.5\pm1.0^k$  &  $31\pm8$ & 1.2  & $22.52\pm0.05$ &Y  & 0.11/0.08$^h$  & $0.6\pm0.1^f$   \\[0.5mm]
     \hline
    \end{tabular}
\begin{flushleft}
    \quad \footnotesize{$^a$ 
    Short GRB with extended emission.} \\
    \quad \footnotesize{$^b$ XRT position error reported at 90\% CL; optical localization error reported at $1\sigma$ (68\%).} \\
    \quad \footnotesize{$^d$ Host galaxy magnitude in $r$-band, and $P_{cc}$ computed using $r$-band magnitude \citep{Berger2010a}, unless otherwise specified.} \\
    \quad \footnotesize{$^e$ Redshift from afterglow (AG) spectroscopy.} \\
    \quad \footnotesize{$^f$ Photometric redshift $z_\textrm{phot}$ based on \texttt{prospector} \citep{Johnson2019} modeling of the host galaxy SED.} \\
    \quad \footnotesize{$^g$ \textit{HST}/$F110W$ magnitude, and $P_{cc}$ computed using IR number counts \citep{Galametz2013}.} \\
    \quad \footnotesize{$^h$ $P_{cc}$ computed using $z$-band number counts \citep{Capak2004}.}\\
     \quad \footnotesize{$^i$ Projected physical offset assuming $z=0.5$.}\\
     \quad \footnotesize{$^j$ Projected physical offset assuming $z=1.0$.}\\
    \quad \footnotesize{$^k$ The uncertainty on the sGRB's offset is computed at the 68\% of the Rayleigh distribution.}\\
\end{flushleft}
\end{table*}



\bibliographystyle{mnras}



\appendix

\section{sGRB Sample Analysis}
\label{sec: appendixsampleanalysis}

\subsection{Optically Localized}


\subsubsection{GRB 091109B}

At 21:49:03 UT on November 9, 2009, GRB 091109B triggered \textit{Swift}/BAT \citep{Oates091109B} and the \textit{Suzaku} Wide-band All-sky Monitor \citep[WAM;][]{Ohno091109B}. The GRB displayed a single spike with duration $T_{90}=0.27\pm0.05$ s. The X-ray afterglow was localized to RA, DEC (J2000) = $07^{h} 30^m 56^{s}.49$, $\ang{-54;05;24.2}$ with accuracy  \ang{;;2.3} (90\% CL). The optical counterpart was discovered at RA, DEC (J2000) = $07^{h} 30^m 56^{s}.61$, $\ang{-54;05;22.85}$.

We analyzed public archival late-time images of GRB 091109B obtained with the \textit{HST}/WFC3 in the $F110W$ filter. These observations are not contaminated by a diffraction spike at the GRB localization, which was observed in previous \textit{HST}/WFC3 imaging \citep{FongBerger2013} that set a limit $F160W\gtrsim 25.0$ mag on a coincident galaxy. In this new \textit{HST} observation, we do not find a coincident source to depth $F110W\gtrsim 27.2$ mag (corrected for Galactic extinction). 

However, we identify two previously unresolved sources (source A and G1) within $2\arcsec$ of the GRB position (Figure \ref{fig: GalOpt}); all other candidate host galaxies were previously discussed in \citet{FongBerger2013} and \citet{Tunnicliffe2014}. Source A is offset by $1.0\arcsec$ from the GRB position with magnitude $F110W=27.0\pm0.3$. G1 is offset by $1.4\arcsec$ with $F110W=26.51\pm0.16$. The probability of chance alignment is $P_{cc}=0.21$ and $0.27$ for source A and G1, respectively. The other host galaxy candidates discussed by \citet{FongBerger2013} and \citet{Tunnicliffe2014} are located at larger offsets ($\sim$\,$12-23\arcsec$), but are significantly brighter ($F110W\sim$\,$18-20$ mag). We find that each of these sources (Source A and B from \citealt{Tunnicliffe2014}, and G1 and G2 from \citealt{FongBerger2013}) have $P_{cc}>0.2$, based on $H$-band number counts, compared to the previously reported $P_{cc}\approx0.10$ (for both sources) based on galaxy number counts in the optical \citep{FongBerger2013,Tunnicliffe2014}. 
In either case, there are multiple galaxies with similar probabilities of chance coincidence, which complicates the host identification. These results confirm that GRB 091109B is observationally hostless.


Furthermore, we note that \citet{OConnor2020} constrained the density of the GRB environment to $n_\textrm{min}\gtrsim 1.7\times 10^{-5}$ cm$^{-3}$ (see their Appendix A). This density is consistent with an IGM-like environment \citep[i.e., $n<10^{-4}$ cm$^{-3}$;][]{OConnor2020}.

\subsubsection{GRB 110112A}

On January 12, 2011 at 04:12:18 UT, \textit{Swift}/BAT triggered and localized GRB 110112A \citep{Stamatikos110112A}. The GRB displayed a single spike with duration $T_{90}=0.5\pm0.1$ s. The X-ray afterglow was localized to RA, DEC (J2000) = $21^{h} 59^m 43^{s}.75$, $\ang{+26;27;24.1}$ with accuracy  \ang{;;1.7} (90\% CL). The optical counterpart was discovered by WHT, and localized to RA, DEC (J2000) = $21^{h} 59^m 43^{s}.85$, $\ang{+26;27;23.89}$ with uncertainty \ang{;;0.14} \citep{Fong2013}.

Here, we present unpublished archival \textit{HST}/WFC3 imaging obtained on October 13, 2016 in the $F110W$ filter. We uncover multiple extended sources within $5\arcsec$, which were not detected in previous deep ground based imaging \citep[Magellan/Gemini;][]{Fong2013} to $r\gtrsim 25.5$ and $i\gtrsim 26.2$ mag. Due to the high density of sources, in Figure \ref{fig: GalOpt} we label only the sources with the lowest probability of chance coincidence (source A, G1, and G2). The closest source to the GRB position (source A) is offset by $1.6\arcsec$ and has magnitude $F110W=27.2\pm0.3$ mag, yielding $P_{cc}=0.45$. The other nearby candidate hosts are G1 and G2 with offsets of $2.3\arcsec$ and $4.8\arcsec$ and magnitude $F110W=26.25\pm0.15$ and $F110W=24.18\pm0.07$ mag, respectively. These sources likewise have a large $P_{cc}$; 0.49 and 0.65 for G1 and G2. 
We do not identify a source coincident to the optical localization to depth $F110W\gtrsim 27.3$ mag. 
Thus we consider GRB 110112A to be observationally hostless, in agreement with previous work \citep{Fong2013, Tunnicliffe2014}.

The previous analysis by \citet{Fong2013} identified 15 galaxies within $3\arcmin$ of the GRB position with the two galaxies having the lowest probability of chance coincidence located at $4.8\arcsec$ (G2 in our analysis) and $20\arcsec$ with $P_{cc}=0.4$ and $0.5$, respectively.
Therefore, based on both ground based and \textit{HST} imaging, GRB 110112A is an outlier among observationally hostless GRBs \citep[e.g.,][]{Fong2013,FongBerger2013, Tunnicliffe2014} as there were no likely host galaxies (i.e., $P_{cc}<0.2$) identified. Our analysis represents a confirmation of the observationally hostless classification with deep \textit{HST} imaging.

\citet{OConnor2020} derived a lower limit to the density of the GRB's environment $n_\textrm{min}\gtrsim 1.4\times 10^{-3}$ cm$^{-3}$ (see their Appendix A). This density is inconsistent with the GRB being physically hostless \citep[see also Figure 7 of ][]{OConnor2020}, and strongly implies the GRB occurred within a galactic environment (either G1, Source A, or a fainter undetected host). 

\subsubsection{GRB 110402A}



GRB 110402A was detected with \textit{Swift}/BAT on April 2, 2011 at 00:12:57 UT \citep{Ukwatta110402A} with duration $T_{90}=56\pm5$ s. 
Additionally, the GRB triggered the \textit{Fermi} Gamma-ray Burst Monitor \citep[GBM;][]{Meegan2009}, the Konus-\textit{Wind} satellite \citep{Aptekar1995}, and \textit{Suzaku}/WAM  \citep{Yasuda110402A}.
The BAT lightcurve displays five short pulses followed by a longer, softer emission from $\sim 5-78$ s which is interpreted as EE. The initial pulses have a duration $\sim 2-3$ s 
and display negligible spectral lag \citep[i.e., consistent with zero;][]{Barthelmy110402A,Golenetskii110402A}, typical of short GRBs with EE \citep{Norris2006, Gehrels2006}.

\textit{Swift}/XRT localized a fading X-ray source, identified as the afterglow, at RA, DEC (J2000) = $13^{h} 09^m 36^{s}.58$, $\ang{+61;15;09.2}$ with accuracy  \ang{;;1.5} (90\% CL). 
\textit{Swift} also detected the optical afterglow in stacked UVOT exposures with detections in the $\textit{wh}$, $b$, $uvw1$, and $uvw2$ filters, implying a redshift $z\lesssim 1.5$. We use the stacked UVOT $\textit{wh}$-band image to localize the GRB position to RA, DEC (J2000) = $13^{h} 09^m 36^{s}.63$, $\ang{+61;15;09.9}$ with $1$-sigma error (statistical) $\sigma_\textrm{AG}=0.07\arcsec$, consistent with the afterglow position originally reported by \citet{Mundell110402A}. The position error does not include the systematic tie uncertainty between UVOT and USNO, as we utilize relative alignment between UVOT and our late-time imaging to derive a precise offset of potential host galaxies from the GRB.

We obtained observations with the Gemini North telescope on August 3, 2020 in the $r$-band, followed by observations in the $i$ and $z$-bands with the LDT on May 5 and 6, 2021. We further complemented our observations with archival Keck/LRIS images taken on May 27, 2014 in the $B$ and $I$ filters. Our observations unveiled the presence of three galaxies nearby the GRB position (Figure \ref{fig: GalOpt}). The first galaxy (G1) is located at $0.91\pm0.17\arcsec$ from the GRB position  with magnitude $B=24.13\pm0.11$, $r=24.20\pm0.20$, $i=23.32\pm0.09$, and $z=22.98\pm0.16$ mag.
The two other galaxies are located at larger offsets of $6.3\arcsec$ (G2) and $7.4\arcsec$ (G3). G2 has magnitudes $r=23.30\pm0.10$, $i=22.62\pm0.08$, and $z=22.18\pm0.09$ mag, whereas G3 has $r=23.19\pm0.12$, $i=22.22\pm0.05$, and $z=21.21\pm0.05$ mag.
No other sources are identified near the GRB position to depth $r\gtrsim25.2$ mag. The probability of chance coincidence for these galaxies is $P_{cc}=0.03$, $0.36$, and $0.29$ for G1, G2, and G3, respectively. Based on this, we consider G1 the putative host galaxy of GRB 110402A.

We utilized the broadband SED (see Table \ref{tab: observations}) from the Keck, Gemini, and LDT observations to derive a photometric redshift $z_\textrm{phot}=0.9\pm0.1$ and a stellar mass $\log(M_*/M_\odot)=9.5^{+0.4}_{-0.2}$ using the \texttt{prospector} software \citep{Johnson2019} with the methods outlined in \citet{OConnor2021} and \citet{Piro2021}; see also Appendix \ref{sec:prospector} and Figure \ref{fig: SED_fits} for more details. This photometric redshift is consistent with the upper limit to the GRB redshift based on the $uvw2$ detection of the afterglow. 

Additionally, we analyzed Keck/LRIS spectra of G1 taken on May 27, 2014 (Table \ref{tab: SpecObs}). A faint trace is visible above $7000$ \AA\, and we identify a single emission line at 6910 \AA\, which we interpret as [OII]$_{3727}$ at $z=0.854\pm0.001$. This interpretation is supported by the galaxy SED and the photometric redshift from \texttt{prospector}.

Adopting a redshift $z=0.854$, we derive a lower limit on the density of the GRB environment, $n_\textrm{min}\gtrsim 4.0\times 10^{-4}$ cm$^{-3}$, using the early X-ray lightcurve. This limit is consistent with the GRBs moderate offset from G1, $R=7.2\pm1.3$ kpc. We further derive a host-normalized offset of $R_o/R_e\sim 1.3\pm0.3$.



\subsubsection{GRB 130912A}

GRB 130912A was detected with \textit{Swift}/BAT \citep{DElia130912A}, \textit{Fermi}/GBM \citep{Zhang130912A}, and the Konus-\textit{Wind} satellite \citep{Golenetskii130912A} on September 12, 2013 at 08:34:57 UT. As seen by BAT, the GRB was double-peaked with duration $T_{90}=0.28\pm0.03$ s. 
A fading X-ray source was localized to RA, DEC (J2000) = $03^{h} 10^m 22^{s}.14$, $\ang{+13;59;48.1}$ with uncertainty $2.0\arcsec$. 
This was followed by the localization of the optical afterglow to RA, DEC (J2000) = $03^{h} 10^m 22^{s}.23$, $\ang{+13;59;48.7}$ by GROND, P60, and WHT \citep{Tanga130912A,Cenko130912A,Tanvir130912A}. We make use of the WHT imaging for relative astrometry, although we note the detection is marginal and the afterglow is localized with a large statistical uncertainty $\sim$\,$0.3\arcsec$ compared to the rest of our optically localized sample.


We carried out late-time observations of GRB 130912A on February 25, 2014 with LDT/LMI in $r$-band and on October 25, 2014 with Keck/LRIS in the $G$ and $R$ filters. We supplement these observations with archival imaging by \textit{HST}/WFC3 in the \textit{F110W} filter obtained on January 9, 2017. In the \textit{HST} imaging we detect three very faint sources at $<3\arcsec$ from the afterglow location which were not previously detected in the ground-based LDT or Keck imaging, see Figures \ref{fig: 130912A_HST_vs_Keck} and \ref{fig: GalOpt}. 
Sources A and B have magnitudes $\sim 26.8\pm0.3$ and $\sim 26.7\pm0.3$ at offsets $0.7\arcsec$ and $1.2\arcsec$, respectively. This yields chance probability $P_{cc}=0.08$ and $0.21$ using $H$-band number counts. 
The third source, labelled as G1, likewise has a high probability of chance coincidence, $\sim 0.4$.  We do not find any other sources at the GRB's optical localization to $F110W\gtrsim27.0$ mag (corrected for Galactic extinction).
We note that although there are other field galaxies identified at offsets $>6\arcsec$ these sources have $P_{cc}>0.4$. Based on these probabilistic arguments, we consider Source A the host galaxy of GRB 130912A, pending confirmation of the source as a galaxy. Based on the extremely faint nature of Source A, we consider that it likely has a high-$z$ origin, and assume $z=1$ in Table \ref{tab: host properties} to compute the projected physical offset of $5.6\pm2.6$ kpc.

Based on the early X-ray lightcurve, we derive a lower limit to the density of $n_\textrm{min}\gtrsim 2.1\times 10^{-3}$ cm$^{-3}$. This density is consistent with an ISM environment, and suggests that GRB 130912A originated within the confines of a nearby host galaxy.

\subsubsection{GRB 131004A}




On October 4, 2013 at 21:41:03 UT GRB 131004A triggered \textit{Swift}/BAT \citep{Hagen131004A} and \textit{Fermi/GBM} \citep{Xiong131004A}. The BAT burst displayed a single short spike with duration $T_{90}=1.5\pm0.3$ s. 
XRT localized a fading X-ray transient at RA, DEC (J2000) = $19^{h} 44^m 27^{s}.11$, $\ang{-02;57;30.3}$ with $2\arcsec$ uncertainty. Shortly thereafter the optical afterglow was localized to RA, DEC (J2000) = $19^{h} 44^m 27^{s}.10$, $\ang{-02;57;30.46}$.
Follow-up observations by Magellan \citep{Chornock131004A} and TNG \citep{DElia131004A} determined a redshift $z=0.717$ based on the identification of superimposed emission lines in the optical spectrum of the afterglow. The evidence for absorption features was reported to be marginal. 


In order to identify the environment and host galaxy of GRB 131004A, we used archival imaging from Keck/MOSFIRE in the $K_s$-band and \textit{HST}/WFC3 in the \textit{F110W} filter. We note that the field is relatively crowded (Figure \ref{fig: GalOpt}), with many foreground stars within a few arcseconds of the GRB position. However, we detected an extended source (G1) nearby to the GRB's optical localization. This source has magnitude $F110W=25.58\pm0.05$ mag and its centroid is located at an offset of $0.41\arcsec$ from the GRB position. The probability of chance alignment for G1 is $P_{cc}=0.05$. There are a number of other nearby faint sources, which cannot be classified as either stars or galaxies. These are Source A with $F110W=26.6\pm0.3$ mag at $0.6\arcsec$, Source B with $F110W=26.2\pm0.2$ mag at $2.3\arcsec$, and Source C with $F110W=25.91\pm0.13$ mag at $3.3\arcsec$ from the optical localization. These sources have a significantly higher probability of chance coincidence compared to G1 with $P_{cc}=0.14$, $0.50$, and $0.67$ for Sources A, B, and C, respectively.

The closest bright galaxy, besides G1, is located at an offset of $7.8\arcsec$ and has magnitude $F110W=21.19\pm0.01$. We refer to this source as G2, and exclude it as a candidate host due to the high probability of chance coincidence ($P_{cc}=0.22$), as well as the fact that it would be odd to detect emission features at such a large offset from the galaxy ($\sim$\,$60$ kpc at $z=0.717$). 

No other source is found coincident to the GRB localization with a $3\sigma$ upper limit $F110W\gtrsim 27.0$ mag (corrected for Galactic extinction). Given the emission line features coincident with the GRB position in the optical spectrum \citep{Chornock131004A,DElia131004A}, we suggest that the GRB originated from a star forming region within G1. At $z=0.717$, G1 is significantly under-luminous for a sGRB host galaxy ($<$\,$0.1L^*$), and this may suggest that GRB 131004A is an interloping long GRB (which is also possible given the softness of its prompt gamma-ray emission). The GRB may just appear short due to a tip-of-the-iceberg effect \citep{Moss2022} (see also \citealt{Bromberg2013}).   

We compute a lower limit to the circumburst density of $n_\textrm{min}\gtrsim 1.5\times10^{-3}$ cm$^{-3}$ (see Table \ref{tab: XrayAGprop}). We note that the physical offset of the GRB from its host galaxy, assuming the galaxy is the true host and also resides at $z=0.717$ \citep{Chornock131004A}, is $3.1\pm1.3$ kpc. Moreover, the host-normalized offset is $R_o/R_e=1.0\pm0.4$, consistent with the half-light, $R_e$, radius of its host galaxy (Table \ref{tab: host properties}). These two factors (i.e., density and offset) are consistent with the GRB occurring in an ISM environment within its host galaxy.

\subsubsection{GRB 140129B}


On Janaury 29, 2014 at 12:51:09 UT, \textit{Swift}/BAT triggered on GRB 140129B \citep{Bernardini140129B}. The burst displayed a duration $T_{90}=1.35\pm 0.21$ s. A fading X-ray source was localized by the XRT to RA, DEC (J2000) = $21^{h} 47^m 01^{s}.62$, $\ang{+26;12;23.0}$ with error $2.2\arcsec$ (90\% CL). Simultaneously, UVOT identified a bright optical afterglow located at RA, DEC (J2000) = $21^{h} 47^m 01^{s}.66$, $\ang{+26;12;22.95}$. The afterglow was detected in all UVOT filters, including $uvw2$, leading to the conclusion that the redshift of the burst is $z\lesssim1.5$ \citep{Swenson140129B}. We utilize the early UVOT imaging for the relative alignment of our late time images. 

We obtained late-time imaging with the LDT/LMI on June 10, 2014, November 3, 2019, and August 6, 2021 covering the $griz$ filters. At the optical localization, offset by only $\sim$\,$0.5\arcsec$, we identify an extended galaxy, referred to as G1 (Figure \ref{fig: GalOpt}). We derive magnitudes $g=24.22\pm 0.18$, $r=23.30\pm 0.09$, $i=23.37\pm 0.10$, and $z>23.0$ AB mag. This photometry suggests that the 4000 \AA break occurs in the $g$-band, leading to a photometric redshift estimate between $z=0.3-0.6$. We compute the probability of chance alignment for G1 to be $P_{cc}=0.009$ using the $r$-band magnitude. We note that the next closest galaxy candidates are located at offsets $>30\arcsec$ with $P_{cc}>0.25$. We can exclude additional nearby galaxies to depth $r\gtrsim 24.8$ mag (corrected for Galactic extinction, see Table \ref{tab: observations}). Based on this, we consider G1 the putative host of GRB 140129B.

We utilized the broadband SED ($griz$; see Table \ref{tab: observations}) to derive a photometric redshift $z_\textrm{phot}=0.4\pm0.1$ and a stellar mass $\log(M_*/M_\odot)=9.1\pm0.1$ using the \texttt{prospector} software. This photometric redshift is consistent with the upper limit to the GRB redshift ($z<1.5$) based on the $uvw2$ detection of the afterglow.

Using the early X-ray afterglow, we compute a lower limit to the circumburst density of GRB 140129B yielding $n_\textrm{min}\gtrsim 1.0\times 10^{-3}$. This is consistent with the GRB occurring in an ISM environment, as expected based on the small offset of the GRB from its host galaxy. Assuming $z\sim 0.5$, as suggested by the galaxy's SED, the physical offset of the GRB from G1 is $\approx$\,$3.0\pm1.0$ kpc, and the host-normalized offset is $R_o/R_e=1.0\pm0.3$ (see Table \ref{tab: host properties}).

\subsubsection{GRB 140930B}

GRB 140930B was detected with \textit{Swift}/BAT and Konus-\textit{Wind} on September 30, 2014 at 19:41:42 UT. The GRB had a duration $T_{90}=0.84\pm0.12$ s.
\textit{Swift}/XRT localized the X-ray afterglow to RA, DEC (J2000) =  $00^{h} 25^m 23^{s}.40$, $\ang{+24;17;41.7}$ with uncertainty $2.0\arcsec$. The optical counterpart was localized to RA, DEC (J2000) =  $00^{h} 25^m 23^{s}.43$, $\ang{+24;17;39.4}$ \citep{Tanvir140930B}. 
We note that the most up-to-date XRT enhanced position is now shifted away from this optical localization, compared to the originally reported enhanced position \citep{Goad140930B}, but that the positions are still consistent at the 99.7\% confidence level \citep[assuming the XRT position error follows Rayleigh statistics][]{Evans2014,Evans2020}.


On August 1, 2020, we obtained late-time imaging of the field of GRB 140930B with Gemini GMOS-N in \textit{r}-band. We supplemented this with early-time Gemini GMOS-N imaging from October 1 and 2, 2014 which was aimed at identifying the GRB afterglow. The afterglow is clearly detected in these early images, but the position is contaminated by the PSF of a saturated, nearby star ($r\sim 13.1$ mag). Although the afterglow position is contaminated, we uncover a faint source 
with magnitude $r=23.8\pm0.2$ AB mag at an offset of $\sim$\,$1.4\arcsec$ from the afterglow localization. 
The probability of chance coincidence for this source is $P_{cc}=0.02$. However, due to the PSF of the saturated star we cannot confirm whether this is a foreground star or a galaxy, and, therefore, we refer to this as Source A. 
Furthermore, we note that in each of these three Gemini images there is a possible extension of Source A to the northwest, but it is not clear based on this data whether this is due to a secondary source underlying the GRB position or a true extension of Source A.

As Source A is also clearly detected in the early Gemini GMOS-N afterglow imaging from October 1 and 2, 2014, we can determine a precise offset (i.e., without a tie uncertainty $\sigma_\textrm{tie}$) from the GRB position of $R_o=1.4\pm1.1\arcsec$. Assuming $z\sim0.5$, this yields a physical offset of $8.8\pm0.9$ kpc. As there are no other likely hosts for GRB 140930B identified in these Gemini images, we consider Source A to be the candidate host galaxy, although we note that deeper observations are required to determine the extension of Source A and confirm its nature as a galaxy.

Following \citet{OConnor2020}, we further derive a lower limit to the circumburst density of $\gtrsim 1.4\times 10^{-3}$ cm$^{-3}$. This implies that the GRB originated from within a dense galactic environment, consistent with the ISM.

\subsubsection{GRB 150423A}


At 06:28:04 UT on April 23, 2015, GRB 150423A was detected with \textit{Swift}/BAT \citep{Pagani150423A}. The burst had a duration $T_{90}=0.22\pm0.03$. XRT detected the afterglow at RA, DEC (J2000) = $14^{h} 46^m 18^{s}.96$, $\ang{+12;17;00.3}$ with $2.1\arcsec$ uncertainty. Shortly after the GRB trigger ($\sim$\,$30$ m), the optical afterglow was localized to RA, DEC (J2000) = $14^{h} 46^m 18^{s}.86$, $\ang{+12;17;00.7}$ \citep{Varela150423A}. 


We analyzed archival \textit{HST}/WFC3 imaging obtained on February 3, 2017 in the $F110W$ filter. The field is relatively crowded with many 
galaxies located at $<8\arcsec$ from the optical localization of GRB 150423A (Figure \ref{fig: GalOpt}). There are also a few bright ($\sim$\,20 to 21 mag) SDSS galaxies residing at larger offsets $\gtrsim 15\arcsec$ with high probabilities of chance coincidence ($P_{cc}\gtrsim 0.3$). These SDSS galaxies are not displayed in Figure \ref{fig: GalOpt}. 

The closest source to the GRB position is a faint galaxy (G1) offset by $1.6\arcsec$ with magnitude $F110W=25.3\pm 0.07$ mag, yielding $P_{cc}=0.18$ using $H$-band number counts \citep{Metcalfe2006,Galametz2013}. 
The other galaxies displayed in Figure \ref{fig: GalOpt} are located at offsets of $3.8$, $4.7$, $6.2$, and $7.0\arcsec$ with magnitudes $F110W=22.696\pm0.007$, $22.620\pm0.006$, $23.93\pm0.03$, and $22.85\pm0.01$ for G2, G3, G4, and G5, respectively.
These galaxies have a high probability of chance alignment with the GRB position ranging from $P_{cc}=0.15$, $0.2$, $0.6$, and $0.5$ for G2, G3, G4, and G5. We further note that the nearby galaxy G2 has a spectroscopic redshift $z=0.456$ reported by \citet{Perley150423A}.
No coincident source is detected at the GRB position to $F110W\gtrsim 27.2$ mag. We therefore conclude that GRB 150423A is observationally hostless as it is unclear which of these multiple candidates is the true host or whether the BNS system resided within a faint undetected galaxy.


We note that optical spectroscopy of the afterglow starting at $\sim$22 min set a robust upper limit $z<2.5$ to the redshift of GRB 150423A \citep{Malesani150423A}. The same observation marginally detected an MgII absorption doublet at $z=1.394$. However, due to the tentative nature of the detection and lack of other evidence, we do not consider this the conclusive redshift of GRB 150423A.

We set a lower limit $n_\textrm{min}\gtrsim 2.6\times 10^{-4}$ cm$^{-3}$ to the density of the GRB's environment. This suggests that the GRB occurred within a galactic ISM environment, either within one of the nearby candidate galaxies or in a faint galaxy ($z<2.5$) which was not detected with the optical and infrared observations presented in this work.

\subsubsection{GRB 160408A}

GRB 160408A was detected with \textit{Swift}/BAT \citep{Evans160408A} and \textit{Fermi}/GBM \citep{Roberts160408A} on April 8, 2016 at 06:25:43 UT. The duration observed by BAT was $T_{90}=0.32\pm0.04$ s. \textit{Swift}/XRT localized the X-ray afterglow to RA, DEC (J2000) =  $08^{h} 10^m 29^{s}.93$, $\ang{+71;07;41.7}$ with uncertainty $2.2\arcsec$. The optical counterpart was localized to RA, DEC (J2000) =  $08^{h} 10^m 29^{s}.81$, $\ang{+71;07;43.7}$.


We carried out late-time imaging with the LDT in $griz$ filters on March 29, 2020. These observations were supplemented by Gemini GMOS-N imaging obtained in \textit{r}-band on April 8 and 9, 2016. In the Gemini imaging we detect two nearby candidate hosts at offsets $1.6\arcsec$ (source A) and $3.8\arcsec$ (G1), see Figure \ref{fig: GalOpt}, whereas in our shallower LDT imaging we detect only G1. Source A has magnitude $r=25.5\pm0.2$ mag and G1 has magnitude $r=23.54\pm0.10$ mag. The probability of chance alignment is $P_{cc}=0.13$ and $0.16$ for source A and G1, respectively. No source is detected coincident with the optical localization to depth $r\gtrsim 25.8$ mag. As both source A and G1 have similar probabilities of chance association, we consider GRB 160408A to be observationally hostless. Moreover, there are no bright galaxies from which it is likely the GRB was highly kicked.

Using the early X-ray afterglow lightcurve, we set a lower limit of $n_\textrm{min}\gtrsim 1.8\times 10^{-4}$ cm$^{-3}$ to the circumburst environment of GRB 160408A. This density implies the GRB occurred within a galactic environment.

\subsubsection{GRB 160410A}
\label{sec: 160410A}

At 05:09:48 UT on April 10, 2016, \textit{Swift}/BAT \citep{Gibson160410A} and Konus-\textit{Wind} \citep{Frederiks160410A} triggered on GRB 160410A. The BAT lightcurve displays an initial short, hard pulse with duration $\lesssim$\,$2$ s. However, there is a clear extended tail of the burst lasting for tens of seconds. The duration reported in the BAT GRB Catalog (Table \ref{tab: XrayAGprop}) is $T_{90}=96\pm50$ s. In addition, \citet{Sakamoto160410A} found that the spectral lag of the initial short pulse is consistent with zero, typical of sGRBEE. The GRB is therefore interpreted as having extended emission.
Shortly after the GRB, \textit{Swift}/XRT localized the X-ray afterglow to RA, DEC (J2000) = $10^{h} 02^m 44^{s}.47$, $\ang{03;28;41.0}$ with $3.2\arcsec$ uncertainty. A more precise localization of the optical counterpart to RA, DEC (J2000) = $10^{h} 02^m 44^{s}.37$, $\ang{03;28;42.4}$ was quickly discovered \citep{Yates160410A}. 

We obtained late-time imaging of the field of GRB 160410A with the LDT/LMI on December 14, 2021 and January 15, 2021 in the $g$ and $r$-bands. These observations were supplemented by public archival imaging with Keck/DEIMOS from April 28, 2016. In order to precisely localize the afterglow location in these late-time images, we utilized the initial detection of the optical counterpart by \textit{Swift}/UVOT \citep{Gibson160410A}. We display a finding chart of the field in Figure \ref{fig: GalOpt}. No source is identified coincident with the optical localization to depth $g\gtrsim 24.9$, $R\gtrsim 25.0$, and $I\gtrsim 24.2$ AB mag. We note that a deeper constraint on an underlying host of $r\gtrsim 27.17$ ($3\sigma$; corrected for Galactic extinction) was presented by \citet{AguiFernandez2021} based on late-time deep GTC imaging. This is in sharp contrast to the results obtained from optical spectroscopy of the afterglow (see below). 
However, we note the presence of two bright SDSS galaxies within $60\arcsec$ of the GRB localization with $r=18.9$ mag at $20\arcsec$ and $r=17.8$ mag at $35\arcsec$ yielding $P_{cc}=0.11$ and $0.14$, respectively. Despite the lower $P_{cc}$ compared to other candidates, the projected physical offset from these galaxies at their estimated photometric redshifts of $z_\textrm{phot}=0.2$ and $z_\textrm{phot}=0.1$ is $\sim$\,69 and 67 kpc, respectively. Furthermore, the photometric redshifts are inconsistent with the measured redshift for GRB 160410A (see below).
The probability of chance coincidence for any other extended object at larger offsets is $P_{cc}\gtrsim 0.5$ due to their faintness $R\sim24$ mag. We therefore consider GRB 160410A to be observationally hostless.

We analyzed Keck spectroscopy performed with LRIS on April 10, 2016 targeted at the optical afterglow of GRB 160410A beginning at 84 min after the GRB. The afterglow is detected as a blue continuum from $\sim 3100$ - $5680$ \AA\, with a large number of visible absorption features. The continuum normalized spectrum is displayed in Figure \ref{fig: spectra160410A}. We identify a broad damped Lyman alpha (hereafter, Ly$\alpha$) absorption feature at $\lambda_\textrm{obs}\sim3304$ \AA, which drives the redshift derivation.
In addition, we find a number of absorption features located at $\lambda_\textrm{obs}\approx 3427$, $3547$, $3559$ and $4146$ \AA\, that correspond to [SiII] transitions; see Figure \ref{fig: spectra160410A}. These features, on top of the Ly$\alpha$ trough, allow us to derive a redshift $z=1.717 \pm 0.001$. 
Moreover, we identify absorption features corresponding to two intervening absorbers for which we identify [CIV] at both $z=1.444$ and $z=1.581$. In Figure \ref{fig: spectra160410A}, we mark also tentative detections of [SiII] and [SiIV] at $z=1.444$ and [SiII] and [NII] at $z=1.581$.
The redshifts of these absorbers are consistent with the estimates of \citet{Bloom1997} that the GRB is not residing further than $1.25\times$ the redshift of the intervening system.
Our results are consistent with the analyses presented by \citet{Selsing160410A}, \citet{Cao160410A}, \citet{Selsing2019}, and \citet{AguiFernandez2021}. 

The Ly$\alpha$ trough provides strong evidence that the GRB originated from within a dense galactic environment with a neutral Hydrogen column density of $\log(N_\textrm{HI}/\textrm{cm}^{-2})=21.3\pm0.3$ \citep{Selsing2019,AguiFernandez2021}, see \citet{AguiFernandez2021} for an in depth discussion of the environment of GRB 160410A. Therefore, GRB 160410A is very unlikely to be physically hostless (i.e., occurring in an IGM-like environment outside of its birth galaxy). This is is contrast to the field of the GRB, for which there are no candidate host galaxies identified to deep limits ($r\gtrsim27.17$ mag; \citealt{AguiFernandez2021}). This event delivers the first substantial evidence for a sample of short GRBs located in high$-z$ galaxies, which are not identified through observational follow-up. Furthermore, the two intervening absorbers at $z=1.444$ and $z=1.581$ are likewise not detected in the Keck/DEIMOS or LDT imaging, further emphasizing the possibility of non-detected high$-z$ galaxies coincident to short GRBs. We emphasize that deep nIR imaging (e.g., \textit{HST}, \textit{JWST}) is crucial to the detection of these galaxies.

As further evidence, we utilized the early X-ray lightcurve in order to derive a lower limit to the circumburst density of  $n_\textrm{min}\gtrsim 2.6\times 10^{-3}$ cm$^{-3}$. This value is inconsistent with an IGM-like environment, and provides further evidence that GRB 160410A occurred within a undetected host galaxy at $z=1.717$.


\subsubsection{GRB 160525B}

At 09:25:07 UT, \textit{Swift}/BAT triggered and
located GRB 160525B \citep{Krimm160525B}. The short burst had a duration $T_{90}=0.29\pm0.05$. The XRT localized the X-ray afterglow to an enhanced position RA, DEC (J2000) = $09^{h} 57^m 32^{s}.30$, $\ang{51;12;24.0}$ with $2.1\arcsec$ uncertainty (90\% CL). In an initial finding chart exposure UVOT marginally detected an optical source coincident with the XRT position. The source was located at RA, DEC (J2000) = $09^{h} 57^m 32^{s}.23$, $\ang{51;12;24.9}$ with uncertainty $0.6\arcsec$ (90\% CL). 
The UVOT detection of the afterglow in the \textit{wh} filter sets an upper limit of $z\lesssim 5$ to the redshift of GRB 160525B. We utilize this detection of the optical afterglow for relative astrometry with our late-time images.


We performed optical imaging with the LDT/LMI on January 29, 2020, February 29, 2020, and December 15, 2021 covering $griz$ wavelengths. We identified a host galaxy candidate coincident with the UVOT localization of GRB 160525B (Figure \ref{fig: GalOpt}). This galaxy, G1, has magnitudes $g=23.27\pm0.15$, $r=23.27\pm0.09$, $i=23.28\pm0.18$, and $z= 23.4\pm 0.3$ mag. 
G1 has a probability of chance alignment of $P_{cc}=0.03$. 
In addition to G1, there are a number of other candidate hosts in the field (see Figure \ref{fig: GalOpt}), including two other faint sources within $7\arcsec$ and two bright SDSS galaxies at offsets of $13\arcsec$ and $21\arcsec$. No other sources are uncovered nearby the GRB position to depth $r\gtrsim 24.6$ AB mag. 
The nearby sources, G2 and source A, have magnitudes $r=24.2\pm 0.2$ and $24.3\pm 0.2$ mag with $P_{cc}=0.25$ and $0.6$. The bright SDSS galaxies have magnitude $r=19.43\pm0.03$ and $19.95\pm0.03$ mag for G3 and G4, respectively, yielding $P_{cc}=0.09$ and $0.26$. Based on the significantly smaller $P_{cc}$ for G1 compared to these other candidates, we consider the coincident galaxy G1 to be the putative host of GRB 160525B.

Using the early X-ray lightcurve, we set a lower limit to the density surrounding the GRB's explosion site of $n_\textrm{min}\gtrsim 6.6\times 10^{-3}$ cm$^{-3}$. This density is consistent with the GRB occurring in an ISM environment, which is likely given the very small offset, $0.06\pm0.25\arcsec$, of the GRB from its putative host galaxy (G1). We note that as the half-light radius of G1 is $\sim$\,$1.0\arcsec$ the host-normalized offset is likewise $0.06\pm0.25$.

\subsubsection{GRB 160601A}

GRB 160601A triggered \textit{Swift}/BAT on June 1, 2016 at 14:43:02 UT \citep{Kocevski160601A}. The burst displayed a single pulse with duration $T_{90}=0.12\pm0.02$ s. 
The X-ray afterglow was detected with \textit{Swift}/XRT at RA, DEC (J2000) = $15^{h} 39^m 44^{s}.55$, $\ang{+64;32;28.7}$ with accuracy  \ang{;;4.3} (90\% CL). The optical afterglow was further localized to RA, DEC (J2000) = $15^{h} 39^m 43^{s}.97$, $\ang{+64;32;30.5}$ at 6.5 hr after the BAT trigger \citep{Malesani160601A}. 



We observed the GRB position with Gemini/GMOS-N on August 1, 2020 to search for underlying galaxies. We supplemented this observation with LDT imaging in the $griz$ filters,  
archival GTC/OSIRIS imaging in $r$-band, and archival imaging from Keck/MOSFIRE in the $K_s$-band.
We identify four nearby galaxies with offsets $\sim 4.8\arcsec$ to the West (G1) and North-East (G2), $6.2\arcsec$ to the East (G3), and $6.5\arcsec$ to the South-West (G4) of the GRB position (see Figure \ref{fig: GalOpt}). Their $r$-band magnitudes are $25.1\pm0.15$ mag (G1), $25.4\pm 0.3$ (G2), $22.90\pm0.05$ (G3), and $24.55\pm0.10$ mag (G4). The chance probability, based on $r$-band number counts, for each is $\gtrsim 0.4$, with the exception of G3 which has $P_{cc}=0.24$.
However, we note that G2, G3, and G4 are infrared bright, and detected in the Keck/MOSFIRE imaging with magnitudes $K_s=21.55 \pm 0.09$, $21.50 \pm 0.15$, and $20.90 \pm 0.07$ AB mag, respectively. The $P_{cc}$ based on these infrared magnitudes is $0.11$, $0.13$, and $0.13$ for G2, G3, and G4, respectively. This further complicates the host identification for GRB 160601A, as these three galaxies are equally likely hosts and none has $P_{cc}<0.1$. As no other sources are identified coincident to the optical localization to depth $r\gtrsim 25.9$ mag, we assign it an observationally hostless classification. Moreover, there are no bright galaxies from which it is likely the GRB was highly kicked.

Based on the early X-ray lightcurve, we set a lower limit to the circumburst density surrounding the GRB of $n_\textrm{min}\gtrsim 1.2\times10^{-5}$ cm$^{-3}$. We note that this lower limit is consistent with the GRB occurring in either an ISM or an IGM-like environment.

\subsubsection{GRB 160927A}
\label{sec: 160927A_res}

\textit{Swift}/BAT detected GRB 160927A on September 27, 2016 at 18:04:49 UT \citep{Gibson160927A}. In addition, GRB 160927A was identified by the CALET Gamma-ray Burst Monitor
(CGBM) in a ground-based analysis with significance $\sim$\,$5.3\sigma$ \citep{Moriyama160927A}.
The mask-weighted BAT lightcurve was double peaked with $T_{90}=0.48\pm0.10$ s. 
XRT detected a fading X-ray source at RA, DEC = $17^{h} 04^m 58^{s}.19$, $\ang{+17;19;55.3}$ with uncertainty $2.2\arcsec$. 
Observations with the Russian-Turkish 1.5-m telescope (RTT150) beginning 55m post-trigger detected an uncatalogued optical source within the XRT enhanced position \citep{Tkachenko160927A}. Further observations by TNG and GROND confirmed the fading of the afterglow \citep{DAvanzo160927A,Wiseman160927A}. 
Observations with the GTC at 26.5 hr after the GRB detected the afterglow with $r=25.2\pm0.2$ mag \citep{deUgarte160927A}. Using these observations, we localized the GRB afterglow position to RA, DEC = $17^{h} 04^m 58^{s}.19$, $\ang{+17;19;55.3}$ with statistical uncertainty $\sigma_\textrm{AG}=0.08\arcsec$. We utilize this GTC imaging for relative astrometry with our late-time imaging (see below).

We obtained late-time imaging of GRB 160927A with LDT on May 20, 2018 in \textit{r}-band and with Gemini GMOS-N on August 1, 2020 in \textit{i}-band. We supplemented these observations with archival imaging from the GTC in \textit{r}-band taken on February 23, 2017 and with Keck/LRIS imaging in the $GRZ$ filters from October 6, 2018 and September 4, 2019.
These late-time images do not resolve any source coincident with the position of the optical afterglow to depth $r\gtrsim 26.0$ AB mag (corrected for Galactic extinction). 
The closest source to the GRB position (source A in Figure \ref{fig: GalOpt}) is offset by $\sim3\arcsec$ with magnitudes $r=25.8^{+0.3}_{-0.2}$ and $i=25.6^{+0.3}_{-0.2}$ mag. 
This source is too faint for a conclusive star/galaxy classification, although we note it appears marginally extended.
The chance probability for Source A is $P_{cc}=0.5$.
Additionally, there are a number of SDSS galaxies (G1, G2, G3, and G4) within the field at $>9\arcsec$, but $P_{cc}\gtrsim0.5$ for each of them.
Due to the lack of putative host galaxy, we consider GRB 160927A to be observationally hostless.


We set a lower limit of $n_\textrm{min}\gtrsim 1.1\times10^{-4}$ cm$^{-3}$ to the density of the GRB's environment based on the early X-ray afterglow. This density is consistent with the GRB occurring within the Virial radius of its host galaxy \citep{OConnor2020}, and introduces the possibility that this GRB occurred in a faint, undetected galaxy.

\subsubsection{GRB 170428A}

On April 28, 2017 at 09:14:42 UT, \textit{Swift}/BAT detected GRB 170428A \citep{Beardmore170428A}. The burst was also detected with Konus-\textit{Wind} \citep{Tsvetkova170428A} and the CGBM \citep{Yamada170428A}. The burst had a duration $T_{90}=0.2\pm0.07$ s. The X-ray afterglow was located at RA, DEC = $22^{h} 00^m 18^{s}.76$, $\ang{+26;54;57.1}$ with uncertainty $2.8\arcsec$. The optical counterpart was detected at RA, DEC = $22^{h} 00^m 18^{s}.78$, $\ang{+26;54;57.0}$ \citep{Bolmer170428A}.

We carried out late-time imaging of the field with LDT/LMI on May 21, 2018 in the $griz$ filters. These data were supplemented by early-time LDT imaging from April 29, 2017 ($\sim$ 1 d post-burst) and archival observations by TNG in \textit{i} and \textit{z} from May 1, 2017 ($\sim$ 3 d post-burst). In order to localize the afterglow, we performed image subtraction between these early and late-time images using the \texttt{HOTPANTS} software \citep{Becker2015}. We do not detect the afterglow in either the LDT or TNG images, and instead use the reported position from GROND \citep{Bolmer170428A}. 

In our late-time LDT imaging, we detect a candidate host galaxy (G1) at offset $1.2\arcsec$ from the afterglow localization. The galaxy has magnitudes $g>23.3$, $r=22.09\pm0.10$, $i=21.84\pm0.15$, and $z=21.88\pm0.15$ mag; the galaxy is not detected in the $g$-band due to the $4000$ \AA\, break. The probability of chance coincidence is $P_{cc}=0.01$. 
We report the detection of another extended galaxy (G2) at offset $\sim$\,$13\arcsec$ with $r=21.53\pm0.07$. This galaxy has an $34\%$ probability of chance alignment. There is no source detected at the GRB's optical localization to $i\gtrsim 23.6$ and $z\gtrsim 23.4$ mag (corrected for Galactic extinction).
Based on these arguments, we consider G1 the putative host galaxy for GRB 170428A.


The galaxy G1 has a redshift of $z=0.454$ determined by optical spectroscopy with the GTC \citep{Izzo170428A}. At this redshift, the projected physical offset of the GRB from its host is $7.2\pm1.8$ kpc. The host-normalized offset is $R_o/R_e=1.0\pm0.3$, consistent with the GRB occurring within the half-light radius of G1. We compute a lower limit for the density of the environment surrounding the GRB of $n_\textrm{min}\gtrsim 1.6\times 10^{-5}$ cm$^{-3}$.

\subsubsection{GRB 170728A}

GRB 170728A was detected and localized by \textit{Swift}/BAT on July 28, 2017 at 06:53:28 UT \citep{Cannizzo170728A}. The burst was single pulsed with duration $T_{90}=1.25\pm0.23$ s. 
A fading X-ray source was detected with \textit{Swift}/XRT at RA, DEC (J2000) =  $03^{h} 55^m 33^{s}.21$, $\ang{+12;10;53.2}$ with uncertainty $2.1\arcsec$. 
Shortly thereafter, UVOT discovered an uncatalogued, fading source inside the XRT position at RA, DEC (J2000) =  $03^{h} 55^m 33^{s}.17$, $\ang{+12;10;54.7}$ \citep{Laporte170728A}. 
We used the UVOT \textit{wh}-band afterglow discovery image to localize the GRB in our late-time images.

In order to search for the host galaxy of GRB 170728A, we obtained late-time imaging with the LDT/LMI on January 8, 2019 in $r$-band.
Additionally, we retrieved publicly available late-time images from the Keck Observatory (PI: Fong) taken January 14, 2018 in $G$ and $R$. In these imaging, we uncover four visually extended sources within $15\arcsec$ of the GRB position. However, the PSF of a nearby, very bright star ($r \sim$11.8 mag; SDSS) contaminates the GRB localization in each image. No source is detected coincident to the GRB position with a $3\sigma$ upper limit of $R\gtrsim 24.7$ mag (the shallow limit is due to a diffraction spike from the bright star, and the Galactic extinction, $E(B-V)=0.21$ mag, in the direction of the burst). For the nearby galaxies, we derive magnitudes $R=23.89 \pm 0.12$, $23.31 \pm 0.15$, $23.76 \pm 0.13$, and $22.76 \pm 0.15$ mag for G1, G2, G3, and G4, respectively, with offsets of $4.4\arcsec$, $6.7\arcsec$, $7.4\arcsec$, and $14\arcsec$. We note that the photometry for G3, in particular, is contaminated by the diffraction spike from the bright star. We find a probability of chance coincidence of $P_{cc}=0.23$, $0.32$, $0.49$, and $0.67$ for G1, G2, G3, and G4, respectively. Thus, we find that GRB 170728A is observationally hostless. Future observations at a different position angle can provide deeper constraints on an underlying source.

We compute a lower limit to the circumburst density of the GRB's environment, constraining it to be $\gtrsim 1.2\times 10^{-4}$ cm$^{-2}$. This suggests the GRB originated from within a galactic environment.

\subsubsection{GRB 170728B}

At 23:03:19 UT on July 28, 2017, \textit{Swift}/BAT \citep{Cenko170728B}, \textit{Fermi}/GBM \citep{Stanbro170728B}, \textit{Fermi}/LAT \citep{Yassine170728B}, and Konus-\textit{Wind} \citep{Kozlova170728B} triggered on GRB 170728B. The GRB displayed an initial short pulse with duration $<1$ s, followed by a weak, softer emission until $\sim$50 s. The $T_{90}$ duration observed by BAT in 15-150 keV is $48\pm25$ s. Due to these features, we classify this event as a candidate sGRBEE. The X-ray afterglow was localized to RA, DEC (J2000) = $15^{h} 51^m 55^{s}.44$, $\ang{+70;07;21.4}$ with uncertainty $1.9\arcsec$ (90\%). An optical counterpart was identified shortly after, localizing the GRB to RA, DEC (J2000) = $15^{h} 51^m 55^{s}.47$, $\ang{+70;07;21.1}$ \citep{DAvanzo170728B}.

We carried out late-time observations with the LDT/LMI on November 3, 2019 and December 7, 2019 covering $griz$ wavelengths. At the position of the optical counterpart we identify a bright host galaxy (G1) with magnitudes $g=23.71\pm0.06$, $r=23.06\pm0.06$, $i=22.63\pm0.05$, and $z=22.33\pm0.15$ mag. The SED suggests that the 4000 \AA\, break occurs between the $g$ and $r$-bands, hinting at a photometric redshift in the range $z\sim0.3-0.6$.
The offset of the GRB from this galaxy is $0.8\arcsec$ yielding $P_{cc}=0.014$. There are no other nearby galaxy candidates to magnitude $r\gtrsim 24.6$ mag.
We note the presence of a catalogued galaxy with magnitude $r=20.3$ at offset $\sim$24$\arcsec$, but the $P_{cc}=0.4$ (due to the large offset, this galaxy is not displayed in the finding chart). We therefore consider G1 to be the putative host galaxy of GRB 170728B.

We used \texttt{prospector} to model the SED of G1 (Figure \ref{fig: SED_fits}), and obtain a photometric redshift $z_\textrm{phot}=0.6\pm0.1$ and a stellar mass $\log(M_*/M_\odot)=9.7\pm0.2$. We further derive a density $n_\textrm{min}\gtrsim 7.5\times 10^{-4}$ cm$^{-3}$ for the GRB environment using the early X-ray afterglow lightcurve. This value is inconsistent with the GRB occurring in an IGM-like environment \citep[i.e., $n<10^{-4}$ cm$^{-3}$;][]{OConnor2020}. We note that the host-normalized offset $R_o/R_e=1.1\pm0.3$ is consisent with the GRB occurring within the half-light radius of G1. Assuming a redshift $z\sim 0.64$, we compute the physical offset between the GRB and G1 to be $\approx$\,$5.5\pm1.7$ kpc.

\subsubsection{GRB 180618A}

On June 18, 2018, at 00:43:13 UT  \textit{Swift}/BAT \citep{Sakamoto180618A}, \textit{Fermi}/GBM \citep{Hamburg180618A}, Konus-\textit{Wind} \citep{Svinkin180618A}, and \textit{AstroSat} \citep{Sharma180618A} triggered on GRB 180618A. The BAT lightcurve displayed a short, multi-peaked pulse with duration $<0.5$ s followed by softer emission for tens of seconds. The total duration of the burst detected with BAT is $T_{90}=47.4\pm11.2$ s. In addition, the spectral lag of the initial pulse is negligible. For these reasons we classify GRB 180618A as an sGRBEE. The X-ray afterglow of GRB 180618A was localized to RA, DEC (J2000) = $11^{h} 19^m 45^{s}.94$, $\ang{+73;50;14.3}$ with uncertainty $2.0\arcsec$ (90\%). A more precise localization was derived by UVOT from the bright optical afterglow to RA, DEC (J2000) =  $11^{h} 19^m 45^{s}.87$, $\ang{+73;50;13.5}$ \citep{Siegel180618A}.


We carried out $grizy$ imaging with the LDT/LMI on April 7, 2019, December 7, 2019, and May 5, 2021. We uncovered a faint galaxy at an offset of $\sim$\,$1.6\arcsec$ from the optical localization of the GRB with magnitudes $g=23.89\pm0.12$, $r=22.92\pm0.08$, $i=22.33\pm0.10$, $z=22.26\pm0.12$, and $y\gtrsim 21.5$ AB mag. The probability of chance coincidence is $P_{cc}=0.03$.
Furthermore, we identified three other candidate host galaxies in the vicinity of the GRB: Source A with $r=24.5\pm0.2$ at $\sim$\,$1.6\arcsec$, G2 with $r=23.01\pm0.08$ at $4.1\arcsec$, and G3 with $r=22.29\pm0.06$ at $7.4\arcsec$. The probability of chance coincidence for these sources is $0.08$, $0.15$, and $0.23$ for Source A, G2, and G3, respectively. No other sources are identified near the GRB localization to $r\gtrsim 24.7$ AB mag (corrected for Galactic extinction). Due to the similar probability of chance coincidence for G1 and Source A (0.03 vs. 0.08), we cannot differentiate between which is the more likely host galaxy. However, deeper observations are required to confirm the source classification of Source A, and whether it is a foreground star or a galaxy. Therefore, we tentatively consider G1 the host galaxy of GRB 180805B.

We obtained optical spectroscopy of G1 with Gemini GMOS-N on February 1, 2021. We detect a very weak trace between $\sim$\,$7300$ to $9500$ \AA. There are no obvious emission or absorption features. Therefore, we instead modelled the broadband SED ($grizy$) within \texttt{prospector}. As the spectrum does not show bright emission features, we turned off nebular emission lines within \texttt{prospector}. We found that $A_V\approx 0$ provided the best fit to the SED, due to the near flat slope in the $rizy$ filters. Thus, we fixed the intrinsic extinction to $A_V=0$ in order to allow for minimization of the likelihood function. The MCMC fit resulted in $z_\textrm{phot}=0.4^{+0.2}_{-0.1}$ and a stellar mass $\log(M_*/M_\odot)=9.6\pm 0.3$ (see Figure \ref{fig: SED_fits}). At this redshift, the offset of GRB 180618A from G1 is $8.8\pm1.1$ kpc. The host-normalized offset is $R_o/R_e=1.58\pm0.24$.

Using the early X-ray lightcurve, and assuming $z\approx0.4$, we identified a lower limit of $4.0\times 10^{-3}$ cm$^{-3}$. This supports that the sGRB occurred within an ISM-like environment.

\subsection{XRT Localized}

\subsubsection{GRB 101224A}

GRB 101224A was detected with \textit{Swift}/BAT \citep{Krimm101224A} and \textit{Fermi}/GBM \citep{McBreen101224A} on December 24, 2010 at 05:27:13 UT. The duration observed by BAT was $T_{90}=0.24\pm0.04$ s. The \textit{Fermi}/GBM spectrum and lightcurve displayed similar properties to GRB 170817A \citep{VonKienlin2019}.
\textit{Swift}/XRT localized the X-ray afterglow to RA, DEC (J2000) =  $19^{h} 03^m 41^{s}.72$, $\ang{+45;42;49.5}$ with uncertainty $3.8\arcsec$. 
No optical counterpart was identified.

A candidate host galaxy (G1) was discovered at the edge of the enhanced XRT position, see Figure \ref{fig: GalXray}. This galaxy was previously reported by \citet{Nugent101224A} and \citet{Tunnicliffe2014}. 
We derive magnitudes $g=22.54\pm0.06$, $r=21.99\pm0.06$, $i=21.83\pm0.05$, and $z=21.78\pm0.05$ AB mag. The probability of chance coincidence for G1 is $P_{cc}=0.11$. In addition, we discovered a very faint source, referred to as Source A, within the XRT error circle with magnitude $r=24.7\pm0.2$. Three other candidate host galaxies, visible in Figure \ref{fig: GalXray}, are uncovered at offsets of $4.5\arcsec$, $6.4\arcsec$, and $8.4\arcsec$. The probability of chance coincidence is $>$\,$0.25$ for each of these sources. 
No other sources are identified within the XRT enhanced position to depth $r\gtrsim 24.9$ AB mag ($3\sigma$, corrected for Galactic extinction). Due to this, no other galaxy will have a lower probability of chance coincidence than G1, even if uncovered in deeper observations, making G1 the most likely host galaxy, despite the higher $P_{cc}$ value.


We performed optical spectroscopy of the candidate host galaxy, G1, on October 27, 2014 with Keck/LRIS (see Table \ref{tab: SpecObs}). The resulting spectrum is displayed in Figure \ref{fig: spectra}. We detect multiple emission lines at  $\lambda_\textrm{obs}\approx 5422$, $7067$, $7209$, $7278$, and $9542$ \AA which we associate to the [OII] doublet, H$\beta$, [OIII]$_{4960}$, [OIII]$_{5008}$, and H$\alpha$ transitions at a redshift $z=0.4536 \pm 0.0004$.
We note that at this redshift there is a marginal detection of H$_\gamma$. Although we cannot classify the galaxy type based on morphology, we suggest that the strong emission features are typical of a late-type galaxy. At this redshift, the offset of GRB 101224A from this galaxy is $R=14\pm17$ kpc.

We derive a lower limit, $n_\textrm{min}\gtrsim 3.6\times10^{-5}$ cm$^{-3}$, to the density of the GRBs environment using the early X-ray lightcurve. This density is consistent with an IGM-like environment ($n<10^{-4}$ cm$^{-3}$).

\subsubsection{GRB 120305A}

GRB 120305A was detected with \textit{Swift}/BAT on March 5, 2012 at 19:37:30 UT \citep{Stratta120305A}. The burst displayed a single peak with a fast rise and slower decay. The burst had a duration $T_{90}=0.10\pm0.02$ s. 
A fading X-ray source, identified as the afterglow, was detected at RA, DEC (J2000) =  $03^{h} 10^m 08^{s}.68$, $\ang{+28;29;31.0}$ with uncertainty $2.0\arcsec$. No optical counterpart was identified. The lack of an optical counterpart may be due to the high Galactic extinction $A_V=1.2$ mag \citep{Schlafly2011} from the GRB's localization in the direction of a molecular cloud \citep{Planck2016cloud}, which also leads to an enhanced background in the XRT localization and throughout the image (see Figure \ref{fig: GalXray}). This enhanced background is observed with a consistent pattern in all imaging of this field (e.g., Gemini, Keck, LDT), and leads to a shallower upper limit (see below). 



We performed late-time imaging with the LDT in $r$-band on March 6, 2014 and with Keck on October 25, 2014 in the $G$ and $R$-bands to search for an underlying galaxy. We further supplemented our imaging with archival Gemini observations taken in $i$-band (PI: Tanvir).
We did not discover a source within the XRT enhanced position to depth $G\gtrsim 24.6$ and $R\gtrsim 24.9$ (corrected for Galactic extinction). However, our imaging revealed the presence of three uncatalogued galaxies (G1, G2, and G3) at offsets $<15\arcsec$, see Figure \ref{fig: GalXray}.

The nearest galaxy, G1, has magnitudes $G=21.7\pm0.06$, $R=21.53\pm0.04$, and $i=20.85\pm0.08$ mag. The galaxy is offset by $5.4\arcsec$ from the GRB position, whereas G2 and G3 are fainter ($R=22.4\pm0.06$ and $22.84\pm0.06$ mag) with larger offsets of $9.8\arcsec$ and $12.6\arcsec$, respectively. The probability of chance coincidence for these galaxies is 0.07, 0.36, and 0.65 for G1, G2, and G3. We therefore consider G1 to be the putative host galaxy for GRB 120305A. We note that G1 has a morphology suggestive of a late-type galaxy. The host-normalized offset is $R_o/R_e=4.6\pm1.2$ (see Table \ref{tab: host properties}). Furthermore, the $griz$ magnitudes hint at a 4000 \AA\, break around the $i$-band, suggesting a redshift $z$\,$\sim$\,$0.6-0.9$. 

We derive a lower limit, $n_\textrm{min}\gtrsim 2.0\times10^{-5}$ cm$^{-3}$, to the density of the GRBs environment using the early X-ray lightcurve. This is consistent with the expected density for an IGM-like environment, but does not rule out that the GRB occurred within a higher density galactic environment, such as G1.

\subsubsection{GRB 120630A}

On June 39, 2012 at 23:17:33 UT, GRB 120630A triggered \textit{Swift}/BAT \citep{Sakamoto120630A}. The burst is comprised of a single pulse with duration $T_{90}=0.58\pm0.18$. Observations with \textit{Swift}/XRT localized a rapidly fading X-ray source at RA, DEC = $23^{h} 29^m 11^{s}.07$, $\ang{+42;33;20.3}$ with uncertainty $4.0\arcsec$. This source was identified as the X-ray afterglow, and faded below \textit{Swift} detectability within the first orbit.




Gemini observations were carried out on July 1, 2012 at 0.5 d after the GRB to search for the optical afterglow of GRB 120630A.
No afterglow was detected within the XRT enhanced position to depth $r\gtrsim 25.0$ mag. However, in these Gemini images we identify seven nearby candidate host galaxies for GRB 120630A (see Figure \ref{fig: GalXray}). We therefore carried out follow-up imaging at late-times with the LDT/LMI on September 5, 2014 in $\textit{riz}$ and Keck/LRIS on October 25, 2014 in the \textit{GR} filters to better identify the putative host.

Within the XRT enhanced position we detect two extremely faint sources (Source A and B) which, due to their faintness, we cannot confirm are extended. Source A has magnitudes $G=25.7\pm0.2$, $R=25.4\pm0.3$, $i=24.7\pm0.3$, $z=24.8\pm0.3$, whereas Source B has $G=25.5\pm0.2$, $R=25.5\pm0.3$.
Due to the large XRT position error ($4.0\arcsec$), these sources have a significant probability ($P_{cc}$\,$\sim$\,$0.8$) of random alignment with the GRB localization. We therefore exclude these sources as candidate host galaxies. The $3\sigma$ upper limit to any other source within the XRT position is $G\gtrsim 25.7$, $R\gtrsim 25.6$, $i\gtrsim 24.9 $, and $z\gtrsim 24.9$ mag (corrected for Galactic extinction). 

The other five sources identified near the GRB position are detected with a high significance, and easily identified as extended galaxies. The brightest of these sources (G1) is located at an offset $\sim$\,$5.8\arcsec$ with magnitude $G=22.11\pm0.03$, $R=21.42\pm0.04$, $i=21.06\pm0.07$, $z=20.99\pm0.05$ mag. This galaxy is catalogued in both the PS1 ($riz$) and CatWISE2020 \citep{Marocco2020} catalogs. The WISE infrared magnitudes are $W1=19.48\pm0.04$ and $W2=19.61\pm0.08$ AB mag. 
G1 has a significantly lower probability of chance alignment with the XRT position, $P_{cc}=0.07$, compared to Sources A and B, especially in the redder filters. 
In comparison to this source the other candidate host galaxies (G2, G3, G4, G5) in Figure \ref{fig: GalXray}, which are much fainter ($r\lesssim 23$ mag), have a large $P_{cc}\gtrsim 0.4$. 
Therefore, we consider the bright galaxy G1 to be the putative host.

We modeled the broadband SED (covering optical wavelengths $GRizW1W2$) of G1 with \texttt{prospector}, see Figure \ref{fig: SED_fits}. We derive a photometric redshift $z_\textrm{phot}=0.6\pm0.1$ and a moderate stellar mass $\log(M_*/M_\odot)=9.8^{+0.2}_{-0.4}$. 
Adopting $z\sim 0.6$, we derive a physical offset of the GRB from G1 of $R=40\pm20$ kpc, and a host-normalized offset $R_o/R_e=6.4\pm3.2$. 

We derive a lower limit, $n_\textrm{min}\gtrsim 9.0\times10^{-5}$ cm$^{-3}$, to the density of the GRBs environment using the early X-ray lightcurve. This is consistent with the expected density for an IGM-like environment, but does not rule out that the GRB occurred within a higher density galactic environment, such as G1.


\subsubsection{GRB 130822A}

On August 22, 2013 at 15:54:17 UT, GRB 130822A triggered \textit{Swift}/BAT \citep{Kocevski130822A}. The burst displayed single pulse with duration $T_{90}=0.04\pm0.01$. 
XRT observations began at 85 s, and localized a fading X-ray source at RA, DEC = $01^{h} 51^m 41^{s}.27$, $\ang{-03;12;31.7}$ with uncertainty $3.3\arcsec$. 
No source was detected within the XRT enhanced position by optical follow-up observations. 

We obtained late-time imaging with Keck/LRIS in the \textit{G} and \textit{R}-bands on October 25, 2014.
The field of GRB 130822A is crowded with $>$\,$30$ sources within $20\arcsec$ in our Keck imaging. There are 8 SDSS galaxies ($r\sim20.7-21.7$ mag) within $60\arcsec$, one of which is significantly brighter than the rest with $R=18.13\pm0.02$ mag. We label this bright galaxy at offset $22\arcsec$ as G7. G7 has $P_{cc}=0.08$ and redshift $z=0.154$ \citep{Wiersema130822A}. An even brighter SDSS galaxy (referred to as G12) at $z=0.045$ \citep{Wiersema130822A} resides at $84\arcsec$ offset from the GRB position with $R=16.204\pm0.005$ ($P_{cc}=0.23$). In addition to these galaxies, there are a number of $r\gtrsim24$ mag galaxies at offsets $\gtrsim10\arcsec$, with $P_{cc}\gtrsim0.8$. We also identify 4 faint sources, $R\gtrsim25$ mag, within $5\arcsec$ of the XRT position (one of which resides inside the 90\% localization region; Figure \ref{fig: GalXray}). These sources have $P_{cc}=0.25-0.5$. The $3\sigma$ upper limit within the XRT position is $R\gtrsim 25.8$ AB mag.

Due to its lower probability of chance alignment, we consider G7 as the putative GRB host. We note that the morphology of G7 is a face-on late-type galaxy. The projected offset from the GRB position is $22.0\pm2.3\arcsec$, which at $z=0.154$ corresponds to $61\pm6$ kpc. The host-normalized offset is $R_e/R_o=8.1\pm0.9$. Thus, GRB 130822A represents the largest offset of a sGRB from a late-type galaxy (Figure \ref{fig: offset_vs_type}).

Based on the early X-ray afterglow lightcurve, we set a lower limit to the density of the environment surrounding the GRB of $\gtrsim 7.1\times 10^{-4}$ cm$^{-3}$. This value is consistent with the GRB occurring in an ISM-like environment. However, we caution that for this GRB re-binning the XRT lightcurve yields two data points with a very steep decay index, hinting that the observed X-ray emission may not be due to the forward shock. In such a case the formalism to constrain the density is not applicable.
%
%

\subsubsection{GRB 140516A}

At 20:30:54 UT on May 16, 2014, \textit{Swift}/BAT triggered on GRB 140516A. The burst had a duration $T_{90}=0.19\pm0.09$ s.
XRT localized the afterglow to RA, DEC = $16^{h} 51^m 57^{s}.40$, $\ang{+39;57;46.3}$ with $2.7\arcsec$ uncertainty (90\% CL). No optical afterglow was discovered for this event.

We obtained late-time imaging of GRB 140516A with the LDT in $r$-band. This was supplemented with archival Gemini and Keck imaging in $i$ and $K_s$, respectively.
The field surrounding the GRB position is sparse, with the exception of a bright foreground star slightly overlapping the XRT position. However, we uncover an extremely faint candidate host galaxy (referred to as Source A) at the edge of the XRT position that is detected in both the Gemini and Keck imaging. Source A has magnitudes $r\gtrsim 25.0$, $i=25.9\pm0.3$, and $K_s= 23.15\pm0.20$ AB mag, suggestive of a high-$z$ origin.
The probability of chance coincidence is $0.6$ based in the $i$-band magnitude and $0.2$ based on the $K_s$-band.
No other source is uncovered in the XRT position to depth $i\gtrsim 26.1$ AB mag, and there are no other nearby candidate galaxies. We note the presence of a bright $r\sim17.5$ mag galaxy at an offset of $80\arcsec$, however, the $P_{cc}>0.3$.
We, therefore, consider GRB 140516A to be observationally hostless.

Based on the early X-ray afterglow lightcurve, we set a lower limit to the density of the environment surrounding the GRB of $\gtrsim 7.3\times 10^{-4}$ cm$^{-3}$. This value is consistent with the GRB occurring in an ISM-like environment.

\subsubsection{GRB 140622A}


GRB 140622A triggered \textit{Swift}/BAT on June 22, 2014 at 09:36:04 UT \citep{DElia140622A}. The burst had duration $T_{90}=0.13\pm0.04$ s. The X-ray afterglow was localized to RA, DEC (J2000) = $21^{h} 08^m 41^{s}.53$, $\ang{-14;25;09.5}$ with accuracy  \ang{;;2.9} (90\% CL). No optical afterglow was uncovered for this event.

We performed late-time observations with the LDT/LMI on August 6, 2021 in the $griz$ filters. We identify a nearby galaxy (Figure \ref{fig: GalXray}) uncovered at offset $4.6\arcsec$ with magnitudes $g=22.53\pm0.07$, $r=22.28\pm0.07$, $i=21.84\pm0.06$, and $z=21.92\pm0.20$.
The probability of chance coincidence for G1 is $P_{cc}=0.08$ using the $r$-band magnitude. Another galaxy, G2, is detected at an offset of $7.7\arcsec$ with $r=22.66\pm0.07$ yielding $P_{cc}=0.29$. In addition, no source is detected within the XRT position to depth $r\gtrsim 24.1$ mag (a previous limit of $r\gtrsim 25.8$ mag was reported by \citealt{Pandey2019} using GTC). We note that any source fainter ($r\gtrsim 24.1$ mag) than this residing with the XRT error circle would have $P_{cc}\gtrsim 0.25$. These arguments lead us to classify G1 as the putative host of GRB 140622A. 


In order to derive the redshift of this galaxy, we carried out optical spectroscopy with Keck/LRIS on October 27, 2014 (see Table \ref{tab: SpecObs}). The spectrum is displayed in Figure \ref{fig: spectra}. We identified emission lines at $\lambda_\textrm{obs}\approx 7304$ and $9810$ \AA which we associate to the [OII] doublet and [OIII]$_{5008}$, respectively. This yields a redshift $z=0.959\pm 0.001$, which is consistent with that reported by \citet{Hartoog140622A}. At this redshift there is a very marginal detection of both $H\beta$ and [OIII]$_{4960}$. 
In our LDT imaging we cannot classify the galaxy type based on morphology, but the emission features are suggestive of a late-type galaxy. At this redshift the offset of the galaxy from the GRB position is $38\pm17$ kpc, towards the high end of the short GRB offset distribution. The host-normalized offset is $R_o/R_e=3.8\pm1.7$.

We derive a lower limit, $n_\textrm{min}\gtrsim 1.8\times10^{-5}$ cm$^{-3}$, to the density of the GRBs environment using the early X-ray lightcurve. This is consistent with the GRB occurring at an offset of $\sim$\,$38$ kpc from G1, and does not exclude the association.

\subsubsection{GRB 150831A}

GRB 150831A triggered \textit{Swift}/BAT on August 31, 2015 at 10:34:12 UT \citep{Lien150831A}. The GRB was also detected with the \textit{Integral} \citep{Mereghetti150831A} and Konus-\textit{Wind} \citep{Golenetskii150831A} satellites. 
The burst had duration $T_{90}=1.15\pm0.22$ s as observed by BAT. The X-ray afterglow was localized to RA, DEC (J2000) = $14^{h} 44^m 05^{s}.84$, $\ang{-25;38;06.4}$ with accuracy  \ang{;;2.2} (90\% CL). No optical counterpart was uncovered for this event.


We analyzed public archival imaging obtained with Gemini/GMOS-S on July 29, 2020 in \textit{i}-band, and from VLT/FORS2 in $R$-band and $I$-band from September 1, 2016 and March 7, 2017, respectively. We identify a galaxy within the XRT enhanced position with magnitude $R=24.95\pm0.10$ and $i=25.1 \pm 0.3$ mag. Due to its faintness, this source has a $\sim$\,32\% probability of chance alignment with the XRT position. There are no other sources detected within the XRT position to depth $R\gtrsim 25.6$ and $i\gtrsim 25.6$ mag. We identify two other galaxies within $15\arcsec$ of the GRB localization (Figure \ref{fig: GalXray}): G2 has magnitude $i=23.45\pm0.09$ at offset $10.9\arcsec$, and G3 with $i=22.14\pm0.05$ at $12.1\arcsec$ 
These sources have $P_{cc}=0.5$ and $0.25$ for G2 and G3, respectively. There are no other bright galaxies within $60\arcsec$ of the GRB localization. Consequently, there is no putative host galaxy for GRB 150831A, and we consider the GRB to be observationally hostless.

Using the early X-ray afterglow lightcurve, we set a lower limit of $n_\textrm{min}\gtrsim 2.4\times 10^{-5}$ cm$^{-3}$ to the circumburst environment of GRB 150831A. This density is consistent with that expected for an IGM-like environment, but does not exclude a higher density.

\subsubsection{GRB 151229A}

GRB 151229A triggered \textit{Swift}/BAT \citep{Kocevski151229A} and \textit{Fermi}/GBM \citep{vonKienlin151229A} on December 29, 2015 at 06:50:27 UT. The duration of the GRB is $T_{90}=1.44\pm0.45$ s and $3.5\pm1.0$ s as seen by BAT and GBM, respectively. 
\textit{Swift}/XRT discovered fading X-ray source was discovered at RA, DEC (J2000) = $21^{h} 57^m 28^{s}.78$, $\ang{-20;43;55.2}$ with accuracy  \ang{;;1.4} (90\% CL). No optical counterpart was discovered.  


We carried out late-time imaging of GRB 151229A with the LDT/LMI in the $r$ and $i$-bands, Gemini/GMOS-N in $r$-band, Gemini/GMOS-S in $i$-band, and Gemini/Flamingos-2 (hereafter F2) in the $J$ and $K_s$-bands. We supplemented these observations with archival $z$-band imaging with Gemini/GMOS-S (PI: Fong) and $Y$-band imaging with Keck/MOSFIRE (PI: Terreran). In these observations we uncover an extended source (G1) coincident with the XRT enhanced position. We derive magnitudes $r=25.75\pm0.2$, $i=25.41\pm0.15$, $z=24.47\pm0.10$, $Y=24.0\pm 0.2$, $J=23.10\pm0.18$, and $K_s=22.78\pm0.2$ AB mag. We note that the probability of chance coincidence (using the $r$-band magnitude) for this galaxy is large, $P_{cc}=0.25$. However, the probability of chance coincidence for G1 based on the redder $z$ and $Y$ magnitudes is significantly lower with $P_{cc}\approx$\,$0.1$\,$-$\,$0.15$. Moreover, the field of GRB 151229A is sparse, and no other candidate hosts were identified to depth $r\gtrsim 26.1$ mag. Therefore, we consider G1 as the putative host galaxy of GRB 151229A.

We analyzed archival Keck/LRIS spectroscopy of this galaxy (see Table \ref{tab: SpecObs}), but did not identify a trace or any emission lines. Instead, we modeled the broadband SED ($rizYJK_s$) of G1 within \texttt{prospector} in order to derive a photometric redshift. We found that in order for the code to achieve a good fit to the SED, we had to turn off nebular emission lines within \texttt{prospector}.
Finally, we obtain $z_\textrm{phot}=1.4\pm 0.2$ and a stellar mass $\log(M_*/M_\odot)=10.3\pm 0.2$ (Figure \ref{fig: SED_fits}).
At this redshift, the physical offset of the GRB is $9
\pm 9$ kpc.
We further derive a host-normalized offset of $R_o/R_e=2.5\pm2.5$.



Adopting $z\approx1.4$, as suggested by the galaxy's SED, we set a lower limit to the density of the GRBs environment $n_\textrm{min}\gtrsim 1.2\times10^{-1}$ cm$^{-3}$. These limits suggest the GRB occurred within a high density galactic environment, and support the association with G1.


\subsubsection{GRB 170127B}

\textit{Swift}/BAT triggered and localized GRB 170127B on January 27, 2017 at 15:13:28 UT \citep{Cannizzo170127B}. The burst was also detected with \textit{Fermi}/GBM \citep{Veres170127B}. As seen by BAT, the burst was single pulsed with duration $T_{90}=0.51\pm0.14$ s.
\textit{Swift}/XRT discovered the X-ray afterglow at RA, DEC (J2000) = $01^{h} 19^m 54^{s}.47$, $\ang{-30;21;28.6}$ with accuracy  \ang{;;2.6} (90\% CL). No optical counterpart was uncovered for this GRB.


We obtained late-time imaging of GRB 170127B on January 30, 2021 from Gemini South in $z$-band (PI: Troja). We also include in our analysis public archival Gemini South observations in $g$-band (PI: Fong) as well as public archival Keck imaging (LRIS/MOSFIRE; PIs: Miller, Terreran) in the $G$, $R$, $I$, and $J$ filters. The field is very sparse, with no bright candidate host galaxies. Nevertheless, in the Keck imaging we identify a faint, extended source (G1 in Figure \ref{fig: GalXray}) within the XRT enhanced position, which is not detected in the Gemini images. This source has magnitudes $G=25.7\pm0.2$, $R=25.5\pm0.2$, $I=25.5\pm0.2$, $z\gtrsim23.9$ and $J\gtrsim24.1$ AB mag. 
The probability of chance coincidence using $r$-band number counts is $P_{cc}=0.55$.
No other source is identified within the XRT position to a $3\sigma$ upper limit $R\gtrsim 26.0$ mag. 
We note there are also two faint ($R\sim 24.5-25.0$ mag) galaxies (G2 and G3), which we refer to as G2 and G3, at offsets $\sim$\,$6\arcsec$ and $9\arcsec$ with a similarly large chance probability $P_{cc}=0.54$ and $0.82$, respectively. Due to these high probabilities, we find that GRB 170127B is observationally hostless.

Using the early X-ray afterglow lightcurve from \textit{Swift}/XRT, we set a lower limit to the density of the GRB's environment of $n_\textrm{min}\gtrsim 7.3\times 10^{-4} $ cm$^{-2}$. This density implies that the GRB originated within a galactic environment.

\subsubsection{GRB 171007A}

At 11:57:38 UT on October 7, 2017, \textit{Swift}/BAT \citep{Cannizzo171007A} and \textit{Fermi}/GBM \citep{Bisaldi171007A} triggered and located GRB 171007A. The burst displayed a single pulse with duration $\sim 3 s$ followed by weaker, softer emission which is characterized as EE. The total duration of the GRB is $T_{90}=105\pm45$ s. In this work, we classify GRB 171007A as a candidate sGRBEE. XRT observations localized an uncatalogued, fading X-ray source to RA, DEC (J2000) = $09^{h} 02^m 24^{s}.14$, $\ang{+42;49;08.8}$ with uncertainty  \ang{;;2.5} (90\% CL) which was identified as the afterglow. No optical or infrared counterpart was identified.



We obtained late-time imaging with LDT on January 9, 2020 in $r$-band and the Gemini North telescope on February 1, 2021 in $i$-band. We uncovered two extremely faint sources in our Gemini imaging at the edge of the XRT enhanced position, see Figure \ref{fig: GalXray}. Due to their faintness we cannot determine whether these sources are extended. The first source, referred to as Source A, has magnitude $i=25.1\pm0.2$ and the second source (Source B) has magnitude $i=26\pm0.4$. Source A is also detected in our LDT imaging with $r=24.8\pm0.3$, whereas Source B is not detected to depth $r\gtrsim 24.9$ mag. The probability of chance coincidence for either source is quite large, $P_{cc}\gtrsim 0.5$.
Therefore, due to the large XRT localization we cannot confidently associate either source to the GRB. No other sources are detected to $i\gtrsim 26.1$ mag within the XRT localization. In addition, there are no other sources with lower $P_{cc}$ outside of the XRT error circle, leading to an observationally hostless classification as it is not clear if either of these sources is the host. We note that any fainter sources identified in deeper imaging would similarly be difficult to confirm a physical association to GRB 171007A due to the high $P_{cc}$.

Using the early X-ray lightcurve, we derive a lower limit to the circumburst density of $\gtrsim 2.0\times10^{-6}$ cm$^{-3}$. We note that this lower limit is not very constraining to the density due to the plateau and early steep decline phase of the X-ray lightcurve, leading us to apply a late time X-ray data point in order to compute the lower limit.

\subsubsection{GRB 180727A}

On July 27, 2018 at 14:15:28 UT, GRB 180727A was detected with \textit{Swift}/BAT \citep{Beardmore180727A} and \textit{Fermi}/GBM \citep{Veres180727A}. 
The duration of the GRB as observed by BAT is $T_{90}=1.05\pm0.22$ s.
XRT observations localized the afterglow to RA, DEC (J2000) = $23^{h} 06^m 39^{s}.86$, $\ang{-63;03;06.7}$ with uncertainty  \ang{;;2.3} (90\% CL) which was identified as the afterglow. No optical counterpart was detected.

We analyzed public archival observations obtained with Gemini/GMOS-S in the $griz$ filters.
We identify an extremely faint source (Source A) within the XRT error circle with magnitudes $g=26.1\pm0.3$, $r=25.9\pm0.3$, $i=25.5\pm0.3$, and $z=25.5\pm0.3$ mag. The probability of chance coincidence for this source is $\sim$\,0.6. 
The upper limit to other sources in the XRT position is $r>26.1$.
We detect three other sources within $10\arcsec$ of the XRT position (Figure \ref{fig: GalXray}).
These sources have $P_{cc}>0.3$, and all other galaxies in the field 
have $P_{cc}>0.5$. We, therefore, consider GRB 180727A to be observationally hostless.

We derive a density $n_\textrm{min}\gtrsim 3.0\times 10^{-5}$ cm$^{-3}$ for the GRB environment. This value is consistent with the GRB occurring in an IGM-like environment \citep[i.e., $n<10^{-4}$ cm$^{-3}$;][]{OConnor2020}.

\subsubsection{GRB 180805B}

At 13:02:36 UT on August 5, 2018, \textit{Swift}/BAT \citep{DAvanzo180805B} and \textit{Fermi}/GBM \citep{Hamburg180805B} triggered on GRB 180805B. The burst displayed an initial short pulse with duration $<1$ s followed by a softer, weak emission for over a hundred seconds. The total duration of the burst detected with BAT is $T_{90}=122\pm18$ s. This lightcurve displays characteristics common to other sGRBEE, and we therefore classify GRB 180805B as an sGRBEE. The X-ray afterglow for this event was localized to RA, DEC (J2000) = $01^{h} 43^m 07^{s}.59$, $\ang{-17;29;36.4}$ with uncertainty \ang{;;2.1}. There was no optical counterpart discovered for this event.


We obtained late-time imaging of the field of GRB 180805B with the LDT/LMI on January 16, 2021 in $z$-band. We supplemented this with archival Keck imaging obtained with LRIS on September 10, 2018 and September 4, 2019 in $G$, $V$, $I$, and $Z$ and with MOSFIRE in $K_s$ from October 15, 2019. We uncover four galaxies nearby to the GRB's XRT position, but no source is identified within the XRT localization to depth $G\gtrsim 26.0$, $V\gtrsim 25.6$, $I\gtrsim 25.4$, $Z\gtrsim 24.4$, $K_s\gtrsim 24.1$ AB mag ($3\sigma$; corrected for Galactic extinction). These four galaxies surround the GRB localization on all sides, with offsets ranging from $2.8$ to $4.2\arcsec$ for G1 and G4, respectively. The brightest galaxy, G3, is located North of the GRB position with magnitudes $G=23.46\pm0.07$, $V=22.79\pm0.09$, $I=22.31\pm0.12$, $Z=21.99\pm0.14$, and $K_s = 21.22\pm 0.15$ AB mag. G3 is offset by $3.4\pm1.0\arcsec$ from the XRT position, yielding a probability of chance alignment of $P_{cc}=0.07$.
The other galaxies have magnitudes $V=24.6\pm 0.2$, $25.2 \pm 0.2$, and $24.5 \pm 0.2$ yielding $P_{cc}=0.19$, $0.36$, and $0.33$ for G1, G2, and G4, respectively. In addition to these, we note that there is a bright SDSS galaxy ($r$\,$\sim$\,$15.5$ mag with $z_\textrm{phot}=0.029\pm0.006$) at an offset of $\sim$\,$90\arcsec$ with $P_{cc}=0.15$.
Based on these probabilistic arguments we consider G3 to be the putative host galaxy for GRB 180805B.

We analyzed optical spectroscopy of G1 taken with Keck/LRIS on September 10, 2018 in order to identify the redshift of the galaxy. The spectrum is shown in Figure \ref{fig: spectra}. We identified emission lines at $\lambda_\textrm{obs}\approx 6190$, $7210$, $8076$, $8238$, and $8318$ \AA which we associate to the [OII] doublet, H$\gamma$, H$\beta$, [OIII]$_{4960}$, and [OIII]$_{5008}$, respectively. This yields a redshift $z=0.6609\pm 0.0004$. 
In the photometry of G1 we observe the 4000 \AA break at this redshift.

Based on the early X-ray afterglow lightcurve, we derive a density of $n_\textrm{min}\gtrsim 3\times 10^{-6}$ cm$^{-3}$ for the environment surrounding GRB 180805B. This is consistent with the projected physical offset, $R=25\pm11$ kpc, of G3 from the GRB position. The host-normalized offset is $R_o/R_e=5.6\pm2.4$.

\subsubsection{GRB 191031D}

On October 31, 2019 at 21:23:31 UT, GRB 191031D triggered \textit{Swift}/BAT \citep{DElia191031D}, \textit{Fermi}/GBM \citep{Maliyan191031D2}, Konus-\textit{Wind} \citep{Frederiks191031D}, \textit{AstroSat} \citep{Gaikwad191031D}, AGILE/MCAL \citep{Ursi191031D}, \textit{INTEGRAL}/SPI-ACS \citep{DElia191031D}, and the \textit{CALET} Gamma-ray Burst Monitor \citep[CGBM][]{Shimizu191031D}. The burst was multi-peaked with a duration $T_{90}=0.28\pm0.05$ s. 
\textit{Swift}/XRT identified the X-ray afterglow at RA, DEC (J2000) = $18^{h} 53^m 09^{s}.57$, $\ang{+47;38;38.8}$ with accuracy  \ang{;;2.3} (90\% CL).





We observed the field of GRB 191031D on November 2, 2019 at 1.3 d after the GRB to search for the optical afterglow. No optical source was detected within the XRT position to depth $r \gtrsim 25.0$ mag \citep{Dichiara191031D}. However, we identified two candidate host galaxies for GRB 191031D, see Figure \ref{fig: GalXray}. In order to better characterize the galaxy SEDs we carried out additional LDT observations in the $gizy$ filters.

The first source, referred to as Source A, is offset by $3.9\arcsec$ from the GRB position and has magnitude $r=24.49\pm0.15$ mag. We cannot determine whether or not the source is extended, and this source is not detected in our LDT $izy$ imaging.
The second source (G1) is a clear galaxy with magnitudes $g=22.47\pm0.07$, $r=21.64\pm0.05$, $i=21.2\pm0.2$, $z=21.2\pm0.3$, and $y=21.0\pm0.3$. 
This galaxy is offset by $7.4\arcsec$ from the GRB position. Using $r$-band number counts we derive $P_{cc}=0.12$ and $0.3$ for G1 and Source A respectively. G1 is also detected in PS1 with smaller errors on the $i$ and $z$-band (as at the time of our LDT observations the conditions were extremely poor). We make use of the PS1 magnitudes in our SED modeling (see below). We further note that G1 is also observed in the ALLWISE catalog \citep{Cutri2021} with magnitudes $W1=19.60\pm0.15$ and $W2=20.16\pm0.30$ AB mag. 
These magnitudes suggest that the 4000 \AA\, break lies above the $r$-band. Therefore, if instead we compute the probability in the redder $i$ and $z$ filters, where the magnitude is significantly brighter, we find $P_{cc}=0.05-0.08$. Based on these arguments, we identify G1 as the putative host galaxy of GRB 191031D.

On November 3, 2019, we carried out optical spectroscopy (Table \ref{tab: SpecObs}) of G1 with Gemini GMOS-N. A trace is visible from $\sim$\,$6400$ to $9500$ \AA, although there are no obvious absorption or emission features. Therefore, we instead modelled the broadband SED ($grizyW1W2$) within \texttt{prospector}. As the spectrum does not show bright emission features, we turned off nebular emission lines within \texttt{prospector}. We derive a photometric redshift of $z_\textrm{phot}=0.5\pm0.2$ and a stellar mass $\log(M_*/M_\odot)=10.2\pm0.2$ (see Figure \ref{fig: SED_fits}).

At redshift $z$\,$\approx$\,$0.5$, we set a lower limit to the circumburst density of the GRB $n_\textrm{min}=7.9\times 10^{-4}$ (see Table \ref{tab: XrayAGprop}) using the X-ray lightcurve.

\subsubsection{GRB 200411A}


GRB 200411A triggered \textit{Swift}/BAT \citep{Tohuvavohu200411A} and \textit{Fermi}/GBM \citep{Fermi200411A} on April 11, 2020 at 04:29:02 UT. The burst was double peaked with duration $T_{90}=0.33\pm0.10$ s, as seen by BAT. 
The X-ray afterglow was detected with at RA, DEC (J2000) = $3^{h} 10^m 39^{s}.39$, $\ang{-52;19;03.4}$ with accuracy  \ang{;;1.4} (90\% CL). No optical or infrared counterpart was detected. 



 We performed late-time imaging with the Gemini/GMOS-S telescope on January 25, 2021 in the \textit{r}-band. We identified two potential host galaxies near to the XRT position (see Figure \ref{fig: GalXray}). The first source (Source A) lies within the XRT enhanced position, and has magnitude $r=25.5\pm0.3$. Due to its faint nature we cannot conclude whether the source is extended. The upper limit to any additional source within the XRT enhanced position is $r\gtrsim 25.8$ mag. The second source (G1) is located at an offset of $4.5\arcsec$ and displays a morphology suggestive of a late-type galaxy. In our Gemini imaging we derive a magnitude $r=22.52\pm0.03$ mag. Based on their $r$-band magnitudes, the probability of chance coincidence for these sources is $P_{cc}=0.21$ and $0.11$ for Source A and G1, respectively. However, G1 is also visible in the DES, Vista Hemisphere Survey \citep[VHS;][]{McMahon2013}, and ALLWISE \citep{Cutri2021} catalogs with AB magnitudes: $g=23.6\pm0.2$, $r=22.6\pm0.1$, $i=21.9\pm0.1$, $z=21.3\pm0.1$, $J=20.9\pm0.2$, $W1=20.0\pm0.1$, and $W2=20.2\pm0.3$ mag. The probability of chance coincidence for G1 is significantly smaller in these redder filters with $P_{cc}=0.08$ using $z$-band number counts \citep{Capak2004}.
 Based on these probabilistic arguments and the lack of other candidates, we consider G1 to be the putative host galaxy of GRB  200411A.
  
 Additionally, we utilized the broadband SED (Figure \ref{fig: SED_fits}) from these archival observations to derive a photometric redshift $z_\textrm{phot}=0.6\pm0.1$ and a moderate stellar mass $\log(M_*/M_\odot)=10.4\pm0.1$ using the \texttt{prospector} software. At this redshift, we derive a lower limit $n_\textrm{min}\gtrsim 2.3\times 10^{-4}$ cm$^{-3}$ to the circumburst density using the early X-ray lightcurve. Adopting $z\sim 0.6$, the physical offset of G1 from the GRB position is $R=31\pm8$ kpc and the host-normalized distance is $R_o/R_e=3.9\pm0.9$. The gas density at this distance is $\rho_g\sim7\times10^{-4}$ cm$^{-3}$, assuming the density profile outlined in \citet{OConnor2020}, which is consistent with the lower limit implied by the early X-ray afterglow.
 

\section{Derivation of the Circumburst Density}
\label{appendix: densitycalc}

Following \citet{OConnor2020}, we compute a lower limit to the circumburst density using constraints on the deceleration time of the GRB jet based on early \textit{Swift}/XRT follow-up. 
The parameters required to compute the circumburst density $n_\textrm{min}$, namely an upper limit to the time of deceleration of the GRB's jet $t_o$ and a lower limit to the peak X-ray flux $F_{X,o}$ are tabulated in Table \ref{tab: XrayAGprop}. In order to calculate the density we adopt the fiducial parameters: the fraction of the burst kinetic energy residing in electrons $\varepsilon_e=0.1$ and magnetic fields $\varepsilon_B=10^{-2}$, a bulk Lorentz factor $\Gamma=300$, and a gamma-ray efficiency $\varepsilon_\gamma=0.15$. The lower limit on circumburst density is then derived using Equation 17 of  \citet{OConnor2020}. We record this value for each GRB in Table \ref{tab: XrayAGprop}.
Due to the different selection criteria in \citet{OConnor2020} (i.e., requiring $T_{90}$\,$<$\,$0.8$ s), 17 events in our sample were not included in their work (i.e., those with extended emission or $0.8$\,$<$\,$T_{90}$\,$<$\,$2$ s). 

We remind the reader that in order for a sGRB to be considered physically hostless (or consistent with the scenario) the density must be $<$\,$10^{-4}$ cm$^{-3}$ \citep{OConnor2020}. In the case of these lower limits, if $n_\textrm{min}$\,$>$\,$10^{-4}$ cm$^{-3}$ then the sGRB is inconsistent with being physically hostless, whereas a smaller value of $n_\textrm{min}$ only implies that the sGRB could be physically hostless and is not conclusive one way or another.


\begin{table*}
 	\centering
 	\caption{Gamma-ray and X-ray properties of sGRBs in our sample. The parameters $t_o$, $F_{X,o}$, and $n_\textrm{min}$ are defined as in \citet{OConnor2020}. 
 	}
 	\label{tab: XrayAGprop}
 	\begin{tabular}{lccccccc}
   \hline
   \hline
   \\[-2.5mm]
      & \multicolumn{4}{c}{\textbf{Prompt gamma-ray properties}} & \multicolumn{2}{c}{\textbf{X-ray afterglow properties}} & \\
      \cmidrule(lr){2-5} \cmidrule(lr){6-7}
    \multirow{2}{*}{\textbf{GRB}} & \textbf{$T_{90}$} & \textbf{$\phi_\gamma$} & \textbf{Hardness ratio}$^c$ & \textbf{Photon index $\Gamma$} & \textbf{$t_o$} & \textbf{$F_{X,o}$} & \textbf{$n_\textrm{min}$} \\[0.9mm]
   & \textbf{(s)} & \textbf{($10^{-7}$ erg cm$^{-2}$)} & & &  \textbf{(s)}  & \textbf{($10^{-11}$ erg cm$^{-2}$ s$^{-1}$)} & \textbf{(cm$^{-3}$)} \\[0.9mm]
    \hline
\multicolumn{8}{c}{\textbf{Short GRBs with $T_{90}$\,$<$\,$2$ s}} \\
 \hline
 091109B & 0.3 & $1.9\pm1.5$ & $2.4\pm0.3$ & $0.7\pm0.1$ & $160^{+120}_{-70}$ & $1.0\pm0.2$ & $1.7\times 10^{-5}$ \\[0.9mm]
 101224A & 0.2 & $0.6\pm0.1$ & $1.9\pm0.5$ & $1.1\pm0.3$  & $630^{+780}_{-540}$ & $0.02\pm0.01$ & $3.6\times 10^{-5}$ \\[0.9mm]
 110112A & 0.5 & $0.3\pm0.1$ & $0.9\pm0.3$ & $2.1\pm0.5$ &$190\pm100$  & $1.4\pm0.4$ & $1.4\times 10^{-3}$ \\[0.9mm]
120305A  & 0.1 & $2.0\pm0.1$ & $2.0\pm0.2$ & $1.0\pm0.09$  &$350^{+250}_{-90}$   & $1.1\pm0.2$ & $2.0\times 10^{-5}$ \\[0.9mm]
 120630A & 0.6 & $0.6\pm0.1$ & $2.0\pm0.6$ & $1.0\pm0.4$ &$175_{-80}^{+120}$ & $0.48\pm0.11$ & $9.0\times 10^{-5}$\\[0.9mm]
 130822A & 0.04 & $0.12\pm0.03$ & $1.3\pm0.3$ & $1.7\pm0.3$ & $140^{+1500}_{-60}$  & $0.06\pm0.01$ & $7.1\times10^{-4}$ \\[0.9mm]
 130912A & 0.3 & $1.7\pm0.2$ & $1.7\pm0.3$ & $1.2\pm0.2$  &$160\pm20$ & $24\pm5$ & $2.1\times 10^{-3}$ \\[0.9mm]
 131004A &1.5 & $2.8\pm0.2$ & $1.1\pm0.1$ & $1.8\pm0.1$  &$120^{+40}_{-50}$ & $3.8\pm0.8$ & $1.5\times10^{-3}$ \\[0.9mm]
140129B  &1.35 & $0.7\pm0.1$ & $0.9\pm0.2$ & $2.2\pm0.3$  &$330\pm10$   & $5.6\pm1.3$ & $1.0\times 10^{-3}$ \\[0.9mm]
 140516A & 0.2& $0.30\pm0.07$ & $1.1\pm0.3$ & $1.9\pm0.3$ &$200^{+2400}_{-110}$  & $0.06\pm0.02$ & $7.3\times 10^{-4}$ \\[0.9mm]
140622A  & 0.13 & $0.13\pm0.04$ & $0.5\pm0.2$ & $3.1\pm0.3$ &$300^{+1100}_{-100}$ & $0.04\pm0.01$ & $1.8\times 10^{-5}$ \\[0.9mm]
 140930B & 0.8 & $4.2\pm0.4$ & $2.6\pm0.4$ & $0.6\pm0.2$  &$187\pm4$  & $40\pm5$ & $1.4\times 10^{-3}$ \\[0.9mm]
 150423A & 0.08 & $0.7\pm0.1$ & $2.3\pm0.5$ & $0.8\pm0.2$  &$110\pm30$  & $1.6\pm0.4$ & $2.6\times 10^{-4}$ \\[0.9mm]
 150831A & 1.15 & $3.6\pm0.3$ & $2.4\pm0.3$ & $0.7\pm0.2$ &$240^{+90}_{-40}$ & $1.5\pm0.4$ & $2.4\times 10^{-5}$ \\[0.9mm]
 151229A &1.4 & $5.9\pm0.4$ & $1.1\pm0.1$ & $1.8\pm0.1$ &$76\pm4$  & $33\pm4$ & $1.2\times 10^{-1}$ \\[0.9mm]
 160408A & 0.3& $1.6\pm0.2$ & $2.2\pm0.4$ & $0.9\pm0.2$ &$300\pm20$ & $3.2\pm0.7$ & $1.8\times 10^{-4}$ \\[0.9mm]
 160525B & 0.3 & $0.3\pm0.1$ & $1.1\pm0.3$ & $1.9\pm0.4$ &$73\pm3$   & $10.5\pm3.0$ & $6.6\times 10^{-3}$ \\[0.9mm]
 160601A & 0.12 & $0.7\pm0.1$ & $2.0\pm0.4$&$1.0\pm0.2$ &$270^{+420}_{-190}$  & $0.10\pm0.02$ & $1.2\times10^{-5}$ \\[0.9mm]
160927A  & 0.48 & $1.4\pm0.2$ & $1.8\pm0.4$ & $1.1\pm0.3$ &$130^{+30}_{-50}$  &$1.9\pm0.4$ & $1.1\times10^{-4}$ \\[0.9mm]
 170127B &  0.5 & $1.0\pm0.2$ & $2.2\pm0.4$ & $0.9\pm0.3$ &$130^{+50}_{-30}$  & $6.0\pm1.0$ & $7.3\times 10^{-4}$ \\[0.9mm]
  170428A &  0.2& $2.8\pm0.2$ & $2.4\pm0.3$ & $0.8\pm0.1$  &$800^{+150}_{-100}$ & $0.25\pm0.06$ & $1.6\times 10^{-5} $\\[0.9mm]
 170728A & 1.3 & $0.8\pm0.2$ & $1.0\pm0.3$ & $2.0\pm0.3$ &$250^{+70}_{-50}$ & $1.0\pm0.2$ &  $1.2\times 10^{-4}$ \\[0.9mm]
 180727A & 1.1 & $2.9\pm0.2$ & $1.3\pm0.1$ & $1.6\pm0.1$  &$230^{+70}_{-50}$  & $1.5\pm0.3$ & $3.0\times 10^{-5}$ \\[0.9mm]
 191031D & 0.3 & $4.1\pm0.4$ & $2.3\pm0.3$ & $0.8\pm0.2$  &$120\pm20$ & $2.9\pm0.7$ & $7.9\times10^{-4}$ \\[0.9mm] 
 200411A & 0.3 & $0.9\pm0.1$ & $2.1\pm0.3$ &  $1.0\pm0.2$ &$280^{+90}_{-80}$ & $1.0\pm0.2$ & $2.3\times10^{-4}$ \\[0.9mm]
 \hline 
 \multicolumn{8}{c}{\textbf{Short GRBs with extended emission}} \\
 \hline
 110402A$^a$  & $56\pm5$ & $32\pm3$ & $1.4\pm0.2$  & $1.6\pm0.2$ &$590\pm40$  & $1.0\pm0.2$ & $4.0\times10^{-4}$ \\[0.9mm]
  160410A$^a$ & $96\pm50$ & $12\pm2$ & $1.9\pm0.4$ & $1.1\pm0.3$ &$360\pm8$  & $8.0\pm2.0$ & $2.6\times10^{-3}$ \\[0.9mm]
   170728B$^a$ & $48\pm30$ & $17\pm4$ & $1.1\pm0.3$ & $1.9\pm0.3$ &$460\pm2$  & $37\pm7$ & $7.5\times 10^{-4}$ \\[0.9mm]
  171007A$^a$ & $68\pm20$  & $2.6\pm0.9$ & $1.6\pm0.6$ & $1.4\pm0.5$ &$370^{+1500}_{-130}$  & $0.17\pm0.03$ & $2.0\times10^{-6}$ \\[0.9mm]
 180618A$^a$ & $47\pm11$ & $6.8\pm1.0$ & $1.5\pm0.3$ & $1.4\pm0.3$  &$69.1\pm0.6$  & $130\pm20$ & $4.0\times10^{-3}$ \\[0.9mm]
  180805B$^a$ & $122\pm18$ & $8.6\pm1.6$ & $1.9\pm0.4$ & $1.1\pm0.3$ & $1300^{+90}_{-130}$  & $0.25\pm0.07$ & $3.0\times 10^{-6}$ \\[0.9mm]
     \hline
    \end{tabular}
\begin{flushleft}
    \quad \footnotesize{$^a$ Short GRB with extended emission.} \\
    \quad \footnotesize{$^b$ The early X-ray lightcurve of this GRB does not fit the criteria outlined by \citet{OConnor2020}.} \\
    \quad \footnotesize{$^c$ The hardness ratio, HR, is defined as $S_{50-100\,\textrm{keV}}/S_{25-50\,\textrm{keV}}$ where $S$ represents the gamma-ray fluence in a given energy range as defined in the Swift/BAT GRB Catalog \citep{Lien2016}.} \\
\end{flushleft}
\end{table*}

\bsp	
\label{lastpage}
\end{document}